\def\spose#1{\hbox to 0pt{#1\hss}}

\def\multleft#1{\hbox to size{\vbox {\halign {\lft{##}\cr #1}}\hfill}\par}
\def\multright#1{\hbox to size{\vbox {\halign {\rt{##}\cr #1}}\hfill}\par}

\def\today{\ifcase\month\or January\or February\or March\or April\or May\or
      June\or July\or August\or September\or October\or November\or December\fi
      \space\number\day, \number\year}

\def\asec{$^{\prime\prime}$}

\def\Msun{\hbox{$\rm\thinspace M_{\odot}$}}

\def\H2{\hbox{H$_{2}$}}

\newcommand{\gtsim}{\mbox{{\raisebox{-0.4ex}{$\stackrel{>}{{\scriptstyle\sim}}
$}}}}

\newcommand{\fesc}{f_{\mathrm{esc}}}

\documentclass{mn2e}
\usepackage{times}
\usepackage{amssymb}
\usepackage{epsfig}
\usepackage{lscape}
\begin{document}
\hsize=6truein
          
\title[A robust sample of galaxies at redshifts 6.0$<$z$<$8.7]
{A robust sample of galaxies at redshifts $\bmath{6.0<z<8.7}$:\\ stellar
populations, star-formation rates and stellar masses}

\author[R.J.~McLure et al.]
{R. J. McLure$^{1}$\thanks{Email: rjm@roe.ac.uk}, J. S. Dunlop$^{1}$,
L. de Ravel$^{1}$, M. Cirasuolo$^{1,2}$, R.S. Ellis$^{3}$, \and
M. Schenker$^{3}$, B.E. Robertson$^{3}$, A.\,M.~Koekemoer$^{4}$, D.\,P.~Stark$^{5}$,
R.A.A. Bowler$^{1}$\\ 
\footnotesize\\
$^{1}$SUPA\thanks{Scottish Universities Physics Alliance}, Institute
for Astronomy, University of Edinburgh, Royal Observatory, Edinburgh
EH9 3HJ\\ 
$^{2}$Astronomy Technology Centre, Royal Observatory of
Edinburgh, Blackford Hill, Edinburgh EH9 3HJ\\ 
$^{3}$Department of Astronomy,
California Institute of Technology, Pasadena, CA 91125, USA\\
$^{4}$Space Telescope Science Institute, 3700 San Martin
Drive, Baltimore, MD 21218, USA\\
$^{5}$Kavli Institute
of Cosmology, University of Cambridge, Madingley Road, Cambridge, CB3
0HA, UK}

\maketitle

\begin{abstract}
We present the results of a photometric redshift analysis designed to
identify redshift $z\geq6$ galaxies from the near-IR HST imaging in
three deep fields (HUDF, HUDF09-2 \& ERS) covering a total area of 45
sq. arcmin. By adopting a rigorous set of criteria for rejecting
low-redshift interlopers, and by employing a deconfusion technique to
allow the available ultra-deep IRAC imaging to be included in the
candidate selection process, we have derived a robust sample of 70
Lyman-break galaxies (LBGs) spanning the redshift range
$6.0<z<8.7$. Based on our final sample we investigate the distribution
of UV spectral slopes ($f_{\lambda}\propto \lambda^{\beta}$), finding a
variance-weighted mean value of $\langle \beta \rangle=-2.05\pm 0.09$
which, contrary to some previous results, is not significantly bluer than
displayed by lower-redshift starburst galaxies. We confirm the
correlation between UV luminosity and stellar mass reported elsewhere,
but based on fitting galaxy templates featuring a range of
star-formation histories, metallicities and reddening we find that, at
$z\geq 6$, the range in mass-to-light ratio ($M_{\star}/L_{\rm UV}$) at a
given UV luminosity could span a factor of $\simeq 50$. 
Focusing on a sub-sample of twenty-one candidates with IRAC detections at $3.6\mu$m
we find that $L^{\star}$ LBGs at $z\simeq 6.5$ have a median stellar
mass of $M_{\star}=(2.1\pm1.1)\times10^{9}\Msun$ (Chabrier IMF) and a
median specific star-formation rate (sSFR) of $1.9\pm{0.8}$ Gyr$^{-1}$. 
Using the same sub-sample we have investigated 
the influence of nebular continuum and line emission, finding that for the majority of candidates 
(16 out of 21) the best-fitting stellar masses are reduced by less than a factor of 2.5.
However, galaxy template fits exploring a plausible range of
star-formation histories and metallicities provide no compelling
evidence of a clear connection between star-formation rate and stellar
mass at these redshifts. Finally, a detailed comparison of
our final sample with the results of previous studies suggests that,
at faint magnitudes, several high-redshift galaxy samples in the literature
are significantly contaminated by low-redshift interlopers.
\end{abstract}

\section{INTRODUCTION}
The goal of identifying and studying the nature of ultra high-redshift
galaxies remains one of the most important challenges in observational
cosmology, and holds the key to furthering our understanding of the
earliest stages of galaxy evolution and unveiling the nature of the
sources responsible for cosmic reionization. 

Observational constraints provided by the Gunn-Peterson trough in the
spectra of high-redshift quasars (e.g Fan et al. 2006) 
suggest that reionization was coming to an end at $z\simeq 6$ (Becker
et al. 2007). Moreover, optical  polarization measurements from the WMAP
experiment indicate that reionization began at $z\simeq 11$ if it is assumed to be a
single, rapid event (Dunkley et al. 2009).  Consequently, it is now apparent that to improve our
understanding of cosmic reionization, and to unveil the earliest epoch
of galaxy formation, it is necessary to extend studies of
high-redshift galaxies into the redshift $7<z<10$ r\'{e}gime (e.g. Robertson et al.~2010). 

Given our existing knowledge of the evolution of the galaxy
luminosity function in the redshift interval $5.0<z<6.5$ (e.g. Bouwens et al. 2007;
McLure et al. 2009) it is clear that achieving this aim requires
ultra-deep near-IR imaging, reaching detection limits of
$26<m_{AB}<30$. At the bright end of this range, wide-field,
ground-based imaging has a unique contribution to make, and has
recently allowed the luminosity function, the clustering
properties and the stellar populations of luminous Lyman-break galaxies (LBGs:
McLure et al. 2009; Ouchi et al. 2009; Grazian et al. 2010) and Lyman
alpha emitters  (LAEs: Ouchi et al. 2010; Ota et al. 2010; Ono et
al. 2010; Nakamura et al. 2010) to be studied in detail. Indeed, the
importance of ground-based imaging and spectroscopy has recently been
highlighted by the spectroscopic confirmation of two LBGs at $z=7.01$
and $z=7.11$ by Vanzella et al. (2010). However, in advance of 30-m
class ground-based telescopes, it is clear that routinely identifying
and studying sub-L$^{\star}$ galaxies at $z\geq7$ is only possible using
space-based imaging.

Consequently, the unparalleled near-IR sensitivity provided by the
new WFC3 camera, installed on the Hubble Space Telescope (HST) in late
2009, has proven to be a crucial breakthrough in  high-redshift galaxy
studies. Indeed, despite only covering an area of $\simeq 4.5$
sq. arcmin, the unprecedented depth ($m_{AB}\simeq 29$, $5\sigma$)  of the first tranche of WFC3/IR imaging of the
Hubble Ultra Deep Field (HUDF; GO-11563)  led to a raft of early science
papers investigating the number densities, luminosity functions,
stellar masses and stellar populations of $6.5<z<8.5$ galaxies
(e.g. Bouwens et al. 2010ab; Oesch et al. 2010; McLure et al. 2010;
Yan et al. 2010; Labb\'{e} et al. 2010; Finkelstein et al. 2010; Bunker et
al. 2010).  Interestingly, Lehnert et al. (2010) have recently claimed
the tentative detection of Ly$\alpha$ emission  at $z=8.56$ in a
WFC3/IR candidate in the HUDF, originally identified by Bouwens et
al. (2010a) and McLure et al. (2010). Although the large rest-frame
equivalent width (EW$\simeq 210$\AA) of the Ly$\alpha$ emission  line
suggests that, if confirmed, this object must be a decidedly atypical
example of a $z\geq6$ LBG (Stark, Ellis \& Ouchi 2010), the location of
the claimed Ly$\alpha$ emission line is in good agreement with
the original photometric redshifts derived by McLure et al. (2010) and
Finkelstein et al. (2010); $z_{phot}=8.45\pm0.50$ and
$z_{phot}=8.61\pm{0.35}$ respectively.

In addition to the WFC3/IR imaging of the HUDF, another key dataset
has  been the WFC3/IR imaging taken as part of the Early Release
Science extra-galactic programme (ERS; GO-11359)
which, although substantially shallower than the HUDF WFC3/IR imaging
($m_{AB}\simeq 27.5$, $5\sigma$), covers an area approximately ten
times larger. By combining the HUDF and ERS datasets to
obtain greater dynamic range in UV luminosity, Bouwens et al. (2010b)
and Labb\'{e} et al. (2010) investigated the relationship between the UV
spectral slope ($f_{\lambda}\propto\lambda^{\beta}$) and UV
luminosity. Both studies find a correlation, with $\beta$ changing
from $\beta\simeq -2$ (a typical value for lower-redshift starburst galaxies)
at $M_{1500}\simeq -20.5$, to extremely blue values of $\beta \simeq -3$ at $M_{1500}\simeq -18.5$. 
As discussed by Bouwens et al. (2010b), although dust-free, low metallicity models can
produce slopes of $\beta \simeq-3$, they can only do so under the
assumption that the ionising photon escape fraction is high
($f_{esc}>0.3$) and that, correspondingly, the contribution from nebular continuum emission is low. 

Based on stacking the ACS+WFC3/IR+IRAC photometry of $z\simeq 7$ LBG candidates in the 
HUDF and ERS datasets, Labb\'{e} et al (2010) find the same correlation
between $M_{1500}$ and spectral slope $\beta$ as Bouwens et
al. (2010b). However, Labb\'{e} et al. (2010) conclude that
it is not possible to reproduce both the blue spectral slopes and
significant $\lambda_{rest}\simeq 4000$\AA\, spectral breaks displayed by the faintest $z\simeq 7$
LBG candidates, without recourse to episodic starformation
histories and/or a significant contribution from nebular line emission. 
Indeed, Ono et al. (2010) also conclude that nebular line emission may 
be necessary to reproduce the observed $J-$m$_{3.6}$ colour in a
stack of $z\simeq 6$ LAE photometry.

Despite the uncertainties, one observational result that has received
significant attention recently is the apparent relationship between star-formation rate
and stellar mass. Both Labb\'{e} et al. (2010) and Gonz\'{a}lez et al. (2010) find an
approximately linear correlation between stellar mass and
star-formation rate at $z\simeq 7$, consistent with the results derived by Stark et
al. (2009) for LBGs in the redshift range $4<z<6$. As a result,
Labb\'{e} et al (2010) and Gonz\'{a}lez et al. (2010) conclude that the
specific star-formation rate (sSFR) of $z\simeq 7$ LBGs is remarkably
constant (sSFR $\simeq 2$ Gyr$^{-1}$), and consistent
with the value of sSFR $\simeq 2.5-4.5$ Gyr$^{-1}$ observed in star-forming galaxies at $z\simeq 2-3$ 
(e.g. Daddi et al. 2007; Magdis et al. 2011). As
previously discussed by Stark et al. (2009), a natural explanation of
this observation would be to invoke a star-formation rate which
exponentially increases with time although, as shown by Finlator, Oppenheimer \& Dav\'{e} (2011), 
star-formation histories of this type may have difficulty in reproducing some of the most 
extreme Balmer breaks reported in the literature at $z\geq6$.

The majority of previous studies which have investigated the
high-redshift galaxy population using WFC3/IR imaging have relied on 
traditional colour-cut, or ``drop-out'', selection techniques. In
contrast, the principal motivation for this paper is to investigate what can be
learned about the $z\geq 6$ galaxy population by fully exploiting the 
excellent multi-wavelength (ACS+WFC3/IR+IRAC) data which is now
available over an area of $\simeq 45$ sq. arcmin. Rather than
applying standard ``drop-out'' criteria, in this work
we continue to pursue the strategy we have previously adopted (McLure
et al. 2006; 2009; 2010) and employ a template-fitting, photometric
redshift analysis to select our final high-redshift galaxy sample. 
A key new element in this strategy is our development of a deconfusion
algorithm capable of providing the robust IRAC photometry necessary
for improved photometric redshift and stellar-mass estimates.

In principle this technique should have several advantages over the standard LBG
``drop-out'' selection. Firstly, by employing all of the
available multi-wavelength data, including the IRAC photometry, 
it is possible to make optimal use of the available information. Secondly, by 
avoiding any colour pre-selection this approach should be less biased 
towards simply selecting the very bluest galaxies at
high-redshift. This second point is potentially crucial within the
context of investigating the claims of ultra-blue UV spectral slopes 
for LBG candidates at $z\geq 6$ (see Section 4). Finally, an SED fitting analysis also provides an estimate of the 
photometric redshift probability density function, $P(z)$, and
therefore allows the prevalence, and significance, of competing photometric redshift solutions 
at low redshift to be transparently investigated.

The primary motivation of this paper is therefore to construct
the most robust sample possible using the available data and
techniques reviewed above, in order to critically address
some of the claims about the properties of the $z>6$ population
newly-found with HST. The structure of the paper is as follows. In Section 2 we 
describe the available data in each of the three fields, including a
brief description of our IRAC deconfusion algorithm. In Section 3 we
describe our initial candidate selection, photometric redshift
analysis and the construction of the final catalogue of robust
candidates. In Sections 4 and 5 we investigate the UV spectral slopes, stellar masses and
star-formation rates of the final robust sample. In Section 6 we
perform a detailed comparison of our final robust sample with samples
previously derived in the literature, exploring the reasons behind any apparent
discrepancies. In Section 7 we provide a summary of our main
conclusions. In the Appendix we provide a full
description of our IRAC deconfusion procedure, full photometry and
grey-scale postage stamp images for each high-redshift candidate and
individual plots illustrating the results of our SED
fitting. All magnitudes are quoted in the
AB system (Oke \& Gunn 1983) and all calculations assume
$\Omega_{0}=0.3, \Omega_{\Lambda}=0.7$ and $H_{0}=70$
kms$^{-1}$Mpc$^{-1}$. 

\section{DATA}
The analysis in this paper relies on the publicly available optical,
near-IR and mid-IR imaging data covering the HUDF, HUDF09-2 and ERS fields. In
this section we briefly describe the relevant details of the various imaging
datasets, and the depth analysis which was performed in order to
attribute accurate error estimates to the candidate photometry. The
basic properties of the three fields are listed in Table 1.

\begin{figure*}
\centerline{\psfig{file=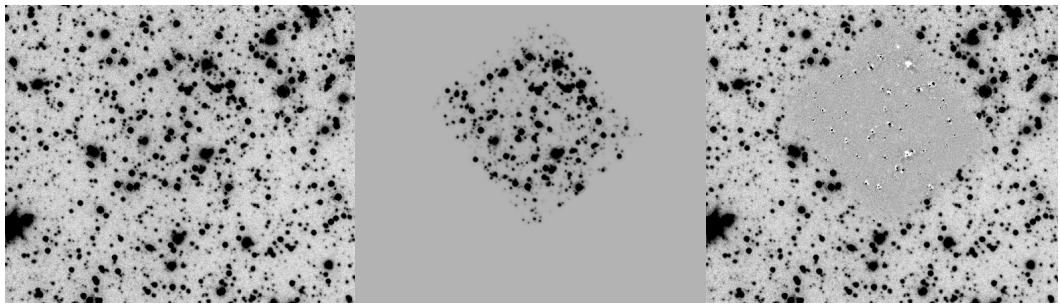,width=14.0cm,angle=0}}
\caption{Illustration of the IRAC deconfusion algorithm. The left-hand panel shows the inverse-variance weighted stack of the epoch1+epoch2 $4.5\mu$m imaging covering the HUDF. The middle panel shows the best-fitting model of the IRAC data, based on using the $H_{160}$ WFC3/IR imaging to provide model templates, and a matrix inversion procedure to determine the 
best-fitting template amplitudes (see Appendix A for full details). The right-hand panel shows the model subtracted image (note that the WFC3/IR imaging does not cover the full area of the HUDF).}
\end{figure*}

\subsection{WFC3/IR imaging}
The WFC3/IR imaging of both the HUDF and the HUDF09-2 fields was
taken as part of the public treasury programme GO-11563
(P.I.=Illingworth)\footnote{We do not consider the third WFC3/IR
pointing obtained as part of GO-11563, HUDF09-1, because deep IRAC
imaging of this field is not currently available.} and consists of
single pointings ($\simeq 4.5$ sq. arcmin) of the WFC3/IR camera in
the F105W, F125W and F160W filters (hereafter referred to as $Y_{105},
J_{125}$ and $H_{160}$). The WFC3/IR dataset in the ERS field was taken
as part of the public programme GO-11359 (P.I.=O'Connell) and consists
of a mosaic of 10 pointings of the WFC3/IR camera in the F098M
($Y_{098}$), $J_{125}$ and $H_{160}$ filters (Windhorst et
al. 2011) \footnote{All of the WFC3/IR data utilised in this paper
conforms to the nominal flight zeropoints, i.e. $Y_{098M}=25.68,
Y_{105W}=26.27, J_{125W}=26.25,H_{160W}=25.96$. }. The WFC3/IR data
were calibrated using {\sc calwf3} and subsequently combined
using MultiDrizzle (Koekemoer et al. 2002) as summarized in
McLure et al. 2010); full details are presented in Koekemoer et
al. (2011). The final mosaics
have PSFs with FWHM in the range $0.15$\asec$-0.18$\asec depending on
the filter, and were drizzled onto a final grid of 0.06\asec/pix. In
the case of the HUDF and ERS mosaics, the final astrometry was matched to
that of the publicly available reductions of the optical ACS imaging
of the UDF (Beckwith et al. 2006) and GOODS-S (GOODSv2.0; Giavalisco
et al. 2004) respectively. The astrometry for the final mosaics of the
HUDF09-2 field was matched to the $J-$band imaging of GOODS-S taken
as part of the MUSYC survey (Cardamone et al. 2010), with a typical
r.m.s. accuracy of $\simeq 0.1$\asec. The WFC3/IR imaging of the ERS
analysed in this paper consists of the data comprising the completed
programme. However, for the HUDF09-2 and HUDF fields we make use of
the epoch 1 observations, which consist of the data publicly available
as of February 2010 and  August 2010 respectively.

\subsection{ACS imaging}
For the HUDF and ERS fields the ACS data used in this study consists
of the publicly available reductions of the F435W, F606W, F775W and
F850LP (hereafter $B_{435}, V_{606}$, $i_{775}$ and $z_{850}$) imaging
of the HUDF (Beckwith et al. 2006) and GOODS-S (v2.0; Giavalisco et
al. 2004).  The ACS imaging covering the HUDF09-2 field is our own
reduction (based on {\sc calacs} and MultiDrizzle) of the $V_{606},
i_{775}$ and $z_{850}$ imaging obtained as part of the UDF05 programme (Oesch et
al. 2007). All of the optical ACS imaging was re-sampled to a
0.06\asec/pix grid to match the WFC3/IR data, and the astrometry of
the ACS data covering HUDF09-2 was also registered to match the MUSYC imaging
of GOODS-S.

\subsection{IRAC imaging}
For the HUDF and ERS fields we make use of the publicly available
reductions (v0.30) of the $3.6\mu$m and $4.5\mu$m IRAC imaging
obtained as part of the GOODS survey (proposal ID 194, Dickinson et
al.; in preparation). The IRAC data covering the ERS consists
of approximately 23 hours of on-source integration in both the
$3.6\mu$m and $4.5\mu$m bands. For the HUDF we performed an inverse
variance weighted stack of the overlapping region of the epoch 1 and epoch 2
imaging, producing final mosaics consisting of approximately 46 hours
of on-source integration at $3.6\mu$m and $4.5\mu$m. For the HUDF09-2
we re-registered and stacked the {\sc mopex} reductions of the $3.6\mu$m
imaging obtained via proposal ID 30866 (P.I. R. Bouwens) producing a
final $3.6\mu$m mosaic with an on-source integration time of
approximately 33 hours. The consistency of the IRAC photometry 
was checked via reference to the SIMPLE (Damen et al. 2010)
imaging at $3.6\mu$m and $4.5\mu$m of the E-CDFS, which overlaps all
three fields. The astrometry of the IRAC imaging in all three fields
was registered to that of the corresponding $H_{160}$ WFC3/IR imaging.

\subsubsection{IRAC deconfusion}
A key new feature of the analysis undertaken in this paper is the
inclusion of the IRAC photometry in the candidate high-redshift galaxy
selection procedure. Due to its depth, and comparatively
broad PSF (FWHM $\simeq 1.5$\asec), the $3.6+4.5\mu$m IRAC imaging
covering the three fields of interest is heavily confused, making the
process of obtaining aperture photometry matched to the optical/nearIR
HST imaging non trivial. In order to achieve this aim it is therefore
necessary to pursue some form of deconfusion process, which allows the
IRAC imaging to be utilised beyond the natural confusion
limit. Although there are several techniques which can be
used to deconfuse IRAC imaging (see Appendix A)
we have developed our own software which uses the WFC3/IR imaging data
to provide normalized templates for each object in the field and then,
via a transfer function, produces synthetic IRAC images on the native
0.6\asec/pix plate scale. Through a matrix inversion procedure, the
amplitude (or total flux) of each template can be simultaneously fitted
to produce the optimal reproduction of the observed IRAC image (see
Fig. 1). As a result of this procedure it is effectively possible to extract
accurate aperture photometry from the $3.6\mu$m and $4.5\mu$m IRAC
imaging at the spatial resolution of the WFC3/IR imaging. An
additional advantage of this approach is that it naturally provides
robust uncertainties on the delivered flux measurements, which depend
both on the signal-to-noise ratio of the IRAC detection {\it and} the local
level of confusion in the IRAC image.

\begin{figure}
\centerline{\psfig{file=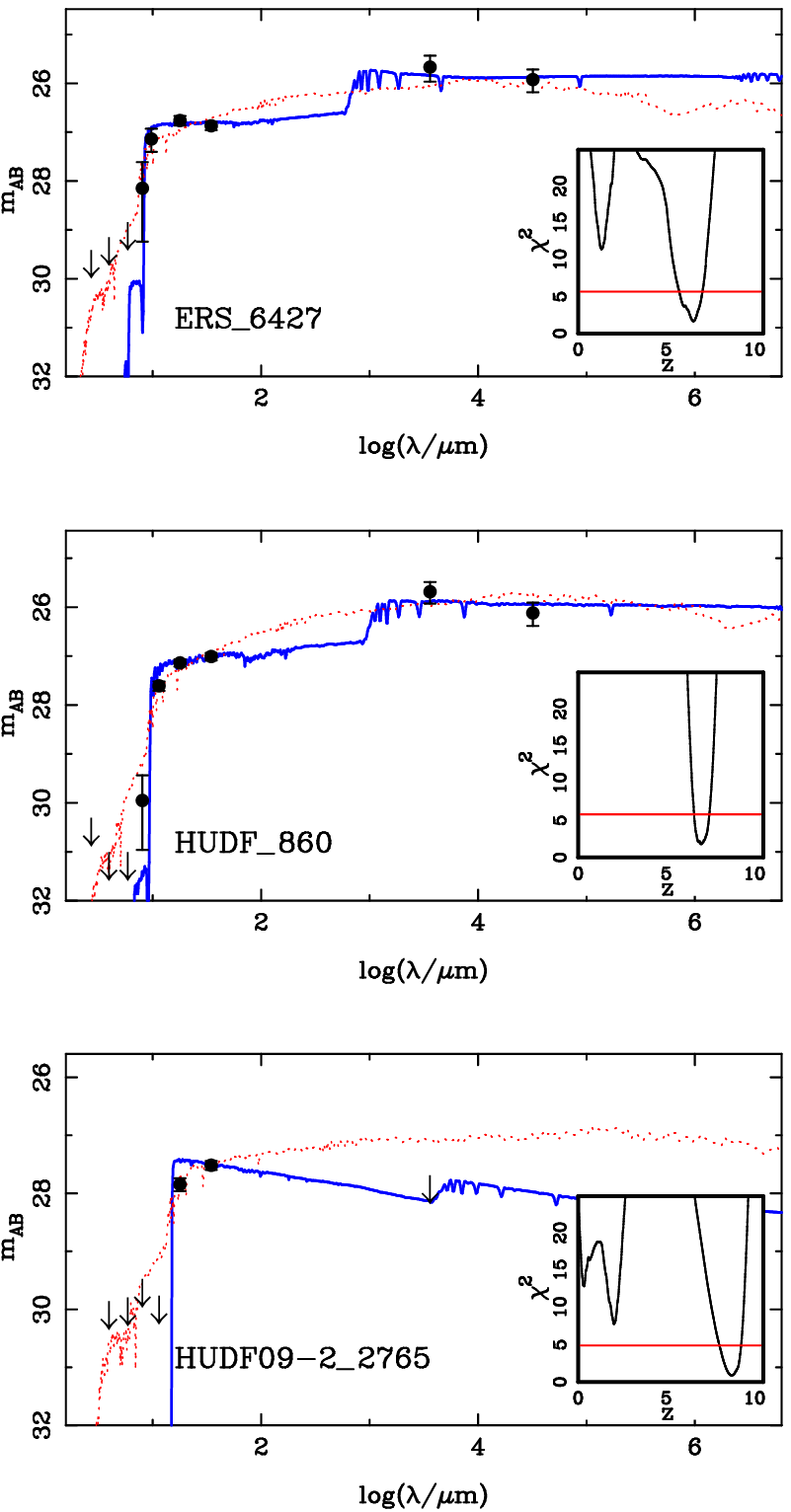,width=7.0cm,angle=0}}
\caption{Example plots showing the results of our SED-fitting 
for three objects from our final robust sample at redshifts $z_{phot}~=~6.5, 7.0\, \&\, 8.7$ (top
to bottom). In each plot the thick (blue) line shows the best-fitting
high-redshift galaxy template, and the thin dotted (red) line shows
the best-fitting alternative solution at low redshift. In each panel
the inset shows the value of $\chi^{2}$ as a function of redshift 
(marginalized over all other free parameters). The horizontal line in
each inset panel highlights the location of ($\chi^{2}_{min}+4$) which
corresponds to our requirement that a candidate is considered
``robust'' only if any alternative low-redshift solutions can be ruled-out at the $\geq 95\%$ confidence
level. In each example,
the combination of image depth and wavelength coverage allows us to
rule-out any alternative low-redshift solutions with high
confidence. The bottom panel (HUDF09-2$\_2765$) provides a good
example of a situation where even a robust upper limit to the IRAC $3.6\mu$m flux provides a powerful redshift discriminant.}
\end{figure}

\begin{table*}
\caption{The results of the image depth analysis described in Section
3.2. Columns 1-4 list the survey field names, central coordinates and
areas (in sq. arcmin). The areas listed in column 4 correspond to the
areas of WFC3/IR imaging which were actually useful for the analysis performed in
this paper (i.e excluding array edges and areas which are not covered
by ACS optical imaging). The remaining columns list the median
$5\sigma$ depths in each available ACS+WFC3/IR+IRAC filter. The depths are all referenced 
to a 0.6\asec diameter aperture, and are not aperture corrected. Due to the fact that the IRAC depths 
are determined via a deconfusion
process (see Section 2.3.1), the local IRAC depths can differ
significantly from the median values listed here.}
\begin{tabular}{lccrcccccccccc}
\hline
Field & RA(J2000)&DEC(J2000)&Area&
$B_{435}$&$V_{606}$&$i_{775}$&$z_{850}$&$Y_{098}$&$Y_{105}$&$J_{125}$&$H_{160}$&
$3.6\mu$m & $4.5\mu$m\\
\hline
HUDF  &03:32:38.5&$-27$:46:57.0&  4.5 & 29.04 & 29.52 & 29.19 & 28.54
& $-$  & 28.59 & 28.67 &28.73&26.3  & 25.9 \\ 

HUDF09-2&03:32:23.4&$-27$:42:52.0&  4.5 & $-$  & 28.49 & 28.22 &
28.06 & $-$& 28.24 & 28.60 &28.49&26.2  & $-$ \\ 
ERS  &03:33:05.5&$-27$:51:21.6& 36.5 & 27.68 & 27.87 & 27.29 & 27.06 & 27.26 & $-$  & 27.66
&27.40&26.0  & 25.6  \\
\hline
\end{tabular}
\end{table*}

\section{Candidate selection}
The process of candidate selection can be broken down into three
separate stages: object detection and photometry, photometric redshift analysis and
sample cleaning. Each stage in the process is described below.

\subsection{Object detection and photometry}
The initial catalog construction process was identical in each of the
three fields, and relied on {\sc sextractor} v2.5.0 (Bertin et al. 1996). 
Preliminary catalogues were constructed in which object detection was
performed in the $Y_{105W}/Y_{098M}$, $J_{125W}$ and $H_{160W}$ bands, using an aggressive set of {\sc sextractor}
parameters, with matched photometry extracted from the corresponding
ACS imaging by running {\sc sextractor} in dual image mode. The
separate catalogues were then concatenated to produce a master catalog 
of unique objects in each of the three fields.

In order to avoid biases which can be introduced by adopting small
photometric apertures, all of the analysis in this paper is based on
0.6\asec diameter aperture photometry. For the purposes of the photometric redshift
analysis, the fluxes from the 0.6\asec diameter aperture photometry
are not corrected to total, but the WFC3/IR and IRAC fluxes are
corrected by small amounts (2\%-10\%) to account for aperture losses {\it relative} to the ACS imaging.

\subsection{Depth Analysis}
A crucial part of the analysis necessary to identify robust
high-redshift candidates is the derivation of accurate photometric
uncertainties in each band. This was achieved by first producing a
so-called $\chi^2$ image (Szalay, Connolly \& Szokoly 1999) of the registered
optical+nearIR images of each field to identify which pixels are genuine ``blank
sky''. Secondly, a grid of 0.6\asec diameter apertures was placed in
the blank sky regions on each image. Thirdly, in order to determine
the local image depth for each candidate, in each filter, the
r.m.s. aperture-to-aperture variation was determined, by examining the
distribution of the nearest 50 blank apertures. In this fashion, we
are able to determine a local depth measurement for each individual
candidate. For information, the median $5\sigma$ depths for each
field are listed in Table~1.

\subsection{Photometric redshift analysis}
To perform the SED analysis necessary for this study we have developed
a new, bespoke, template-fitting code. The primary motivation for developing this new 
code was to provide the freedom to explore the relevant multi-dimensional parameter space in
detail, investigating the impact of different SED templates, IMFs,
dust attenuation prescriptions and IGM absorption recipes. Moreover, by employing our 
own software it is possible to have full control over which derived quantities are provided 
as output, and the exact details of how the template fitting is performed. 
For example, our new code performs the SED fitting based on flux
densities ($f_{\nu}$), rather than magnitudes, 
which has the advantage of allowing the flux errors to be dealt with
in a rigorous manner. Moreover, if necessary, the new code offers the possibility of 
fitting the input photometry with multi-component stellar populations, each with separate metallicities and/or 
dust attenuation prescriptions. 

For the purposes of this study we employed the Bruzual \& Charlot (2003)
and Charlot \& Bruzual 2007 (priv. comm) stellar evolution models (hereafter BC03 \& CB07), considering
models with metallicities ranging from solar ($Z_{\odot}$) to $1/50$th
solar ($0.02Z_{\odot}$). 
Models with instantaneous bursts of star-formation, constant
star-formation and star-formation rates exponentially declining with
characteristic timescales in the range 50~Myrs~$<~\tau~<$~10~Gyrs were all considered. 
The ages of the stellar population models were allowed to range from
10 Myrs to 13.7 Gyrs, but were required to be less than the age of the
Universe at each redshift. Dust reddening was described by the Calzetti et al. (2003)
attenuation law, and allowed to vary within the range
$0.0<A_{V}<2.5$ magnitudes. Inter-galactic medium absorption short-ward of Ly$\alpha$
was described by the Madau et al. (1995) prescription, and a Chabrier
(2003) IMF was assumed in all cases \footnote{Derived quantities such
as stellar masses and star-formation rates can be converted to a
Salpeter (1955) IMF by multiplying by a factor of 1.8.}. In Fig. 2 we
show example SED fits for three objects (one from each field) covering
the redshift range $6.5<z_{phot}<8.7$.

\subsection{Sample cleaning}
Based on the results of the photometric redshift fitting, all objects
which displayed a statistically acceptable solution at $z_{phot}\geq
4.5$ were retained,  while those with no acceptable solution at high
redshift were excluded. In each of the three fields, this initial
screening process removed more than 90\% of the original input
catalogues. Following the initial photometric  redshift fitting, the
remaining samples of potential high-redshift candidates were manually
screened to remove artefacts (e.g. diffraction spikes), edge effects
and spurious candidates such as high-surface brightness features
within the extended envelopes of luminous low-redshift galaxies.

\begin{figure}
\centerline{\psfig{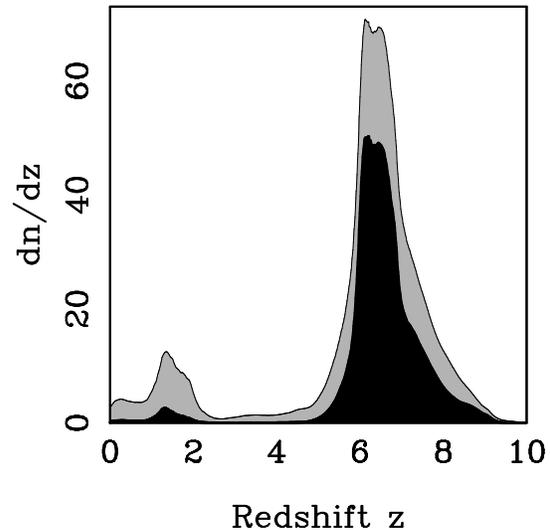}}
\caption{In black we show the redshift distribution of our final
robust sample, which has been calculated by summing the estimated redshift probability density
function of each candidate. This is our best estimate of the redshift
distribution of the N=70 objects which satisfy all three of the criteria listed in Section 3.4.1. 
In grey we show the redshift distribution of the N=130 objects which
satisfy the first criterion listed in Section 3.4.1 (i.e. a
statistically acceptable redshift solution at $z_{phot}\geq 6.0$) but
fail the other two criteria (see text for details).}
\end{figure}

\subsubsection{Final candidate sample}
From a practical perspective, the primary goal of this study is to 
produce a robust sample of high-redshift galaxy
candidates at $z~\geq~6$. In order to achieve this aim, three
criteria were applied to the remaining potential high-redshift candidates:
\begin{itemize}
\item{Statistically acceptable redshift solution at $z_{phot}\geq 6.0$}
\item{Secondary redshift solution excluded at $\geq 95\%$ confidence}
\item{Integrated probability $\int^{z=10}_{z=6}P(z^{\prime})\delta z^{\prime} \geq 0.5$} 
\end{itemize}
\noindent
The first criterion simply restricts our final sample to those objects
for which the best-fitting SED template  lies at $z_{phot}\geq6$. The second
criterion rejects those objects for which the competing low-redshift
solution cannot  be excluded at high confidence. Specifically, this
criterion is enforced by insisting that
the  $\Delta \chi^2$ between the primary and secondary photometric
redshift solution (following marginalization  over
all other relevant parameters) is $\Delta \chi^2 \geq 4$. The final criterion is designed to
exclude a small number of candidates with relatively flat $P(z)$
distributions for which, despite having a primary photometric redshift
solution at $z_{phot}\geq6$, the majority  of their integrated
probability density distribution function lies at $z_{phot}\leq 6$. We note
that this final criterion is very similar to that employed by
Finkelstein et al. (2010) in their HUDF analysis.

Our final robust sample of $z\geq 6$ galaxy candidates consists of
N=70 objects, spanning the redshift range $6.0<z<8.7$ and covering
more than a factor of ten in intrinsic UV luminosity, from $-18.2<M_{1500}<-21.2$ \footnote{The final sample
consists of 73 objects if 3 additional objects are included which only satisfy
our selection criteria when Ly$\alpha$ emission is included in the SED
templates.}. It is perhaps worth pointing out that if we had only
insisted on a statistically acceptable primary photometric redshift
solution at $z_{phot}\geq6.0$, the final sample would contain N=130
candidates. It should be stressed that it is likely that a
significant fraction of the excluded objects are
indeed $z\geq 6$ galaxies (see Fig. 3), it is simply that with the data in-hand, it
is not possible to consider them as robust candidates. 

Due to the fact that LBGs at $z\geq 6$ are 
necessarily young galaxies, the differences between the SED fits provided by the BC03 and CB07 models are negligible. Consequently, to ease the comparison with previous studies, throughout the rest of the paper 
we adopt the results of the SED-fitting analysis based on the BC03 models.
The final robust samples in each of the three fields, along with the
best-fitting photometric redshift solutions and various other derived
parameters are presented in Tables 2, 3 \& 4. Note that, although we
make no further use of this information throughout the rest of the paper, in Tables
2, 3 \& 4 we also list the best-fitting photometric redshift for each
candidate if Ly$\alpha$ emission with a rest-frame EW$_{0}$ in the range
0\AA\,$<$ EW$_{0}<$ 240\AA\, is included as an extra free parameter
in the SED-fitting procedure. This information is provided to indicate
the maximum plausible redshift for each candidate. The three candidates
which are listed separately at the bottom of Tables 3 \& 4 only
pass our criteria as robust $z\geq 6$ candidates if Ly$\alpha$
emission is included in the SED-fitting procedure, and are not
included in any of the subsequent analysis. The 0.6\asec diameter
aperture photometry for each 
candidate is listed in  Tables B1, B2 \& B3 of Appendix B, along with plots of the best-fitting SED templates and 
grey-scale optical-nearIR postage stamps.

\begin{figure}
\centerline{\psfig{file=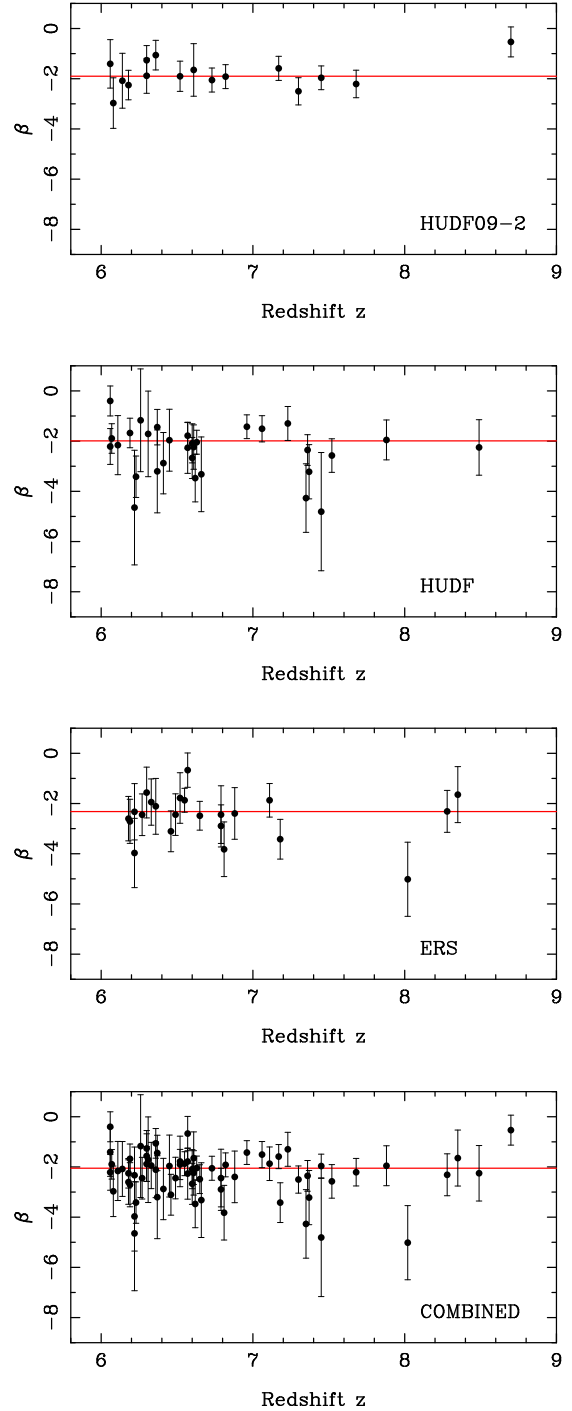,width=7.3cm,angle=0}}
\caption{Plots of UV spectral slope ($\beta$) versus redshift for the
final robust sample of seventy objects at $z_{phot}\geq 6$. The top three panels
show $\beta$ versus redshift for the three separate survey fields,
while the bottom panel shows $\beta$ versus redshift for the full
combined sample. In each panel the horizontal line shows the variance-weighted mean value of $\beta$ (see Table 5).}
\end{figure}

\begin{table*}
\caption{Details of the final high-redshift sample in the HUDF field. The first three columns list the ID number and 
coordinates of each candidate. Columns four, five and six list the
photometric redshifts, the $1\sigma$ uncertainty on the photometric
redshifts and the values of $\chi^{2}$ for the best-fitting SED
template. Column seven lists the difference in $\chi^{2}$ between the
best-fitting high-redshift SED template and the alternative
photometric redshift solution at low redshift (typically at
$1.0<z_{phot}<2.5$). Column eight lists the absolute magnitude of the
best-fitting SED template, where $M_{1500}$ is measured using a 100\AA\,-wide filter 
centred on a rest-frame wavelength of 1500\AA. 
Column nine lists the value of the UV spectral slope
($f_{\lambda}\propto \lambda^{\beta})$ for each candidate, derived using
the formulae listed in Section 4. Column ten lists the best-fitting photometric
redshift derived including Ly$\alpha$ emission as an additional
free parameter, and column eleven lists the logarithm of luminosity of
the corresponding best-fitting Ly$\alpha$ emission line (in ergs
s$^{-1}$). Column twelve lists the results of cross-checking each
candidate against existing literature studies of the WFC3/IR data in
the HUDF field. Matches were found with the following papers: M
(McLure et al. 2010); B (Bouwens et al. 2011; robust), Bp (Bouwens et al. 2011; potential), F (Finkelstein et
al. 2010), W (Wilkins et al. 2010), Y (Yan et al. 2010) and L (Lorenzoni et al. 2011).}
\begin{tabular}{lccccrrccccl}
\hline
ID & RA(J2000)&DEC(J2000)&z&$\Delta
   z$&$\chi^2$&$\Delta \chi^{2}$&M$_{1500}$ &$\beta$&z$_{Ly\alpha}$&
   $\log(L_{Ly\alpha})$   &Literature\\ 
\hline
HUDF\_1344 &03:32:36.63&$-27$:47:50.1&6.06 & 5.89$-$6.17& 1.1 &27.6&$-19.3$ & $-2.2\pm0.7$ & 6.16 &42.2&M\\

HUDF\_1016 &03:32:35.06&$-27$:47:40.2&6.06 & 5.97$-$6.15& 4.3 &32.0&$-19.6$ & $-0.4\pm0.6$& 6.33 &42.4&M\\

HUDF\_522  &03:32:36.47&$-27$:46:41.4&6.07 & 5.98$-$6.15& 4.5 &151.3&$-20.8$ &$-1.9\pm0.6$& 6.07 &$-$ &$-$ \\

HUDF\_2622 &03:32:36.64&$-27$:47:50.2&6.11 & 5.95$-$6.38& 1.3 & 13.4&$-18.8$ &$-2.2\pm1.2$& 6.43 &42.3 &M\\

HUDF\_796  &03:32:37.46&$-27$:46:32.8&6.19 & 5.86$-$6.31&1.4&46.0&$-19.9$&$-1.7\pm0.6$&6.50&42.7&M,F\\

HUDF\_2836 &03:32:35.05&$-27$:47:25.8&6.22 & 5.97$-$6.43& 0.8 & 7.4 &$-18.4$ &$-4.6\pm2.3$& 6.51 &42.1&M\\

HUDF\_1692 &03:32:43.03&$-27$:46:23.6&6.23 & 6.11$-$6.34& 3.1 & 38.5&$-19.3$ &$-3.4\pm0.8$& 6.49 &42.4&M\\

HUDF\_2743 &03:32:36.52&$-27$:46:42.0&6.26 & 5.80$-$6.72& 0.5 & 4.0  &$-18.2$ &$-1.2\pm2.0$& 6.40 &41.7&M,Y\\

HUDF\_2316 &03:32:44.31&$-27$:46:45.2&6.31 & 6.03$-$6.54& 1.2 & 9.1  &$-18.7$ &$-1.7\pm1.7$&6.30&$-$&M\\

HUDF\_2281 &03:32:39.79&$-27$:46:33.7&6.37 & 6.11$-$6.57& 0.3&5.7    &$-18.6$ &$-3.2\pm1.7$&6.35&$-$&M\\

HUDF\_1442 &03:32:42.19&$-27$:46:27.8&6.37 & 6.17$-$6.55& 6.4&11.2   &$-19.2$ &$-1.4\pm0.7$&6.43&41.7&M,F,W,B\\

HUDF\_2324 &03:32:41.60&$-27$:47:04.5&6.41 & 6.18$-$6.60& 0.6&5.5    &$-18.6$ &$-2.9\pm1.4$&6.40&$-$&B\\

HUDF\_2672 &03:32:37.80&$-27$:47:40.4&6.45 & 6.14$-$6.67& 0.5&7.9    &$-18.6$ &$-2.0\pm1.2$&6.81&42.4&M\\

HUDF\_1818 &03:32:36.38&$-27$:47:16.3&6.57 & 6.35$-$6.72& 2.1&17.3   &$-19.1$ &$-2.3\pm1.0$&7.05&42.7&M,F,W,B,Y\\

HUDF\_1473 &03:32:36.77&$-27$:47:53.6&6.57 & 6.42$-$6.71& 2.1&24.8   &$-19.2$ &$-1.8\pm0.5$&6.99&42.6&M,F,W,B\\

HUDF\_1730 &03:32:43.78&$-27$:46:33.7&6.60 & 6.37$-$6.84& 0.5&11.5   &$-19.1$ &$-2.7\pm0.8$&6.59&$-$&M,F,W\\

HUDF\_1632 &03:32:37.44&$-27$:46:51.2&6.60 & 6.40$-$6.74& 0.7&13.3   &$-19.1$ &$-2.1\pm0.8$&6.60&$-$&M,F,W,B,Y\\

HUDF\_2084 &03:32:40.57&$-27$:46:43.6&6.61 & 6.39$-$6.80& 2.6&11.9   &$-18.8$ &$-2.2\pm0.9$&7.13&42.6&M,F,W,B,Y\\

HUDF\_1995 &03:32:39.58&$-27$:46:56.5&6.62 & 6.31$-$6.91& 4.2&6.2    &$-18.9$ &$-3.5\pm0.9$&6.60&$-$&M,F,B,Y\\

HUDF\_658  &03:32:42.56&$-27$:46:56.6&6.63 & 6.53$-$6.79& 1.4&81.9   &$-20.5$ &$-2.0\pm0.5$&6.85&42.5&M,F,W,B,Y\\

HUDF\_2701 &03:32:41.82&$-27$:46:11.3&6.66 & 6.35$-$6.91& 2.3&4.5    &$-18.5$ &$-3.3\pm1.5$&6.88&42.0&F,W,Bp,Y \\

HUDF\_860  &03:32:38.81&$-27$:47:07.2&6.96 & 6.72$-$7.23& 1.8&31.8   &$-20.0$ &$-1.4\pm0.5$&6.96&$-$&M,F,W,B,Y\\

HUDF\_1102 &03:32:39.55&$-27$:47:17.5&7.06 & 6.75$-$7.42& 2.5&7.1    &$-19.7$ &$-1.5\pm0.5$&7.06&$-$&M,F,B,Y\\

HUDF\_1419 &03:32:43.13&$-27$:46:28.5&7.23 & 6.80$-$7.48& 5.9&9.1    &$-19.2$ &$-1.3\pm0.7$&7.95&42.8&M,F,W,B,L,Y\\

HUDF\_2641 &03:32:39.73&$-27$:46:21.3&7.35 & 6.97$-$7.76& 1.2&7.2    &$-18.7$ &$-4.3\pm1.4$&8.06&42.6&M,F,B,Y\\

HUDF\_1962 &03:32:38.36&$-27$:46:11.9&7.36 & 6.80$-$7.73& 1.0&5.5    &$-19.1$ &$-3.2\pm1.1$&7.27&$-$&B,F,Y\\

HUDF\_1173 &03:32:44.70&$-27$:46:44.3&7.36 & 7.07$-$7.72& 4.5&9.0    &$-19.8$ &$-2.4\pm0.6$&7.36&$-$&M,F,B,Y\\

HUDF\_2664 &03:32:33.13&$-27$:46:54.5&7.45 & 6.98$-$7.89& 1.9&4.2    &$-18.6$ &$-4.8\pm2.4$&8.08&42.5&M,B,L\\

HUDF\_1660 &03:32:37.21&$-27$:48:06.2&7.52 & 7.24$-$7.76& 0.9&14.4   &$-19.3$ &$-2.6\pm0.7$&7.98&42.5&M,F,B,Y\\

HUDF\_1679 &03:32:42.88&$-27$:46:34.5&7.88 & 7.51$-$8.11& 1.7&5.5    &$-19.1$ &$-2.0\pm0.8$&8.80&42.7&M,F,B,L,Y\\

HUDF\_2003 &03:32:38.13&$-27$:45:54.0&8.49 & 8.08$-$8.75& 0.9&7.5    &$-19.1$ &$-2.3\pm1.1$&8.89&42.6&M,F,B,L,Y\\
\hline\hline
\end{tabular}
\end{table*}

\begin{table*}
\caption{Details of the final high-redshift sample in the ERS
field. Columns one to eleven list the same quantities as in Table 2. Column twelve lists the results of cross-checking each
candidate against existing literature studies of the WFC3/IR data in
the ERS field. Matches were found with the following papers:  B (Bouwens et al. 2011; robust), Bp (Bouwens et al. 2011; potential), W
(Wilkins et al. 2010) and L (Lorenzoni et al. 2011).}
\begin{tabular}{lccccrrccccl}
\hline
ID & RA(J2000)&DEC(J2000)&z&$\Delta
   z$&$\chi^2$&$\Delta \chi^{2}$&M$_{1500}$ &$\beta$&z$_{Ly\alpha}$& $\log(L_{Ly\alpha})$   &Literature\\ 
\hline
ERS\_7086 & 03:32:34.75 & $-27$:40:35.1 & 6.18 & 6.02$-$6.35 & 1.4 & 11.2& $-$20.2&$-2.4\pm0.6$&    6.36 & 42.5 & $-$\\

ERS\_6066 & 03:32:07.86 & $-27$:42:17.8 & 6.19 & 5.87$-$6.41 & 2.7 & 15.8& $-$20.3&$-2.5\pm0.6$&    6.66 & 43.1 & $-$\\

ERS\_9869 & 03:32:15.40 & $-27$:43:28.6 & 6.21 & 6.01$-$6.41 & 0.4 & 8.0 & $-$19.8&$-3.4\pm1.0$&    6.44 & 42.5 & Bp\\

ERS\_8668 & 03:32:27.96 & $-27$:41:19.0 & 6.22 & 5.88$-$6.55 & 2.0 & 7.6 & $-$19.9&$-2.2\pm0.8$&    6.22 & $-$ & $-$\\

ERS\_9100 & 03:32:20.24 & $-27$:43:34.3 & 6.27 & 5.95$-$6.50 & 1.4 & 6.6 & $-$19.8&$-2.3\pm0.6$&    6.45 & 42.5 & Bp\\

ERS\_7225 & 03:32:36.31 & $-27$:40:15.0 & 6.30 & 6.02$-$6.73 & 4.3 & 11.4& $-$20.1&$-1.7\pm0.7$&    7.16 & 43.1 & $-$\\

ERS\_6438 & 03:32:25.28 & $-27$:43:24.2 & 6.33 & 6.14$-$6.70 & 6.7 & 9.5 & $-$20.3&$-2.0\pm0.7$&    7.26 & 43.2 & W\\

ERS\_6263 & 03:32:06.83 & $-27$:44:22.2 & 6.36 & 6.14$-$6.59 & 4.9 & 10.0& $-$20.3&$-2.1\pm0.8$&    6.40 & 41.9 & B\\

ERS\_7776 & 03:32:03.77 & $-27$:44:54.4 & 6.46 & 6.15$-$6.66 & 3.9 & 6.5 & $-$20.0&$-1.6\pm0.8$&    6.58 & 42.3 & $-$\\

ERS\_5847 & 03:32:16.00 & $-27$:43:01.4 & 6.49 & 6.31$-$6.59 & 2.7 & 17.2& $-$20.5&$-2.4\pm0.8$&    6.90 & 43.1 & W\\

ERS\_8987&03:32:16.01&$-27$:41:59.0&  6.52   &6.06$-$6.84& 2.5 & 5.9&$-19.7$&$-1.8\pm1.0$ &     6.72&42.5& B\\

ERS\_3679 & 03:32:22.66 & $-27$:43:00.7 & 6.55 & 6.42$-$6.71 & 5.5 & 14.5& $-$21.2&$-1.9\pm0.5$&    6.55 & $-$ & W,Bp\\

ERS\_7412 & 03:32:09.85 & $-27$:43:24.0 & 6.57 & 6.37$-$6.77 & 9.1 & 8.8 & $-$20.2&$-0.7\pm0.7$&    7.58 & 43.3 & $-$\\

ERS\_6427 & 03:32:24.09 & $-27$:42:13.9 & 6.65 & 6.37$-$6.88 & 1.2 & 10.3 & $-$20.3&$-2.5\pm0.6$&    6.64 & 42.2 & W,B\\

ERS\_8858 & 03:32:16.19 & $-27$:41:49.8 & 6.79 & 6.33$-$7.08 & 1.2 & 6.8 & $-$20.0&$-2.9\pm0.8$&    6.77 & $-$ & B\\

ERS\_7376 & 03:32:29.54 & $-27$:42:04.5 & 6.79 & 6.50$-$6.98  & 0.6 & 5.3 & $-$20.2&$-2.4\pm1.1$&   7.27 & 43.0 & W,B\\

ERS\_8176 & 03:32:23.15 & $-27$:42:04.7 & 6.81 & 6.62$-$6.98 & 4.4 & 13.6& $-$20.1&$-3.8\pm1.1$&    7.73 & 43.3 & W\\

ERS\_7672 & 03:32:10.03 & $-27$:45:24.6 & 6.88 & 6.64$-$7.05 & 3.8 & 7.7 & $-$20.3&$-2.4\pm1.0$&    7.77 & 43.3 & $-$\\

ERS\_7475 & 03:32:32.81 & $-27$:42:38.5 & 7.11 & 6.83$-$7.31 & 5.2 & 6.2 & $-$20.3&$-1.9\pm0.7$&    7.76 & 43.2 & $-$\\

ERS\_7236 & 03:32:11.51 & $-27$:45:17.1 & 7.18 & 6.99$-$7.35 & 5.0 & 5.4 & $-$20.3&$-3.4\pm0.8$&    7.74 & 43.0 & $-$\\

ERS\_9041 & 03:32:23.37 & $-27$:43:26.5 & 8.02 & 7.61$-$8.20 & 6.2 & 9.1 & $-$20.0&$-5.0\pm1.5$&    8.16 & 43.0 & L\\

ERS\_10288 & 03:32:35.44 & $-27$:41:32.7 & 8.28 & 7.59$-$8.56 & 2.1 & 7.8 &$-$20.1&$-2.3\pm0.8$&    9.50 & 43.2 & B\\

ERS\_8584 & 03:32:02.99 & $-27$:43:51.9 & 8.35 & 7.63$-$8.74  & 2.9 & 4.7 &$-$20.4&$-1.6\pm1.1$&    9.37 & 43.3 & B,L\\

\hline\hline
ERS\_8496 &03:32:29.69&$-27$:40:49.9&6.07   &5.49$-$6.56& 4.6 &9.0      &$-19.7$&$-1.8\pm1.1$ &     6.87&42.9& $-$\\

ERS\_9923 & 03:32:10.06 & $-27$:45:22.6 & 6.59 & 6.37$-$6.78 & 6.8 & 7.7 & $-$20.0&$-1.7\pm1.3$&    7.59 & 43.3 & $-$\\

\hline\hline
\end{tabular}
\end{table*}

\begin{table*}
\caption{
Details of the final high-redshift sample in the HUDF09-2
field. Columns one to eleven list the same quantities as in Table 2. Column twelve lists the results of cross-checking each
candidate against existing literature studies of the WFC3/IR data in
the HUDF09-2 field. Matches were found with the following papers: B
(Bouwens et al. 2011; robust), Bp (Bouwens et al. 2011; potential) and
W (Wilkins et al. 2010).}
\begin{tabular}{lccccrrccccl}
\hline
ID & RA(J2000)&DEC(J2000)&z&$\Delta
   z$&$\chi^2$&$\Delta \chi^{2}$&M$_{1500}$ &$\beta$&z$_{Ly\alpha}$&
   $\log(L_{Ly\alpha})$   &Literature\\ 

\hline
 HUDF09-2\_2459 &03:33:06.30&$-27$:50:20.2&   6.06  & 5.92$-$6.17& 3.2 & 20.8& $-19.2$&$-1.4\pm1.0$ &6.37 &42.5& $-$\\

 HUDF09-2\_2613 &03:33:06.52&$-27$:50:34.6&   6.08  & 5.90$-$6.22& 1.0 & 13.2& $-19.2$&$-3.0\pm1.0$ &6.44 &42.5& $-$\\

 HUDF09-2\_2638 &03:33:06.65&$-27$:50:30.2&   6.14  & 5.76$-$6.46& 0.1& 4.5 & $-19.0$&$-2.1\pm1.1$ &6.44 &42.2& $-$\\
 
 HUDF09-2\_1543 &03:33:01.18&$-27$:51:22.3&   6.18  & 6.05$-$6.26& 0.3 &22.3 & $-20.4$&$-2.3\pm0.6$ &6.12 &$-$& $-$\\
 
 HUDF09-2\_605  &03:33:01.95&$-27$:52:03.2&   6.30  & 6.06$-$6.49& 0.1& 6.0 & $-19.5$&$-1.9\pm0.7$ &6.30 &$-$& $-$\\ 
 
 HUDF09-2\_2587 &03:33:04.20&$-27$:50:31.3&   6.30  & 6.11$-$6.39& 3.3 &27.9 & $-20.3$&$-1.3\pm0.6$ &6.28 &41.8& $-$\\ 

 HUDF09-2\_1660 &03:33:01.10&$-27$:51:16.0&   6.36  & 6.15$-$6.47& 3.8 & 7.9 & $-20.3$&$-1.1\pm0.6$ &6.26 &$-$& $-$\\ 

 HUDF09-2\_1745 &03:33:01.19&$-27$:51:13.3&   6.52  & 6.22$-$6.82& 0.3& 7.9 & $-19.4$&$-1.9\pm0.6$ &6.98 &42.7& W,B\\ 

 HUDF09-2\_1620 &03:33:05.40&$-27$:51:18.8&   6.61  & 6.30$-$6.93& 1.7 & 7.0 & $-19.1$&$-1.7\pm1.0$ &7.39 &42.9& W,B\\ 

 HUDF09-2\_1721 &03:33:01.17&$-27$:51:13.9&   6.73  & 6.39$-$7.05& 2.5 & 4.3 & $-19.8$&$-2.1\pm0.5$ &6.78 &42.0& $-$\\

 HUDF09-2\_2455 &03:33:09.65&$-27$:50:50.8&   6.82  & 6.73$-$6.89& 1.9 &34.4 & $-20.6$&$-1.9\pm0.5$ &7.12 &43.1& W,Bp\\ 

 HUDF09-2\_1584 &03:33:03.79&$-27$:51:20.4&   7.17  & 6.79$-$7.36& 0.7 &15.6 & $-20.6$&$-1.6\pm0.5$ &8.03 &43.3& W,B\\

 HUDF09-2\_2814 &03:33:07.05&$-27$:50:55.5&   7.30  & 6.90$-$7.66& 0.5 & 5.7 & $-19.7$&$-2.5\pm0.6$ &7.26 &$-$& Bp\\ 

 HUDF09-2\_1596 &03:33:03.76&$-27$:51:19.7&   7.45  & 7.06$-$7.62& 6.0 &15.8 & $-20.4$&$-2.0\pm0.5$ &7.95 &43.1&B\\ 

 HUDF09-2\_2000 &03:33:04.64&$-27$:50:53.0&   7.68  & 7.30$-$7.90& 1.7 &15.5 & $-19.7$&$-2.2\pm0.5$ &8.01 &42.7& B\\

 HUDF09-2\_2765 &03:33:07.58&$-27$:50:55.0&   8.70  & 8.37$-$9.05& 0.9 & 6.9 & $-20.0$&$-0.5\pm0.6$ &8.76 &41.8& B\\ 
\hline\hline
 HUDF09-2\_799  &03:33:09.15&$-27$:51:55.4&   6.88  & 6.70$-$7.00& 9.1 &15.5 &$-19.5$&$-1.6\pm0.6$ &7.67&43.1& B,W\\
\hline\hline
\end{tabular}
\end{table*}

\section{The UV spectral slopes}
As discussed in the introduction, one of the most interesting, and
controversial, results to emerge from the new WFC3/IR-selected LBGs has been the claim that faint LBGs ($M_{1500}\simeq-~18.5$) at $z\geq6$ display extremely blue ($\beta\simeq -3$) UV spectral slopes
(e.g. Bouwens et al. 2010b; Labb\'{e} et al. 2010). Given the
relatively small  areas which have currently been imaged with WFC3/IR
(i.e. $\simeq 50$ sq. arcmin), the brightest WFC3/IR-selected
$z\simeq7$ LBGs have absolute UV luminosities of $M_{1500}\simeq
-21$. It is widely agreed in the literature that at these absolute
magnitudes ($L\simeq 2L^{\star}$), $z\simeq 7$ LBGs display the same
UV spectral slopes ($\beta\simeq -2$) as observed for young  ($\simeq
100$ Myr) starbursts at redshifts $3<z<5$. However, in
contrast, it has been claimed that the faintest LBGs at $z\simeq7$ 
($M_{1500}\simeq -18.5$) display much bluer spectral slopes; $\langle\beta\rangle=-3.0\pm0.2$ (Bouwens et al. 2010b). 

Although this may appear to be a relatively small difference in
spectral slope, it is potentially of great interest. The reason is
very straightforward. While UV spectral slopes of $\beta\simeq-2$ can
be comfortably reproduced by standard simple stellar population models
(without recourse to ultra-young ages or ultra-low metallicities),
spectral slopes of $\beta\simeq-3$ cannot, and probably require a
combination of  zero reddening, very young ages (i.e. $\simeq 10-30$
Myrs) and a high escape fraction of photons short-ward of Ly$\alpha$
(e.g. Bouwens et al. 2010b; Labb\'{e} et al. 2010). Given the
potential importance of this result, not least for  studies of
reionization, it is clearly of interest to investigate the UV spectral
slopes displayed by the sample of  high-redshift LBGs derived
here. 

The individual values of $\beta$ measured for each candidate are listed in Tables 2, 3 \& 4. The 
$\beta$ values have been calculated using the following formulae:
\begin{equation}
\beta=4.43(J_{125}-H_{160}) -2.0
\end{equation}
\begin{equation}
\beta=5.47(Y_{105}-J_{125})\,-2.0
\end{equation}
\begin{equation}
\beta=3.91(Y_{098}-J_{125})\,-2.0
\end{equation}
\noindent
depending on the available filters and the redshift of the
candidate. To derive the above formulae we have adopted the following pivot wavelengths for
the $Y_{098}, Y_{105}, J_{125}$ \& $H_{160}$ filters: 0.9864$\mu$m, 1.0552$\mu$m, 1.2486$\mu$m \& 1.5369$\mu$m 
(WFC3 Instrument Handbook for Cycle 19). In order to sample as similar a range of rest-frame
wavelengths as possible, and to ensure no potential contamination from
Ly$\alpha$ line emission, the values of $\beta$ have been calculated
using equations 2 or 3 for candidates at $z_{phot}\leq 6.4$ and equation 1
for those candidates at $z_{phot}>6.4$. In Fig. 4 we plot the estimated UV spectral slopes versus photometric redshift
for the final robust sample, split by field. Several features of this
plot are worthy of comment and are briefly discussed below.

\subsection{Uncertainties on derived UV spectral slopes}
As can readily be seen from Fig. 4, the uncertainties on measuring
$\beta$ are typically large. This is
simply a consequence of attempting to determine a spectral slope using
two filters which are not well separated in wavelength. As an
illustration, consider a galaxy at $z=7$ with a canonical UV spectral
slope of $\beta=-2$, which is detected at $5\sigma$ significance in
both the $J_{125}$ and $H_{160}$ filters. The corresponding estimate
of the UV spectral slope is $\beta=-2.0\pm 1.3$, where the
error simply reflects the photometric uncertainty. Clearly, deriving
meaningful estimates of $\beta$ on an individual object-by-object
basis requires significantly better than $5\sigma$ photometry in both
filters. One obvious method of overcoming this problem is to assume
that each $\beta$ measurement, although inaccurate, is at least unbiased. In
which case, one can proceed to bin the data and attempt to estimate
the mean value of $\beta$. However, even when adopting this approach, 
it is necessary to account for the wide range in $\beta$ uncertainties displayed by the
objects in a typical sample, by calculating a properly weighted mean: 
\begin{equation}
\langle\beta\rangle= \frac{\displaystyle\sum_{i=1}^{n}\frac{\beta_i}{\sigma_{i}^{2}}}{ \displaystyle\sum_{i=1}^{n} \frac{1}{\sigma_{i}^{2}}}
\end{equation}
where $\beta_i$ represents an individual $\beta$ measurement for a
single candidate, and $\sigma_{i}^2$ is the corresponding
variance.

\subsection{Average values of UV spectral slopes}
The variance-weighted values of $\langle\beta\rangle$ for each
sub-sample, and the full combined sample are listed in Table 5 where, for
comparison, we also list the straight arithmetic means and standard
errors. It can be seen from Table 5 that for the HUDF and ERS
sub-samples (and for the full combined sample), the variance-weighted mean results in a significantly 
redder estimate of the typical value of the UV slope than the straight arithmetic mean. 
Interestingly, for the HUDF09-2 sub-sample, where the photometry is
most robust (see discussion below), the difference between the two estimates is negligible. 

The results listed in Table 5 indicate that the ERS sub-sample contains a higher
percentage of objects with $\beta\leq-2$ than the other two fields
(based on the variance-weighted means), although the difference is not
significant. However, it is worth noting that any suggestion that the ERS candidates display bluer UV
spectral slopes cannot be due to a trend for increasingly blue UV
spectral slopes with decreasing UV luminosity, given that the
median absolute magnitude of the ERS sample is $M_{1500}=-20.2$ compared
to $M_{1500}=-19.1$ for the HUDF. Overall, our results provide no
evidence that the members of the $z\geq 6.5$ LBG population display values of
$\beta$ significantly different from those seen in comparably luminous
LBGs in the redshift interval $3<z<5$.

\begin{table}
\caption{Estimates of the typical value of the UV spectral slope
($\beta$) for our final robust sample. The first two columns list the
names and sizes of the different samples being considered. Columns three and four list
the variance-weighted and arithmetic mean values of $\beta$
respectively, together with their corresponding uncertainties. Note
that candidate HUDF09-2$\_$2765 has been excluded from these
calculations because it provides a biased estimate of $\beta$ due to
its high redshift ($z_{phot}=8.7\pm0.3$).}
\begin{center}
\begin{tabular}{lccc}
\hline
Sample   &N &$\langle\beta_{\rm var}\rangle$&$\langle\beta_{\rm arith}\rangle$ \\
\hline

HUDF     &31&$-1.99\pm{0.14}$&$-2.40\pm{0.18}$\\

HUDF09-2 &15&$-1.90\pm{0.15}$&$-1.91\pm{0.13}$\\

ERS     &23&$-2.32\pm{0.18}$&$-2.52\pm{0.19}$\\

COMBINED     &69&$-2.05\pm{0.09}$&$-2.33\pm{0.11}$\\
\hline\hline
\end{tabular}
\end{center}
\end{table}

\subsection{Potential for bias}
It can be seen from Fig. 4 that the HUDF09-2
sub-sample seems to display a particularly tight distribution of UV
slopes, whereas the HUDF and ERS sub-samples show considerably more
scatter. At least part of the explanation for this is that the
HUDF09-2 sub-sample has the most robust WFC3/IR photometry. The reason
is that, although the WFC3/IR imaging of HUDF09-2 is deep
(particularly the $J_{125}$ data), the supporting data at other
wavelengths is not, in a relative sense, as good (e.g. no $B_{435}$
data, relatively shallow $V_{606}$+$i_{775}$ data, and no $4.5\mu$m
data). As a consequence, candidates in the HUDF09-2 field are required
to be somewhat brighter in the near-IR in order to pass our
robustness criteria (see photometry in Appendix B).

Another noteworthy point is that the bluer mean UV slope in the ERS is
probably connected to the relative depths of the WFC3/IR 
imaging in this field. Due to the fact that the $J_{125}$ imaging in the
ERS is significantly deeper than the accompanying $Y_{098}$ and
$H_{160}$ imaging, the ERS sub-sample is the closest of the three to
being purely $J_{125}-$selected. It is clear that when estimating the
UV spectral slope from the $J_{125}-H_{160}$ colour, selecting the
sample largely on the apparent $J_{125}$ magnitude must introduce the
potential for biasing the sample towards objects with blue values of
$\beta$. A proper investigation of the sources of bias, and the
potential for constraining the true underlying distribution of UV
spectral slopes, requires detailed simulation work which, although
beyond the scope of this paper, is investigated in detail by Dunlop et
al.~(2011).

\begin{figure*}
\centerline{\psfig{file=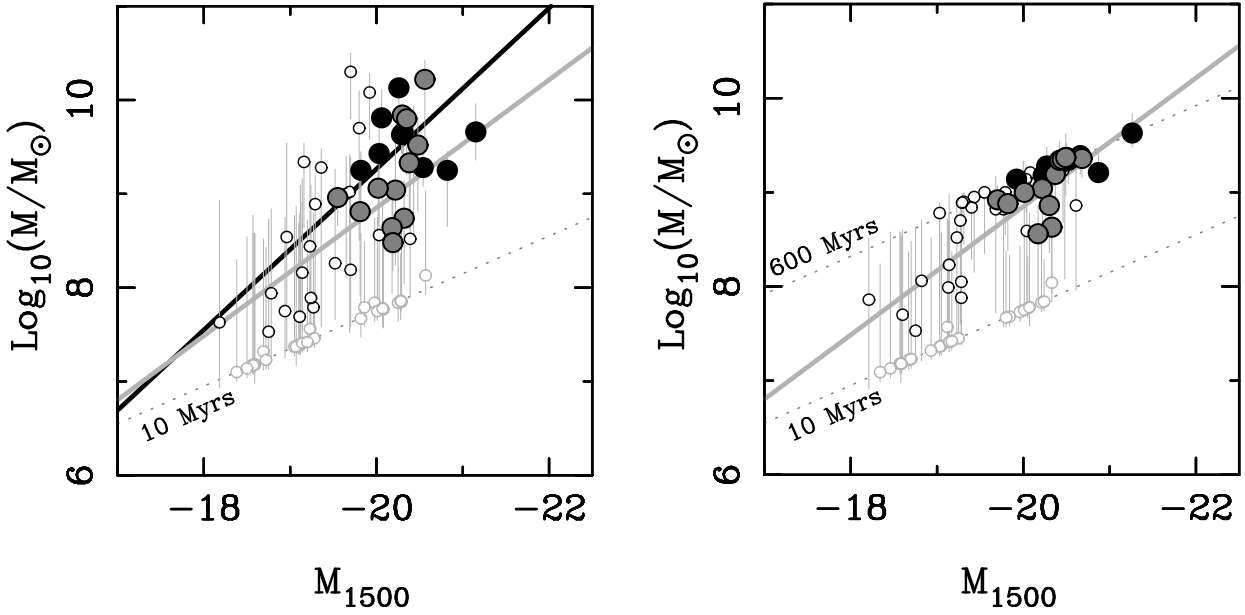,width=16.5cm,angle=0}}
\caption{Stellar mass versus absolute UV magnitude ($M_{1500}$) for the final robust sample of seventy $z\geq 6$
LBGs, where $M_{1500}$ is derived from the best-fitting SED models using a 100\AA\,-wide filter centred on a rest-frame wavelength of
1500\AA. In each panel, the small open circles are objects with non-detections (i.e. $\leq
2\sigma$) at 3.6$\mu$m, the large grey circles are objects which are detected at
$3.6\mu$m and the large black circles are objects detected at both $3.6\mu$m and $4.5\mu$m. 
In the left-hand panel the stellar-mass measurements are based on our
SED-fitting analysis using the full range of star-formation histories,
metallicities and reddening described in Section 3.3. In the
right-hand panel the stellar-mass measurements are based on a single set of SED templates with a constant star-formation rate,
$Z=0.2Z_{\odot}$ and zero reddening (CSF model, see text for details).
In the left-hand panel the thick black line is our best-fitting
$M_{\star}-L_{\rm UV}$ relation (Equation 5) and the dotted line indicates
the lower limit enforced by insisting that each candidate has an age
$\geq10$ Myrs. Those objects which lie on the lower limit (small grey
open circles) have very poorly constraint stellar masses and were not included in the derivation of the best-fitting $M_{\star}-L_{\rm UV}$ relation.
In both panels the thick grey line is the $M_{\star}-L_{\rm UV}$ relation (corrected to Chabrier IMF) derived by Gonz\'{a}lez et al. (2011)
using a large sample of LBGs at $z\simeq 4$. In the right-hand panel the upper dotted line indicates the expected $M_{\star}-L_{\rm UV}$
relation for a CSF model which has been forming stars for $\simeq 600$ Myrs
(i.e. since $z\simeq 20$ for an object at the mean redshift of the final robust
sample; $z=6.8\pm0.1$).}
\end{figure*}

\begin{table*}
\caption{The best-fitting parameters returned by our SED-fitting
analysis of the twenty-one objects in our final robust sample with
detections at either $3.6\mu$m or $3.6+4.5\mu$m. The first column lists the objects IDs. The second column
lists the best-fitting star-formation history (SFH) which is either:
an instantaneous burst (Burst), constant star-formation rate (Const) or
exponentially decaying star-formation rate (E). For the those objects where the best-fitting SFH is exponentially decaying (i.e. SFR $\propto e^{(-t/\tau)}$), 
the proceeding number indicates the characteristic star-formation timescale in
Gyrs (i.e. E0.2 $\Rightarrow \tau=0.2$ Gyrs). Columns 3-8 list the metallicity, age, reddening, stellar
mass, star-formation rate and $\chi^{2}$ of the best-fitting SED template.
For those objects where the best-fitting SFH
is a Burst, the SFR listed in column 7 is derived from the best-fitting
model with a constant or exponentially decaying SFH. Note that the ages listed in columns 4 \& 10 refer to the total age of the system, rather than a luminosity-weighted age. Columns 9-12 list
the $\chi^{2}$, age, stellar mass and SFR of the best-fitting constant
SFR model with a metallicity of $0.2Z_{\odot}$ and 
$A_{\rm v}=0.0$ (CSF model, see text for details). Based on this restricted set of SED parameters those objects highlighted with a $\dagger$ symbol in column 9 would have been rejected because their best-fitting SED templates have an unacceptably high $\chi^{2}$. The final column lists an estimate of the SFR based on the UV
luminosity of each object ($M_{1500}$), with no correction for dust
attenuation, which has been derived using the Madau, Pozzetti \& Dickinson (1998) formula (corrected to
a Chabrier IMF).}

\begin{tabular}{llcccrrccccccc}
\hline
ID &SFH & $Z$ &Age   & A$_{\rm v}$ & $M_{\star}$\phantom{00} & SFR\phantom{000} &
   $\chi^{2}_{b}$&$\chi^{2}_{c}$&Age&$M_{\star}$&SFR&SFR$_{\rm UV}$ \\
   &    &    ($Z_{\odot}$)          &(Myrs) &            &($10^{9}\Msun$)    &($\Msun$yr$^{-1}$)          &          &                       &(Myrs)&($10^{9}\Msun$) &  ($\Msun$yr$^{-1}$)  & ($\Msun$yr$^{-1}$)           \\
\hline
HUDF$\_1016$& Burst  & $0.02$ &\phantom{0}10&1.1&\phantom{0}0.9$^{+0.3}_{-0.5}$&40.0$^{+13.8}_{-11.3}$& 4.3  & 19.3$^{\dagger}$ &575& 0.9$^{+0.5}_{-0.4}$& 2.1$^{+0.1}_{-0.1}$ &\phantom{0}3.8 \\[1ex]
HUDF$\_522$& Burst&$0.02$&\phantom{0}50&0.1&\phantom{0}1.8$^{+0.9}_{-1.0}$ &15.0$^{+6.8}_{-1.9}$&4.5& \phantom{0}6.9 &365& 1.6$^{+1.0}_{-0.4}$& 6.4$^{+0.3}_{-0.4}$ & \phantom{0}11.3\\[1ex]
HUDF$\_658$& Burst   & $0.50$&\phantom{0}65&0.0&\phantom{0}1.9$^{+1.9}_{-1.0}$&\phantom{0}2.5$^{+3.3}_{-0.3}$&1.4  &\phantom{0}5.2  &725& 2.4$^{+0.6}_{-0.9}$ & 5.1$^{+0.1}_{-0.4}$ & \phantom{0}8.6 \\[1ex]
HUDF$\_860$& Burst   & $0.50$ &100          &0.1&\phantom{0}2.6$^{+2.6}_{-0.5}$&\phantom{0}1.8$^{+2.0}_{-0.1}$& 1.8   & 17.8$^{\dagger}$ &645& 1.5$^{+0.4}_{-0.5}$ & 3.4$^{+0.3}_{-0.3}$ & \phantom{0}5.4 \\[1ex]
HUDF$\_1173$& Burst  & $0.02$ &\phantom{0}55&0.0&\phantom{0}0.6$^{+1.4}_{-0.5}$&\phantom{0}1.1$^{+2.1}_{-0.4}$& 4.5   & \phantom{0}4.7&455 & 0.8$^{+0.8}_{-0.5}$ & 2.4$^{+0.5}_{-0.3}$  & \phantom{0}4.5 \\
\hline
HUDF09-2$\_1543$  & Const  & $0.02$ & 725 & 0.1 &\phantom{0}2.1$^{+3.1}_{-1.3}$&\phantom{0}4.5$^{+1.3}_{-2.6}$& 0.3   & \phantom{0}0.4 &575& 1.5$^{+0.9}_{-0.3}$ & 4.0$^{+0.4}_{-0.4}$ & \phantom{0}7.8 \\[1ex]
HUDF09-2$\_2587$  & Const  & $0.02$ & 645 & 0.6 &\phantom{0}6.3$^{+3.6}_{-3.4}$& 13.8$^{+4.4}_{-7.6}$&3.3 &\phantom{0}9.5$^{\dagger}$ &815& 2.3$^{+0.6}_{-0.9}$ &4.1$^{+0.1}_{-0.3}$ &\phantom{0}7.2 \\[1ex]
HUDF09-2$\_1660$  & E1.0  & $0.02$ & 725& 0.5 &\phantom{0}6.9$^{+1.8}_{-5.1}$ &\phantom{0}10.0$^{+2.8}_{-6.1}$& 3.8 &\phantom{0}8.7$^{\dagger}$ &815& 2.1$^{+1.3}_{-0.5}$ & 4.1$^{+0.3}_{-0.3}$ & \phantom{0}7.2\\[1ex]
HUDF09-2$\_1584$  & E0.2  & $0.20$ & 725& 0.1 &16.5$^{+9.6}_{-13.3}$           &\phantom{0}3.8$^{+0.5}_{-3.6}$& 0.7  &\phantom{0}7.6$^{\dagger}$ &645& 2.3$^{+0.6}_{-0.9}$ & 5.3$^{+0.3}_{-0.3}$ &\phantom{0}9.4 \\
\hline
ERS$\_7086$& Burst  & $0.02$ &\phantom{0}25 & 0.1 &\phantom{0}0.4$^{+0.4}_{-0.1}$ &\phantom{0} 4.9$^{+4.1}_{-2.6}$& 1.4 & \phantom{0}1.7 &130& 0.4$^{+0.4}_{-0.6}$ & 3.6$^{+2.3}_{-0.5}$ &\phantom{0}6.5\\[1ex]
ERS$\_6066$& E0.2 & $1.00$ & 645 & 0.1 &13.4$^{+3.5}_{-8.0}$&\phantom{0}4.4$^{+0.4}_{-0.5}$& 2.7  & 25.4$^{\dagger}$ &815& 2.4$^{+0.6}_{-0.9}$ & 4.4$^{+0.4}_{-0.3}$&\phantom{0}7.2\\[1ex]
ERS$\_9100$& E0.2 & $0.50$ & 405 & 0.0 &\phantom{0}1.8$^{+2.6}_{-0.9}$&\phantom{0}2.0$^{+1.3}_{-0.8}$& 1.4   & \phantom{0}3.2 &815& 1.4$^{+0.4}_{-0.5}$ & 2.6$^{+0.3}_{-0.4}$ &\phantom{0}4.5\\[1ex]
ERS$\_7225$& E0.05  & $0.20$ & 325 & 0.0 &\phantom{0}6.4$^{+6.4}_{-1.3}$&\phantom{0}0.4$^{+1.9}_{-0.1}$& 4.3  & 28.0$^{\dagger}$ & 815&1.9$^{+0.5}_{-0.8}$ & 3.6$^{+0.4}_{-0.4}$ & \phantom{0}6.0\\[1ex]
ERS$\_6438$& Burst  & $0.20$ & 160 & 0.0 &\phantom{0}4.5$^{+2.6}_{-2.3}$&\phantom{0}3.6$^{+0.4}_{-0.4}$& 6.7   & 14.9$^{\dagger}$ &725& 2.1$^{+0.5}_{-0.8}$ & 4.5$^{+0.3}_{-0.5}$ &\phantom{0}7.2\\[1ex]
ERS$\_6263$& Burst  & $0.02$ & 330& 0.0 &\phantom{0}0.5$^{+1.1}_{-0.4}$&\phantom{0}2.5$^{+5.3}_{-0.8}$& 4.9   &\phantom{0}5.3 & 130&0.4$^{+0.6}_{-0.3}$ & 4.3$^{+4.4}_{-0.8}$ & \phantom{0}7.2\\[1ex]
ERS$\_7776$& E0.2 & $0.02$ & 325 & 0.0 &\phantom{0}1.1$^{+1.1}_{-0.8}$&\phantom{0}2.1$^{+1.1}_{-1.3}$& 3.9   &\phantom{0}4.2 & 515&1.0$^{+0.6}_{-0.6}$& 2.9$^{+0.5}_{-0.4}$ &\phantom{0}5.4\\[1ex]
ERS$\_5847$& E1.0 & $0.20$ & 815 & 0.0 &\phantom{0}3.3$^{+4.9}_{-2.4}$&\phantom{0}4.1$^{+1.1}_{-2.8}$& 2.7   &\phantom{0}2.9 &815&2.4$^{+2.4}_{-1.4}$ & 4.4$^{+0.4}_{-0.4}$ & \phantom{0}8.6\\[1ex]
ERS$\_3679$& Burst  & $0.02$ & 725 & 0.2 &\phantom{0}4.5$^{+6.8}_{-2.6}$&15.0$^{+0.9}_{-10.6}$& 5.5  & \phantom{0}8.3 &725& 4.3$^{+1.1}_{-1.5}$ & 8.9$^{+0.3}_{-0.8}$ & 16.4\\[1ex]
ERS$\_7412$& Burst  & $1.00$ & \phantom{0}10 & 0.2 &\phantom{0}0.4$^{+0.8}_{-0.3}$&\phantom{0}11.4$^{+7.1}_{-2.9}$ & 9.1  & 12.2$^{\dagger}$ &255& 0.8$^{+0.8}_{-0.5}$& 3.9$^{+1.1}_{-0.5}$ &\phantom{0}6.5\\[1ex]
ERS$\_6427$& E0.2 & $0.02$ & 575 & 0.0 &\phantom{0}4.4$^{+4.4}_{-1.6}$&\phantom{0}2.1$^{+0.4}_{-0.8}$& 1.2   &\phantom{0}6.6 &725& 2.0$^{+0.5}_{-0.4}$ & 4.1$^{+0.4}_{-0.5}$ &\phantom{0}7.2\\[1ex]
ERS$\_7376$& Const  & $0.20$ & 455 & 0.0 &\phantom{0}1.1$^{+1.6}_{-0.9}$&\phantom{0}3.5$^{+1.1}_{-2.5}$& 0.6   &\phantom{0}0.6 & 455&1.1$^{+1.1}_{-0.8}$ & 3.5$^{+0.9}_{-0.5}$ &\phantom{0}6.5\\

\hline
\hline
\end{tabular}
\end{table*}

\section{Stellar masses and star-formation rates}
A key advantage of employing a template-fitting SED analysis is that 
stellar mass and star-formation rate estimates can be directly derived 
from the best-fitting models. In this Section we use this information
to investigate the relationship between stellar mass and UV luminosity, and try to determine 
the typical specific star-formation rate (sSFR) for $L^{\star}$ LBGs
at $z\geq 6$. Throughout this Section we will repeatedly refer to the
results for a sub-sample of twenty-one objects which have the most reliable star-formation 
rate and stellar-mass estimates because they are detected at either 3.6$\mu$m or $3.6+4.5\mu$m.
In Table 6 we list the best-fitting parameters returned by our
SED-fitting analysis for these objects, based on the best-fitting templates drawn from the full range of
star-formation histories, metallicities and reddening described in Section 3.3.  Based on photometry alone it is
very difficult to accurately constrain the star-formation history and metallicity of
high-redshift galaxies. As a result, it is common in the literature to derive star-formation rate 
and stellar-mass estimates from a much more restricted set of SED templates (typically constant star-formation rate models).
In order to investigate the effect of this approach, in Table 6 we also list the best-fitting parameters derived from an SED template
with a constant star-formation rate, $Z=0.2Z_{\odot}$ and zero reddening (CSF model).

\subsection{Stellar mass - UV luminosity relation}
In Fig. 5 we plot stellar mass versus absolute UV magnitude ($M_{1500}$)
for our final robust sample of seventy $z\geq 6$ LBGs. 
In the left-hand panel we plot stellar-mass estimates based on the best-fitting SED templates 
drawn from the full grid of star-formation histories, metallicities and reddening. 
In the right-hand panel we plot the stellar-mass estimates based on the CSF model alone.
In both panels the small open circles indicate those objects which are formally undetected at $3.6\mu$m. For
these objects the only stellar mass constraints at
$\lambda_{\rm{rest}}\geq 4000$\AA\, come from the upper limits at
$3.6/4.5\mu$m provided by the deconfusion analysis. In contrast, those objects which are detected at 
$3.6\mu$m ($\geq 2 \sigma$) are plotted as large grey circles, and those objects detected at both $3.6\mu$m and 
$4.5\mu$m are plotted as large black circles. 

Based on the data presented in left-hand panel of Fig. 5 we have used the {\sc fitxy} routine (Press et al. 1992) to derive the
following relationship between stellar mass and UV luminosity~($L_{1500}$):
\begin{equation}
\log\left(\frac{M_{\star}}{\Msun}\right) = (2.14\pm0.56)\log\left(\frac{L_{1500}}{{\scriptstyle {\rm WHz}}^{-1}}\right)-37.05\pm 4.52
\end{equation}
\noindent
which is shown as the thick black line in the left-hand panel of
Fig.~5. It is interesting to compare our equation 5 with 
the $M_{\star}~-~L_{\rm UV}$ relation derived by Gonz\'{a}lez et al. (2011) based on $\simeq500$ $B-$drop galaxies at $z\simeq4$. The $M_{\star}-L_{\rm UV}$ relation  derived by Gonz\'{a}lez et al. (2011) is plotted as the thick grey line in both panels of Fig. 5, and has the form: $M_{\star} \propto L_{1500}^{1.7\pm0.2}$. It can be seen from the left-hand panel of Fig. 5 that both relations are clearly consistent, although 
the $M_{\star}-L_{\rm UV}$ relation derived here for LBGs with a mean
redshift of $z=6.8\pm0.1$ is somewhat steeper.

In an earlier study, Stark et al. (2009) also explored the $M_{\star}~-~L_{\rm UV}$ relation based on $4<z<6$ LBGs 
selected from the GOODS N+S fields. At $z\simeq 4$ the data from Stark et al. (2009), based on a
sample of $\simeq 700$ $B-$drop candidates, is entirely consistent with the $M_{\star}-L_{\rm UV}$ relation derived by Gonz\'{a}lez et
al. (2011). At higher redshifts, both Stark et al. (2009) and Gonz\'{a}lez et
al. (2011) investigated the $M_{\star}-L_{\rm UV}$ relation at $z\simeq 5$ and $z\simeq 6$, based on samples of $V-$drop 
and $i-$drop LBG candidates respectively. Interestingly, at $z\simeq 5$ the results of both studies do appear to
be consistent with a steepening of the $M_{\star}-L_{\rm UV}$
relation. In fact, this effect was noted by Stark et al. (2009) but,
based on the available data, both authors 
concluded that there was no strong evidence for redshift evolution.
At $z\simeq 6$ neither study had sufficient dynamic range in $L_{\rm UV}$ to constrain the slope of the $M_{\star}-L_{\rm UV}$ relation. 

At a given UV luminosity, the range of stellar masses displayed by the
LBG candidates in Fig. 5 is simply a function of their
$M_{\star}/L_{\rm UV}$ ratios which, in turn, are largely a function
of their stellar population ages. Unfortunately, those candidates which have neither detections or
meaningful upper limits at IRAC wavelengths inevitably have stellar
ages/masses which are very poorly constrained (small grey open circles
in Fig. 5). These objects (which we have excluded from our determination
of the best-fitting  $M_{\star}-L_{\rm UV}$ relation) can be seen to
congregate close to the lower limit which is imposed during the SED-fitting procedure by insisting
that each candidate must have an age of $\geq 10$ Myr. However, in
reality, the majority of these objects can tolerate  SED
fits with stellar populations as old as $\simeq$ 200 Myrs, at which point their
estimated stellar masses become an order of magnitude
larger. Consequently, the apparent steepening of the $M_{\star}-L_{\rm
UV}$ relation at faint magnitudes must be viewed with considerable caution.

It is clear from Fig. 5 that, based on the current sample, it is not possible to determine if 
the $M_{\star}-L_{\rm UV}$ relation at $z\geq 6$ is
steeper than at $z\simeq 4$. Indeed, our results for the twenty-one objects with the most reliable stellar-mass estimates are entirely
consistent with the conclusion that the slope and normalisation of the
$M_{\star}-L_{\rm UV}$ relation does not change over the redshift
interval $4.0<z<7.0$. However, by restricting ourselves to those
objects with the most reliable stellar-mass estimates, the results
presented in the left-hand panel of Fig. 5 suggest that $L^{\star}$ ($M_{1500}\simeq -20.2$) galaxies at
$z\simeq 6.5$ have a median stellar mass of $M_{\star}~=~(2.1\pm~1.1)\times10^{9}\Msun$. 
Moreover, by deriving stellar-mass estimates using stellar population models
covering a wide range of metallicities, star-formation histories and
reddening, our results indicate that the full range of
$M_{\star}/L_{\rm UV}$ ratios displayed by $L^{\star}$ galaxies at
this epoch could span a factor of $\simeq 50$.

Within this context it is interesting to compare the left-hand and
right-hand panels of Fig. 5 where the limiting effect of restricting the SED-fitting to a constant star-formation rate (CSF) model 
is explored. It can immediately be seen from the right-hand panel that if we adopt the same approach as
Gonz\'{a}lez et al. (2011) and restrict our SED-fitting to the CSF model, our stellar-mass estimates at 
$z\simeq 6.8$ fall into excellent agreement with the $M_{\star}-L_{\rm UV}$ relation they derived
at $z\simeq 4$. Moreover, it is also clear that restricting the
SED-fitting analysis to the CSF model significantly reduces (perhaps unrealistically) the
scatter in the stellar-mass estimates at a given UV luminosity. Finally, as illustrated by the upper dotted line in the
right-hand panel of Fig. 5, the brightest LBGs in our sample
(i.e. $M_{1500}\leq19.0$) are fully consistent with the expected $M_{\star}-L_{\rm UV}$ relation for a galaxy which has been forming
stars at a constant rate for $\simeq 600$ Myrs. Importantly, at the
mean redshift of the final robust sample ($z=6.8\pm0.1$), 600 Myrs
represents $\simeq80\%$ of the age of the Universe. Indeed, the primary cause of the clustering of objects 
around the $M_{\star}-L_{\rm UV}$ relation corresponding to $\simeq 600$ Myrs of constant star-formation is the 
requirement imposed during the SED-fitting that objects must be  younger than the age of the Universe.
The underlying cause is simply that (with no dust reddening) the CSF models are bluer than the observed photometry
unless their age is close to the maximum allowable at this epoch. In summary, although the results shown in both 
panels of Fig. 5 are broadly compatible, it is clear that adopting a 
restricted set of SED templates may well provide a misleadingly low
estimate of the true level of scatter in stellar mass at a given UV luminosity.

Before moving on to consider the relationship between stellar mass and star-formation rate, it is worth
remembering that one of the principal motivations for studying the $M_{\star}-L_{\rm
UV}$ relation at high redshift is to constrain the galaxy stellar-mass
function (e.g. McLure et al. 2009). The results presented in Fig. 5
clearly illustrate that in order to successfully constrain the
stellar-mass function at $z\geq 6.5$ is will be necessary to constrain the
$M_{\star}-L_{\rm UV}$ relation at UV luminosities substantially
fainter than $L^{\star}$. Over the next three years, the new Cosmic
Assembly Near-infrared Deep Extragalactic Legacy Survey (CANDELS;
Co-P.I.s S. Faber \& H. Ferguson; see Grogin et al. 2011 and
Koekemoer et al. 2011) offers the prospect of significant
progress. The deep portion of CANDELS will provide $Y_{105},
J_{125} \,\&\, H_{160}$ WFC3/IR imaging to $m_{AB}\simeq 28 (5\sigma)$ over an area of $\simeq 150$ sq. arcmins in the
GOODS N+S fields. The CANDELS programme should therefore provide a
sample of $\gtsim \,200$ robust $z\simeq 7$ candidates in the
magnitude range $-19>M_{1500}>-20$, all covered by the deep IRAC
imaging available in GOODS N+S. A sample of this size should be
sufficient to obtain robust constraints on the typical
$M_{\star}/L_{\rm UV}$ ratio at $M_{1500}\simeq -19$ by employing a stacking analysis to 
provide the necessary IRAC photometry. Obtaining constraints on the
$M_{\star}-L_{\rm UV}$ relation at magnitudes as faint as
$M_{1500}\simeq -18.5$ (i.e. $\simeq 0.2L^{\star}$ at $z\simeq7$) will
rely on stacking the final epoch2  WFC3/IR imaging of the HUDF into
the forthcoming, ultra-deep, IRAC data being obtained as part of
the Cycle 7 Spitzer warm mission GO-70145 (P.I.~Labb\'{e}).

\begin{figure*}
\centerline{\psfig{file=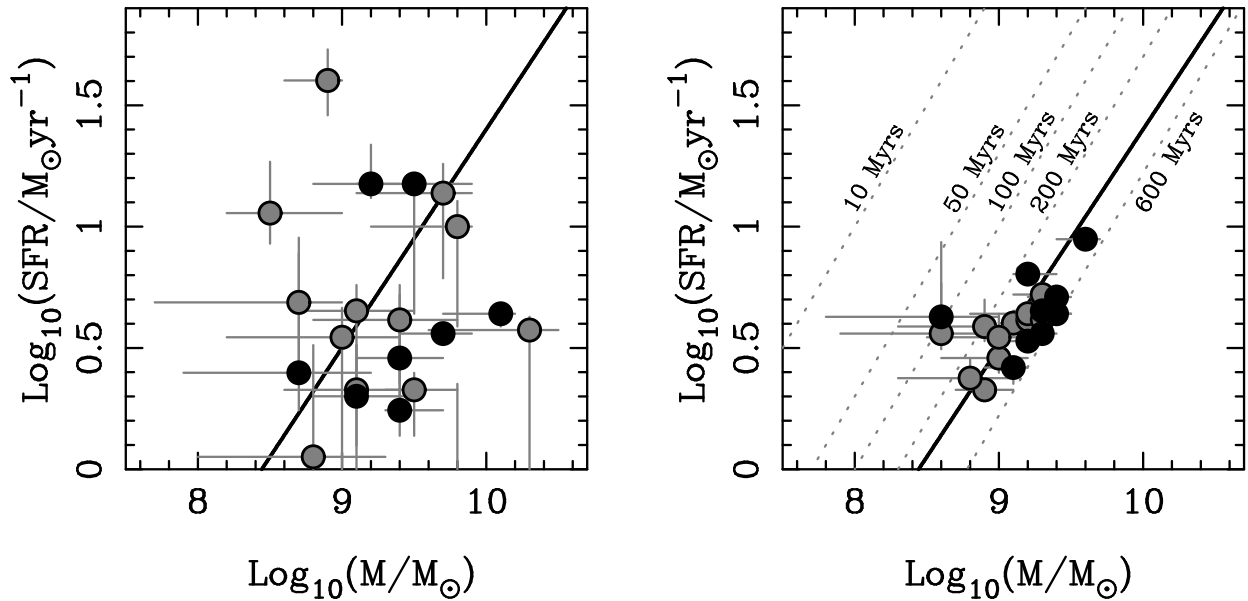,width=16.5cm,angle=0}}
\caption{Star-formation rate versus stellar mass for the twenty-one objects in our final robust 
sample with IRAC detections at either 3.6$\mu$m (grey) or
3.6+4.5$\mu$m (black). In the left-hand panel the star-formation rates
and stellar masses have been measured from the best-fitting SED
template drawn from the full range of star-formation histories,
metallicities and reddening described in Section 3.3. In the
right-hand panel the star-formation rates and stellar masses have been
estimated from the best-fitting CSF model (see text for details). The $1\sigma$ errors on both parameters have been calculated by determining the $\Delta \chi^{2}=1$ interval, after marginalization over all other
free parameters. The solid line in both panels is the SFR$-M_{\star}$ relation derived by Daddi et al. (2007)
for star-forming galaxies at $z\simeq 2$ and corresponds to a
sSFR of $\simeq 2.5$ Gyr$^{-1}$. In the right-hand panel the dotted lines illustrate how the SFR$-M_{\star}$ relation for a galaxy with a constant star-formation rate and zero reddening varies as a function of stellar population age.}
\end{figure*}

\subsection{Star-formation rate versus stellar mass}
Over recent years it has become clear that studying the ratio of the
current star-formation rate to the previously assembled stellar mass
(SFR/$M_{\star}$), the so-called specific star-formation rate (sSFR),
can provide a useful insight into the average star-formation history
of a galaxy population. Studies of star-forming galaxies in the SDSS
at $z\simeq0.1$ (Brinchmann et al. 2004), in the Extended Groth Strip
at $z\simeq 1.0$ (Noeske et al. 2007) and in the GOODS fields at
$z\simeq 1.0$ and $z\simeq 2.0$ (Elbaz et al. 2007; Daddi et al. 2007)
have consistently shown a correlation of the form SFR $\propto$
$M_{\star}^{0.9}$ (sSFR $\propto M_{\star}^{-0.1}$), over a wide
dynamic range in stellar mass. Moreover, these studies have shown that compared to a median value of
sSFR~$\simeq 2.5 $ Gyr$^{-1}$ at $z=2$ (Daddi et al. 2007), the
normalisation of the sSFR$-M_{\star}$ relation has decreased by a
factor of $\simeq 40$ over the last 10 Gyrs.

Indeed, there is some evidence that a sSFR of $\simeq 2.5$ Gyr$^{-1}$
may represent the maximum sustainable star-formation rate for galaxies
at $z\geq 2$. A recent study by Karim et al. (2011) found that high
mass ($M_{\star}\geq 10^{10}\Msun$) star-forming galaxies in the
COSMOS field consistently follow a steeper sSFR-$M_{\star}$ relation
than determined by previous studies (sSFR $\propto
M_{\star}^{-0.4}$). However, by  $z\simeq 3$ the results of Karim et
al. (2011) suggest that the sSFR-$M_{\star}$ relation flattens at
masses of $M_{\star}\leq 10\Msun$, becoming consistent with sSFR
$\propto M_{\star}^{-0.1}$. Karim et al. (2011) interpret this
flattening as a result of a natural limit to the sustainable sSFR of
$\simeq 2.5 $ Gyr$^{-1}$, corresponding roughly to the inverse of the typical
galaxy free-fall timescale. Within this context, it is interesting to note
that Gonz\'{a}lez et al. (2010) recently found that the median sSFR of
a sample of twelve $z-$drop candidates at $z\simeq7$ was also $\simeq
2.5 $ Gyr$^{-1}$.

In Fig. 6 we plot SFR versus stellar mass for the twenty-one objects
which have IRAC detections and therefore the most robust SFR and
stellar-mass estimates. In the left-hand panel the stellar mass and
dust corrected SFR estimates have been taken from the best-fitting SED
templates drawn from the full range of star-formation histories,
metallicities and dust reddening described in Section 3.3. In
contrast, in the right-hand panel the stellar mass and SFR estimates
have been taken from the best-fitting CSF model.  Although the
distribution of objects in the two panels is significantly different,
both provide a consistent estimate for the typical sSFR. In the
left-hand panel the median sSFR is $1.9\pm0.8$ Gyr$^{-1}$, while in
the right-hand panel the median sSFR is $2.6\pm 0.4$ Gyr$^{-1}$. Both
estimates are clearly consistent with the typical sSFR value for
$z\simeq 2$ star-forming galaxies estimated by Daddi et al. (2007). To
illustrate this point, in both panels of Fig. 6 the thick solid line
is the best-fitting SFR-$M_{\star}$ relation from Daddi et al. (2007)
which corresponds to a sSFR of $\simeq2.5$ Gyr$^{-1}$.

Consequently, taken at face value, our results provide additional support to the
conclusion that a direct proportionality between SFR and stellar mass is still
viable at $z\simeq 6.5$, and that the corresponding sSFR of $\simeq 2.5 $
Gyr$^{-1}$ may correspond to a physical limit on the maximum sustainable star-formation rate. 
However, it is clear from the left-hand panel that allowing a
reasonable range of star-formation histories, metallicities and dust
reddening leads to a large scatter in the SFR at a given stellar
mass. Consequently, although the data shown in the left-hand panel are
consistent with SFR and stellar mass being roughly proportional, they
are also entirely consistent with star-formation and stellar mass
being entirely unrelated. In contrast, the results shown in the
right-hand panel suggest that SFR and stellar mass are well
correlated, lying along a SFR$-M_{\star}$ relation with a slope close to
unity and a normalization consistent with a sSFR of $\simeq 2.5$ Gyr$^{-1}$. 

However, it is worth noting that the apparently simple picture
presented in the right-hand panel of Fig. 6 probably reflects limitations of relying on the CSF model, rather than offering genuine physical insight into high-redshift star-formation. The simple reason for this caution is that the agreement is largely inevitable when you only consider SEDs with constant star-formation and no reddening. In this situation, each object is required to lie on a relation with slope of unity, with it's position on the SFR$-M_{\star}$ plane simply determined by the best-fitting age. To illustrate this point we have plotted the expected SFR$-M_{\star}$ relations for CSF models of various stellar population ages as the dotted lines in the right-hand panel of Fig. 6. This demonstrates that, provided the typical stellar population age lies in the 
range 200-600 Myrs, the resulting SFR$-M_{\star}$ relation will automatically have a slope close to unity, and result in a typical sSFR consistent with $\simeq 2.5$ Gyr$^{-1}$.

In summary, although it is possible to constrain the typical sSFR of $L^{\star}$ LBGs at $z\simeq 6.5$, the limitations of the current sample do not allow meaningful constraints to be placed on the form of the SFR$-M_{\star}$ relation. In order to resolve this issue it will be necessary to obtain much larger samples of $z\geq 6$ LBGs with stellar masses $M_{\star}\geq 10^{8.5}\Msun$. Within this context, the new CANDELS WFC3/IR imaging data should prove decisive. The wide portion of CANDELS will
proved $J_{125}+H_{160}$ imaging to a depth of $m_{AB}\simeq 27 (5\sigma)$ over an area of
$\simeq 0.2$ sq. degrees, all of which is covered by deep IRAC imaging
at $3.6+4.5\mu$m ($m_{AB}\simeq 26$, $5\sigma$) provided by the Spitzer
Extended Deep Survey (SEDS; P.I. G. Fazio). The combination of
CANDELS+SEDS should therefore allow the SFR$-M_{\star}$ relation at
$z\simeq 6.5$ to be investigated using a sample of $\gtsim\,250$ LBGs
with reliable stellar-mass estimates of $M_{\star}\geq 10^{8.5}\Msun$.

\subsection{The effect of nebular emission}
Recent work has suggested that nebular continuum and line emission
might contribute to the observed SEDs of $z\sim 6-7$ galaxies
(e.g. Ono et al. 2010; Schaerer et al. 2010). As discussed by
Robertson et al. (2010), galaxies with strong UV continua can
typically be fit using pure stellar populations with ages of a few
hundred Myr, or by much younger populations ($\leq$ few Myr) with
significant nebular contributions and an implied low escape fraction
($\fesc$) of Lyman continuum photons. Such nebular solutions can yield
much lower stellar masses than those in the purely stellar case (Ono
et al. 2010).

In order to quantify this degeneracy and its possible effect on our
derived physical properties, we have examined in more detail the
sub-sample of twenty one galaxies detected in the 3.6 $\mu$m IRAC band (nine of which are also detected at 4.5$\mu$m).
For these objects it is possible to investigate whether the IRAC detections can provide a valuable
discriminant between the nebular and stellar solutions, given the
location of prominent nebular lines, such as H$\beta$ and [OIII]
5007\AA, at the redshifts of interest.
 \begin{figure}
\centerline{\psfig{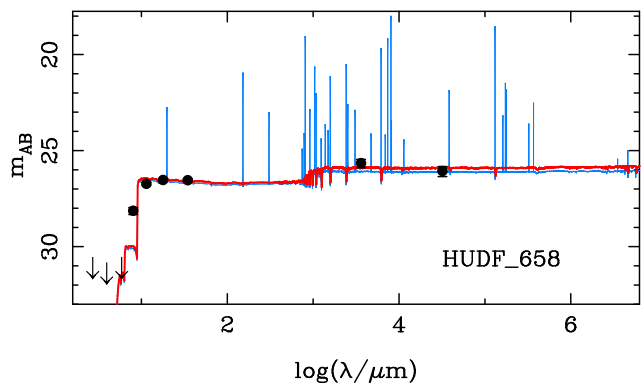}}
\caption{An example of our SED-fitting procedure using models which
  incorporate nebular continuum and line emission. In this
  illustration the pure stellar continuum model ($\fesc=1$) is shown
  as the red (thick) line and the template featuring nebular emission
  ($\fesc=0.2$) is shown as the blue (thin) line. For clarity the
  galaxy template featuring nebular emission has been displaced by 0.1
  magnitudes (both models have a metallicity of Z~=~0.2 Z$_{\odot}$).} 
\end{figure}

\subsubsection{Nebular emission methodology}
As before we use the Bruzual \& Charlot (2003) models to generate a
set of spectral templates based on a Chabrier IMF. For the models presented here, we use a 
representative exponentially decaying star-formation history ($\tau\simeq0.4$ Gyr) and, motivated by the fact that high-redshift galaxies often exhibit low metallicities (e.g. Finkelstein et al. 2011), we consider both solar (Z = Z$_{\odot}$) and fifth solar models (Z = 0.2 Z$_{\odot}$). In addition to exponentially decaying star-formation histories, we also investigated models with
constant star formation, but found that these did not significantly alter our results. The contribution of the
nebular continuum and line emission is computed in the manner of
Robertson et al. (2010), providing nebular emission models similar to
those calculated by Ono et al. (2010). The strength of the nebular
emission is tied to the number of ionizing photons per second ($N_{\mathrm{Lyc}}$), calculated from the stellar population 
model, via the H$\beta$ luminosity~(in~erg~s$^{-1}$):
\begin{equation}
L(H\beta) = 4.78 \times 10^{-13} (1 - \fesc)N_{lyc} 
\end{equation}
\noindent
Other HI line intensities follow from ratios predicted by standard
recombination theory (Osterbrock \& Ferland 2006). Lines from common
metallic species are included using relative intensities given by
Anders et al. (2003) assuming the gas phase metallicity is either Z$_{\odot}$ or 0.2 Z$_{\odot}$.
We use the method of Brown et al. (1970) to calculate the strength of
bound-free and free-free continuum emission, and use results from
Osterbrock \& Ferland (2006) for the two photon emission from H.

When fitting the SED models we consider two fixed values of the escape
fraction, $\fesc = 0.2$ (stellar and nebular emission) and $\fesc = 1$
(purely stellar emission).  The value of $\fesc=0.2$ is motivated by
direct observations of the Lyman continuum in galaxies at $z\sim 3$
(Shapley et al. 2006) and typical values of $\fesc$ required for
star-forming galaxies to maintain reionization at $z\sim 7$ (Robertson
et al. 2010).   As we are primarily interested in how nebular emission
might alter the inferred stellar mass and age, we do not include the
possible effects of Ly$\alpha$ emission or reddening. An example SED fit featuring nebular emission is shown in Fig. 7 and the 0.2 Z$_{\odot}$ nebular SED fits for 
all twenty one objects can be found in Appendix~B.

\begin{figure}
\centerline{\psfig{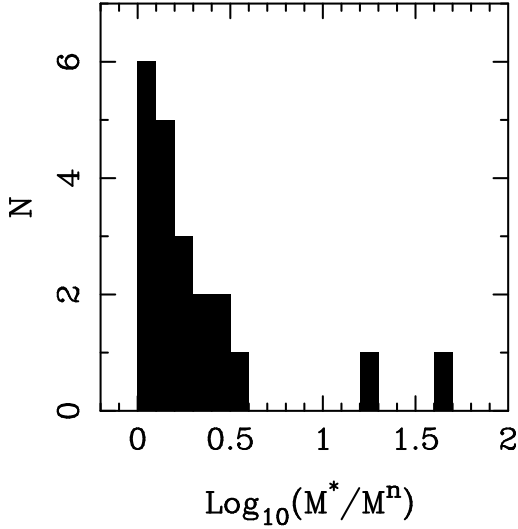}}
\caption{A histogram of the ratio of the best-fitting stellar masses 
returned by fitting Z = 0.2 Z$_{\odot}$ SED models with ($M^{n}$) and without ($M^{\star}$) nebular 
continuum and line emission for the twenty one objects with $3.6\mu$m IRAC detections. 
For 16 out of 21 objects the stellar-mass estimates differ by less than a factor of 2.5. However, it can 
be seen that there are two objects for which the nebular fits return stellar masses which are an 
order of magnitude smaller than the stellar masses returned by fits without nebular emission.}
\end{figure}

\subsubsection{Nebular emission results}
For the models with sub-solar metallicity it is found that, compared to the purely stellar templates ($\fesc =1$), the templates which include nebular emission ($\fesc = 0.2$) provide a better fit to 14 of the 21 objects, and all nine of the objects with detections in both IRAC bands. The data favour the
nebular emission models because of the predominantly blue [3.6] -
[4.5] $\mu$m observed colours that are easily reproduced by including
rest-frame optical line emission. In contrast, the shape of the purely
stellar model SEDs  redward of the Balmer/$4000$\AA\, break cannot
easily accommodate these blue rest-frame optical colours.

For the majority of this sub-sample (16 out of 21) the best-fitting ages
are still $\geq$ 100 Myr with stellar masses reduced by less than a factor of 2.5 (see Fig. 8).
In contrast, the best-fitting models to the remaining five galaxies have significantly lower stellar masses (by more than a factor of ten in two cases). Interestingly, all 
five of these objects are only detected in the 3.6 $\mu$m band and have best-fitting ages 
which are perhaps unphysical (median age 9 Myrs). Using solar 
metallicities, we find eight galaxies with significantly altered parameters, including 
all five found for models with Z = 0.2 Z$_{\odot}$. 

In summary, with the current precision of the HST and IRAC photometry,
we are unable to draw firm conclusions about the possible presence of
nebular emission in these sources.  However, given that the inferred
stellar masses from the models including nebular emission are
generally similar to those inferred from purely stellar models,
we conclude that the results presented in Sections 5.1 and 5.2 appear
to be robust to the inclusion of $\fesc=0.2$ nebular emission, especially
for those objects with detection in both IRAC bands.

\section{Comparison with previous studies}
As discussed in the introduction, the availability of the various new
WFC3/IR datasets has led to a proliferation  of papers focused on
$z\geq6$ LBGs. With authors each applying their
individual candidate selection procedures, and in many cases using their own
independent reductions of the publicly available data, it is difficult
to obtain a clear overview of the subject, and to identify whether
different studies are in good agreement, or not. Consequently, in this
section we compare our final robust sample with those derived
elsewhere in the literature (on a field-by-field basis), highlighting the
objects we have in common, and investigating the properties of previously published 
high-redshift candidates which are not included in our final robust sample.

Given that one of the primary motivations for this study was to derive
a sample of high-redshift candidates which is as robust as possible,
and that previous samples of WFC3/IR high-redshift candidates were
selected for a variety of purposes, it is not the case that we regard
any object not included in our final sample as a low-redshift
interloper. In fact, as the proceeding discussion will demonstrate,
each of the candidates from the literature samples falls into one of
four categories. The first category consists of objects which are in
common with our final sample of seventy $z\geq 6$ LBGs and we
therefore regard as being robust. The second category consists of
objects which were not included in our final sample (because they
failed to meet one or more of our adopted criteria), but which nevertheless our
analysis suggests are likely to be at high redshift. The third category consists of objects which our analysis
suggests are likely to be a low redshift, but do have an acceptable
(albeit lower probability) solution at high redshift. The fourth
category consists of those objects which our analysis suggests are
very unlikely to be at high redshift. Throughout the discussion in
this Section we have attempted to make it as clear as possible which
category each of the candidates falls into. Finally, it should be
noted that where a research group has published a number of studies of
a particular survey field, we only discuss the results from the most
recent study, under the assumption that they supersede any previous
work. 

\subsection{HUDF}

\subsubsection{McLure et al. 2010}
In McLure et al. (2010) we published our initial analysis of the HUDF
WFC3/IR dataset, providing a list of N=49 high-redshift candidates
with $z_{phot}\geq 5.9$. The candidate selection procedure employed in
McLure et al. (2010) was broadly similar to that adopted here, with
the most noteworthy difference between the two analyses being that in
this work we have directly employed deconfused IRAC photometry in the
candidate selection procedure.

Of the N=31 objects identified in the final HUDF sample listed in
Table 2, N=28 are in common with the sample derived in McLure et
al. (2010), demonstrating an excellent level of agreement between the
two studies. However, there are N=21 candidates published in McLure et
al. (2010) which do not feature in the final robust sample derived
here. The reason behind this is that primary aim of McLure et
al. (2010) was to provide an estimate of the $z=7$ and $z=8$ galaxy
luminosity functions. Consequently, the McLure et al. (2010) sample
was designed to be as {\it complete} as possible, and therefore
contained all {\it potential} $z\geq 5.9$ LBGs revealed by the
SED-fitting analysis, irrespective of whether or not they also
displayed an acceptable low-redshift solution \footnote{Note that the
alternative low-redshift solutions were also listed by McLure et
al. (2010).}. In contrast, the principal aim of this study is to derive
a sample of $z\geq 6$ LBG candidates which is as robust as possible,
which means, in effect, requiring that any alternative low-redshift
solutions can be statistically excluded. Although all N=21 of the
additional candidates listed in McLure et al. (2010) also feature in
the initial catalogues derived here, all of them were excluded from
our final robust HUDF sample because the best-fitting alternative
solutions at low-redshift could not be excluded at the
$\Delta \chi^{2}\geq4$ ($95\%$) level.

\subsubsection{Bouwens et al. 2011}
Based on their analysis of the HUDF dataset, Bouwens et al. (2011)
list a total of N=31 robust high-redshift candidates, which are a
mixture of $z-$drops and $Y-$drops. Of these N=31 candidates, eighteen
also featured in our final robust sample of HUDF candidates listed in
Table 2. However, it is clearly of interest to investigate why the
remaining thirteen objects identified by Bouwens et al. (2011) do not
feature in our final robust sample.

Firstly, we should note that six of the thirteen additional objects
({\small UDFz-38537518, UDFy-37588003, UDFy-33446598, UDFy-39347255,
UDFy-40338026 \& UDFy-42406550}) are simply too faint to make it into
our final robust sample. None of these six objects is bright enough
(in a 0.6\asec diameter aperture) to provide a $\geq 5\sigma$
detection in any of the WFC3/IR bands, and all six are fainter than
any of our robust HUDF candidates. Consequently, this leaves a total
of seven high-redshift candidates  listed by Bouwens et al. (2011)
which could, in principle, also feature in our final robust sample. 

Our SED-fitting analysis suggests that two of the additional objects
({\small UDFz-44746449 \& UDFy-43086276}) are likely to be at
high-redshift ($z_{phot}=8.1$ and $z_{phot}=8.3$ respectively), but
just failed to make our final robust sample because the competing
low-redshift solutions could not be ruled-out at $\geq 95\%$
confidence. A further two additional objects ({\small UDFz-42567314 \&
UDFz-42247087}) also have primary photometric redshift solutions at
$z_{phot}\geq 6.0$, but were subsequently rejected because they were
either too close to the WFC3/IR array edge ({\small
UDFz-42567314}\footnote{reported as ID=1144 in McLure et al. (2010)}),
or were deemed to have unreliable photometry due to contamination from
a nearby, bright, low-redshift galaxy ({\small UDFz-42247087}). Of the
remaining three objects, one ({\small UDFy-37796001}) does have an
acceptable solution at $z_{phot}\geq 8$, but was rejected because our
analysis suggests that the alternative solution at $z_{phot}\simeq 2$
is marginally preferred. The other two ({\small UDFz-37296175 \&
UDFy-37636015}) were rejected because our SED fitting
analysis returned a primary photometric redshift solutions at $z_{phot}\simeq 5$.

Finally, it can be seen from Table 2 that our final robust sample
contains thirteen objects which are not featured in the Bouwens et al. (2011) robust candidate
list. However, the noteworthy feature of these objects is that the
vast majority (11/13) are at $z_{phot}\leq6.5$, whereas the Bouwens et
al., colour-colour, selection criteria are tuned to select objects at
$z_{phot}\geq6.5$. The two exceptions ({\small HUDF$\_1730$ \& HUDF$\_2701$}) have been
identified by several different studies (see Table 2 for details) and one ({\small HUDF$\_2701$})
does feature in the Bouwens et al (2011) list of potential high-redshift candidates.

\subsubsection{Finkelstein et al. 2010}
In their analysis of the WFC3/IR HUDF dataset, Finkelstein et
al. (2010) used a similar template-fitting technique to that employed
in both McLure et al. (2010) and this work, and used each
candidate's photometric redshift probability density function in the
construction of their final list of N=31 candidates at
$6.3~<~z_{phot}~<~8.6$. As part of their analysis, Finkelstein et
al. (2010) conducted a detailed comparison between their final list of 
high-redshift candidates and the McLure et al. (2010) sample, finding
a good level of agreement between the two studies.

As might be expected, the overall agreement between the analysis of
Finkelstein et al. (2010) and the final robust HUDF sample derived
here is still good. In the redshift range covered by both studies, our
final robust HUDF sample consists of N=22 candidates at $z_{phot}>6.3$, eighteen of
which are in common with Finkelstein et al. (2010). The four
additional candidates which feature in our final robust sample are:
{\small HUDF$\_2281$, HUDF$\_2324$, HUDF$\_2672$ \& HUDF$\_2664$} (see
Appendix B for plots of the SED fits).

Of the N=31 candidates
in the Finkelstein et al. (2010) sample, N=19 also feature in the
final HUDF sample derived here. However, this still leaves a total of
twelve candidates from Finkelstein et al. (2010) which do not feature
in our final sample. All twelve of these additional candidates do
feature in our original HUDF catalogues, but were excluded from the
final robust sample for a number of different reasons. One object
({\small FID 3022}) was excluded from our sample because it is too
faint ($J_{125}\geq 29$) to provide a robust high-redshift solution,
and a further four objects ({\small FIDs 640, 1818, 2013 \& 2432})
were excluded because they were judged to have photometry which was
potentially contaminated by bright, nearby, low-redshift galaxies. 
For the remaining seven objects ({\small FIDs 200, 213, 567, 653, 1110,
1566, \& 2055}) our SED fitting analysis does indicate that the primary
photometric redshift solution is at $z_{phot}\geq 6.3$. However, all seven
objects were excluded from the final robust sample because our analysis suggested that 
the alternative low-redshift solution could not be ruled-out with $\geq95\%$ confidence.

\begin{figure*}
\centerline{\psfig{file=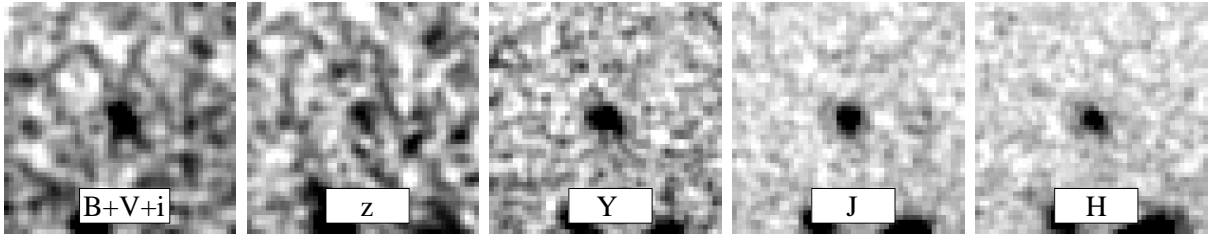,width=16.0cm,angle=0}}
\caption{Stacked postage-stamp images of three $z\simeq 7$ candidates
(ERS.z.80252, ERS.z.47667  \& ERS.z.20851)
published by Wilkins et al. (2010). All three of these objects were rejected as high-redshift candidates by
our SED fitting analysis, principally due to low signal-to-noise
detections in the blue optical bands. This figure demonstrates that
although there is a drop in flux between the $z_{850}$ and $Y_{098}$
filters, these objects are also clearly detected in a stack of the
$B_{435}+V_{606}+i_{775}$ GOODS imaging, suggesting that they are unlikely to be at $z\simeq 7$.}
\end{figure*}

\subsubsection{Yan et al. 2010}
In their analysis of the HUDF, Yan et al. (2010) used $z-$drop and
$Y-$drop criteria to identify a sample of N=35 high-redshift
candidates at $z\simeq 7$ and $z\simeq 8$. Excluding a likely
transient, Yan et al. (2010) list a total of twenty $z-$drop
candidates, fourteen of which are in common with our final robust HUDF
sample. Of the six $z-$drops listed by Yan et al. (2010) which don't
make it into our final robust HUDF sample, two ({\small A046 \& A056})
were excluded because their alternative low-redshift solutions could
not be ruled-out with $\geq95$\% confidence, one ({\small A017}) was
excluded because its photometry was contaminated by a bright,
low-redshift, galaxy and one ({\small A008}) was rejected because it
lies too close to the array edge. The final two $z-$drops ({\small
A055 \& A062}) listed by Yan et al. (2010) do not feature in any of our
catalogues and do not appear to be robust objects based on our reduction
of the epoch 1 HUDF dataset.

Yan et al. (2010) list a total of fifteen $Y-$drop candidates in the
HUDF. Of these fifteen candidates, only two  ({\small B092 \& B115})
make it through to our final robust sample. Of the thirteen $Y-$drops
listed by Yan et al. (2010) which do not feature in our final sample, our analysis suggests that 
five ({\small B041, B088, B114, B117 \& SB27}) do have acceptable
high-redshift photometric redshift solutions, but were excluded
because they all have alternative low-redshift solutions which cannot
be ruled-out at the $\geq$95\% confidence level. Two further objects
({\small B087 \& B094}) also feature in our original catalogues but,
based on our 0.6\asec diameter photometry, are not $Y-$drops and have
primary photometric redshift solutions at $z_{phot}\leq 7$. The remaining six candidates 
({\small SB30, SD02, SD05, SD15, SD24 \& SD52}) do not appear as robust objects in our 
reduction of the epoch 1 HUDF dataset. Finally, we note that Yan et al. (2010) also identify a sample of
twenty three $J-$drops in the HUDF, none of which feature in our final robust HUDF sample.

\subsubsection{Wilkins et al. (2010)}
Wilkins et al. (2010) identify a total of eleven $z-$drop candidates
in the HUDF, nine of which also feature in our final robust sample. Of
the two additional candidates listed by Wilkins et al., our analysis suggests that 
one ({\small HUDF.z.6497}) does have an acceptable solution at $z_{phot}\geq 6$, but was
was excluded because the primary photometric redshift solution lies at $z_{phot}=3.5$. 
The other object ({\small HUDF.z.6433$^{6}$}) was rejected because it lies close to the array edge 
and was therefore deemed to have unreliable photometry.

\subsubsection{Lorenzoni et al. (2011)}
Based on their analysis of the HUDF dataset, Lorenzoni et al. (2011)
identify a sample of six $Y-$drop candidates. Of these six
candidates, three ({\small HUDF.YD1, HUDF.YD3 \& HUDF.YD4}) make it
into our final robust HUDF sample. Of the remaining three candidates,
our SED fitting analysis suggests that two ({\small HUDF.YD2 \&
HUDF.YD8}) have an acceptable $z\simeq 8$ photometric redshift
solution, but were excluded from our final robust sample because they
both have an alternative low-redshift solution which cannot be
securely ruled out (i.e. $\Delta \chi^{2}\leq 4$). The remaining
candidate ({\small HUDF.YD9}) does not appear as a robust object in any of our catalogues.

\subsection{ERS}

\subsubsection{Bouwens et al. (2011)}
The robust ERS sample derived by Bouwens et al. (2011) consists of
N=19 objects in total, thirteen $z-$drops at $z\simeq 7$ and six
$Y-$drops at $z\simeq 8$. Of the thirteen $z-$drops listed by Bouwens
et al. (2011), only five appear in our final robust sample (see Table 3). 
Of the remaining eight additional $z-$drops listed by Bouwens et
al. (2011), one object ({\small ERSz-2352941047}) is too faint
($J_{125}\geq 27.5$) to produce a robust high-redshift solution based
on our criteria, leaving seven additional $z-$drops to account for. Of
these, three ({\small ERSz-2150242362, ERSz-2225141173 \&  ERSz-2354442550 }) have
statistically acceptable photometric redshift solutions at
$z_{phot}\geq 6.5$, but were excluded from the final robust sample
because it was not possible to rule-out the alternative low-redshift
solutions at $\geq 95$\% confidence. One further object ({\small
ERSz-2150943417}) was rejected because based on our photometry it
wasn't possible to obtain a statistically acceptable solution at high-redshift. 
Of the final three objects, two ({\small ERSz-2111644168 \& ERSz-2432842478}) have acceptable solutions at $z_{phot}\geq6$ but were excluded because our primary photometric redshift solution lies at $z_{phot}\leq 2$. The final object
({\small ERSz-2056344288}) does not have an acceptable high-redshift solution based on our 0.6\asec 
diameter aperture photometry.

Of the six $Y-$drops listed by Bouwens et al. (2011), two ({\small
ERSY-2354441327 \& ERSY-2029843519}) make it into are final robust
sample. Of the four additional $Y-$drops listed by Bouwens et
al. (2011), one object ({\small ERSY-2377942344}) is too faint in a
0.6\asec diameter aperture ($J_{125}\geq 27.5$) to produce a robust
high-redshift solution based on our criteria, leaving three additional
$Y-$drops to be accounted for. Of these three objects, two ({\small
ERSY-2399642019 \& ERSY-2251641574}) have acceptable primary
photometric redshift solutions at $z\geq7.5$, but were excluded
because the alternative low-redshift solutions could not be
rule-out. The final object, {\small ERSY-2306143041}\footnote{This
object was highlighted by Bouwens et al. (2011) as being potentially at low redshift.}, was rejected because our primary photometric
redshift solution is at $z_{phot}\leq 2$.

\subsubsection{Lorenzoni et al. (2011)}
Lorenzoni et al. (2011) identify a total of nine $Y-$drop candidates
in the ERS field (five of which, marked with *, are described as
``more marginal candidates'') . Of these nine candidates, only two ({\small
ERS.YD1 \& ERS.YD2*}) make it into our final robust sample. 
Of the remaining seven candidates, our analysis suggests that three
({\small ERS.YD5*, ERS.YD6 \& ERS.YD9*}) have an acceptable solution
at $z_{phot}>7$, but were rejected because the alternative low-redshift solution could not be ruled out 
at $\geq95\%$ confidence. A further two candidates ({\small
ERS.YD7* \& ERS.YD8*}) were excluded because our primary photometric
redshift solution lies at $z_{phot}\simeq 2$. The remaining two
objects ({\small ERS.YD3 \& ERS.YD4}) do not appear as robust objects in any of our catalogues.

\subsubsection{Wilkins et al. (2010)}
Based on their analysis of the ERS field, Wilkins et al. (2010)
identify a sample of eleven $z-$drop candidates, six of which  
also feature in our final robust sample. Of the five additional
candidates listed by Wilkins et al., one object ({\small
ERS.z.26813}) does have an acceptable primary photometric redshift
solution at $z_{phot}=6.6$, but was excluded from our final sample
because it has an equally acceptable solution at $z_{phot}=1.5$. 
A further object ({\small
ERS.z.70546}) was rejected because it was not possible to obtain an acceptable high-redshift SED fit.
The three remaining candidates listed by Wilkins et al. ({\small ERS.z.80252, ERS.z.47667 \& ERS.z.20851}) were rejected as low-redshift interlopers by our SED-fitting analysis due to the presence of consistent, low-level, detections in the bluer optical bands. To illustrate this
point we have stacked the ACS+WFC3/IR data for these three objects and
show the resulting postage-stamp images in Fig. 9. It can clearly be seen
that although there is a drop in flux between the $z_{850}$ and
$Y_{098}$ filters, the significant detection of flux in the stack of the
$B_{435}+V_{606}+i_{775}$ images suggests these objects are unlikely to be at $z\simeq 7$.

\subsection{HUDF09-2}

\subsubsection{Bouwens et al. (2011)}
Bouwens et al. (2011) lists a total of N=35 robust high-redshift
candidates in the HUDF09-2 field, consisting of eighteen $z$-drops and
seventeen $Y$-drops. Only seven of these thirty-five candidates appear in
our final robust sample (including {\small HUDF09-2$\_799$} which
requires a contribution from Ly$\alpha$ line emission), which clearly
requires some explanation. The principal reason for this apparent
discrepancy is that the Bouwens et al. (2011) sample contains many
fainter objects than our final robust sample. Indeed, of the
thirty-five candidates listed by Bouwens et al.,  seventeen are
fainter (in our 0.6\asec diameter aperture photometry) than the
faintest member of our final robust sample. Therefore, based on the
data utilised in this study, and our criteria for isolating robust
candidates, it is likely that these seventeen objects are simply too
faint to make it into our final robust sample. \footnote{Bouwens et al. (2011) exploit deep F814W
imaging which partially covers the HUDF09-2 field and, in some cases,
will allow the selection of fainter high-redshift candidates.}

However, even accounting for the difference in selection depth, there
are still eleven robust candidates identified by Bouwens et
al. (2011) which should, in principle, also appear in our final
robust sample. All eleven of these candidates do feature in our
HUDF09-2 catalogues, but were excluded from the final robust sample for
a number of reasons. Three of the additional candidates ({\small
UDF092z-00811320, UDF092z-07091160 \& UDF092y-07090218}) have
acceptable high-redshift photometric redshift solutions, and were
close to making our final robust candidate list. However, for these
candidates, the difference in $\chi^2$ between the primary photometric
redshift solution and the alternative low-redshift solution
($\Delta \chi^{2}\simeq 3$) did not quite match our adopted criterion
of $\Delta \chi^{2}\geq 4$. Of the remaining eight additional
candidates listed by Bouwens et al. (2011), five ({\small
UDF092y-02731564, UDF092z-09770485, UDF092z-09151531,
UDF092y-06321217 \& UDF092y-06391247}) were rejected because our primary photometric redshift 
solutions lie in the redshift interval
$4.9<z_{phot}<5.9$. The remaining three additional candidates ({\small
UDF092y-04242094, UDF092y-09611126 \& UDF092y-09661163}) were rejected
because our analysis suggests that their primary photometric redshift solutions are at
$z_{phot}\simeq 2.1$.

Finally, we should note that two of the candidates which appear in
our final robust sample ({\small HUDF09-2$\_$2455 \&
HUDF09-2$\_$2814}) feature in the Bouwens et al. (2011) list of
potential, but non-robust, high-redshift candidates. Moreover, our
final robust sample features eight candidates which do not appear in
any of the Bouwens et al. (2011) lists, although 7/8 of these
additional candidates have $z_{phot}\leq 6.3$, where the $z-$drop
criteria applied by Bouwens et al. is less sensitive.

\begin{figure}
\centerline{\psfig{file=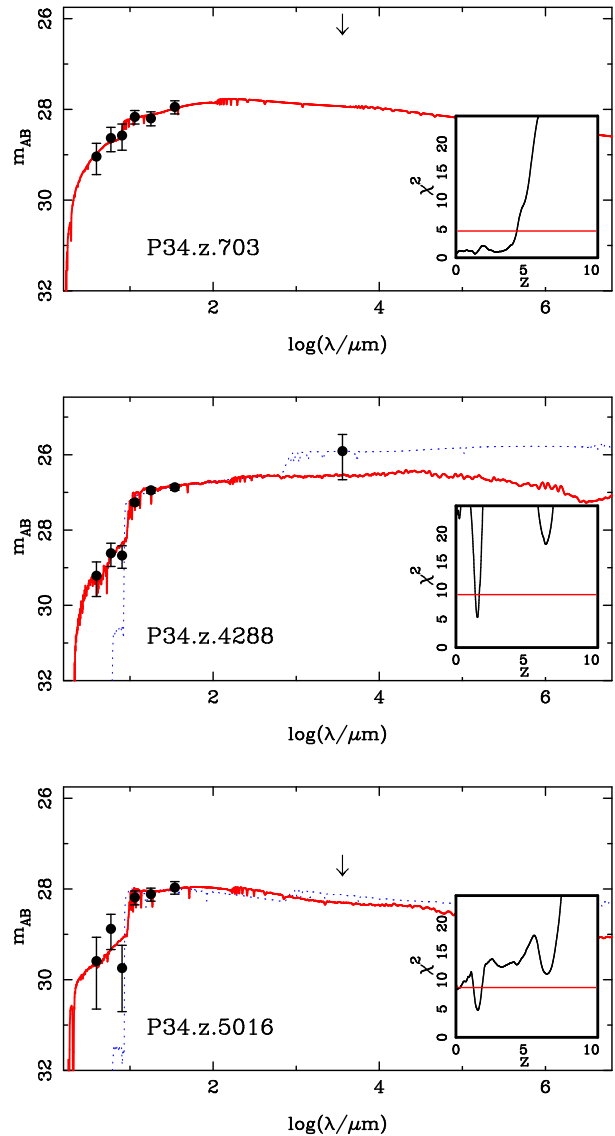,width=8.0cm,angle=0}}
\caption{The results of our SED-fitting analysis for three 
$z\simeq7$ candidates in HUDF09-2 (P34) from Wilkins et al. (2010). 
Each of these objects was rejected from our final robust sample because
our SED-fitting analysis does not return a plausible high-redshift solution.
In each panel the solid (red) line shows the best-fitting galaxy template at low redshift 
while the dotted (blue) line shows the alternative high-redshift
solution (note: there is no high-redshift solution for P34.z.703).
The best-fitting photometric redshifts for these candidates 
are $z_{phot}=1.4, 1.6 \,\& 1.6 $ for P34.z.703, P34.z.4288 and P34.z.5016 respectively.
In each panel the inset showing $\chi^{2}$ as a function of redshift
demonstrates that any solution at $z\simeq7$ has a very low probability.}
\end{figure}

\subsubsection{Wilkins et al. (2010)}
Wilkins et al. (2010) list a total of fifteen $z-$drop candidates in
the HUDF09-2 field, of which six also appear in our final robust
sample. All nine of the additional candidates listed by Wilkins et
al. (2010) feature in our initial HUDF09-2 catalogues, but were
excluded from the final sample for a number of different reasons. One
candidate ({\small P34.z.3996}) does have a valid high-redshift
solution at $z_{phot}=6.55$, and was close to making it into the final
robust sample, but was rejected because the $\Delta \chi^{2}$ between
the high-redshift and low-redshift alternative solution was too small
($\Delta \chi^{2}=3.4$). Of the remaining eight candidates, two
candidates ({\small P34.z.703 \& P34.z.2428}) were excluded because
their primary photometric redshift solutions are in the interval
$4.5<z_{phot}<5.5$. The remaining six candidates ({\small P34.z.2397,
P34.z.3053, P34.z.4288, P34.z.4501, P34.z.5016 \& P34.z.3990}) were
rejected because their primary photometric redshift solutions were all
at $z_{phot}\leq4.5$. As an illustration, in Fig. 10 we show our SED fits for three of the
$z-$drop candidates from Wilkins et al. (2010). In each case our analysis suggests 
that there is a very low probability of the candidate being at $z\simeq 7$.

\subsubsection{Lorenzoni et al. (2011)}
Lorenzoni et al. (2011) list a total of seven $Y-$drop candidates in
the HUDF09-2 field (two highlighted with a * as more marginal candidates). None of these seven candidates make it into our
final robust sample. Four of the seven Lorenzoni et al. $Y-$drop
candidates ({\small P34.YD1, P34.YD2, P34.YD3 \& P34.YD6*}) feature in our
original HUDF09-2 catalogues, and do have acceptable photometric
redshift solutions at $7.0<z_{phot}<8.5$. However, all four objects
have comparable/preferred solutions in the redshift range $1.3<z_{phot}<2.1$ and
were therefore rejected. Of the remaining objects, one ({\small
P34.YD7*}) is not a robust object in our reduction of the HUDF09-2 dataset.
The final two candidates were excluded because they were deemed to have unreliable photometry, 
either due to being within the wings of a bright star ({\small P34.YD4}) or for being too close to the noisy array edge ({\small P34.YD5}).

\section{Conclusions}
We have presented the results of a study designed to identify robust
high-redshift ($z\geq6$) galaxies using the available multi-wavelength
(ACS+WFC3/IR+IRAC) imaging covering the HUDF, HUDF09-2 and ERS
fields. By  exploiting the advantages provided by our SED-fitting
analysis, and by incorporating deconfused IRAC photometry directly
into our candidate selection procedure, we have produced a robust
sample of seventy high-redshift candidates at redshifts
$6.0<z_{phot}<8.7$. Based on this final robust sample we have
investigated the distribution of UV spectral slopes, stellar masses
and star-formation rates. Finally, we have presented the results of a
detailed comparison between our final robust sample and previous
samples of high-redshift candidates derived from the same datasets
using different selection criteria. Our main conclusions can be
summarised as follows:

\begin{enumerate}
\item{Based on our final sample of seventy robust high-redshift
candidates, and employing a variance weighted mean to account 
for the wide range in individual uncertainties, we find that the
typical value of the UV spectral 
slope is $\langle\beta\rangle=-2.05\pm0.09$.
Consequently, in contrast to some previous studies, we find no
evidence that LBGs in the redshift interval $6.0<z_{phot}<8.7$ display
UV spectral slopes which are significantly bluer than seen in
star-forming galaxies at lower redshifts.}

\item{Using the stellar-mass estimates provided by our SED-fitting we have investigated the relationship between stellar mass and UV luminosity at $z\simeq 7$, finding a best-fitting relationship of the form: $M_{\star}\propto L_{1500}^{2.1\pm0.6}$. We note that our best-fitting $M_{\star}-L_{\rm UV}$ relation is consistent with, although somewhat steeper than, the $M_{\star}-L_{\rm UV}$ relations derived by previous authors based on large samples of LBGs at $z\simeq 4$.}

\item{Focusing on a sub-sample of high-redshift candidates with
reliable IRAC photometry, we find that $L^{\star}$ LBGs at
$z~\simeq~6.5$ (i.e. $M_{1500}\simeq -20.2$) have a median stellar
mass of $M_{\star}~=~(2.1~\pm~1.1)\times10^{9}\Msun$. However, by
employing SED templates featuring a range of metallicities,
star-formation histories and reddening we find that the scatter in
stellar mass at a given UV luminosity could span a factor of $\simeq
50$. We caution that deriving stellar-mass estimates solely from
models with a constant star-formation rate may provide a misleadingly
low estimate of the real range in stellar mass at a given UV
luminosity.} 

\item{Based on the same sub-sample of twenty one objects with IRAC
detections, we find that the median specific star-formation rate
(sSFR) of $L^{\star}$ LBGs at $z\simeq 6.5$ is $1.9\pm0.8$ Gyr$^{-1}$, a value which is consistent with the sSFR $\simeq 2.5$ Gyr$^{-1}$ observed in star-forming galaxies at $z\simeq 2$. However, our SED fitting indicates that the range in the star-formation rate at a given stellar mass is potentially large and that adopting constant star-formation rate models with zero reddening may provide an underestimate of the real level of scatter. Moreover, we find that SED fitting based solely on constant star-formation rate models (with zero reddening) will inevitably tend to produce a typical sSFR close to $\simeq 2.5$ Gyr$^{-1}$ for high-redshift galaxies with ages in the range $200-600$ Myrs.}

\item{Using the sub-sample of twenty one objects with IRAC detections we have also investigated
the impact on our stellar-mass estimates of including the effects of nebular continuum and line emission in our galaxy SED templates. Based on SED templates with sub-solar metallicity (Z = 0.2 Z$_{\odot}$) we find that nebular ($\fesc=0.2$) models 
are typically capable of providing a better fit to the predominantly blue [3.6] - [4.5] $\mu$m
colours. However, in the majority of cases (16 out of 21) the best-fitting stellar masses returned by the nebular fits are
less than a factor of 2.5 lower than those returned by the stellar continuum fits. Those objects for which the nebular fits return significantly lower stellar masses (5 out of 21) are found to have a somewhat unphysical median age of 9 Myrs.}

\item{A detailed comparison between our final robust sample and previous high-redshift samples derived using different selection/analysis techniques produces mixed results. When confined to the brightest objects, and the best data, the overlap between our final robust sample and samples derived using traditional ``drop out'' criteria is reasonably good. However, at fainter magnitudes our analysis suggests that some literature samples are heavily contaminated (in some cases at the $\geq 50\%$ level) by objects which are very likely at lower redshifts.}

\end{enumerate}

\section{acknowledgements}
The authors would like to thank the anonymous referee whose comments and suggestions significantly improved the
final version of this manuscript. RJM would like to acknowledge the funding of the Royal Society via the
award of a University Research Fellowship and the Leverhulme Trust via
the award of a Philip Leverhulme research prize.  JSD acknowledges the
support of the Royal Society via a Wolfson Research Merit award, and
also the support of the European Research Council via the award of an
Advanced Grant. LDR acknowledges STFC funding via the award of a STFC
PDRA. MC acknowledges the award of an STFC Advanced Fellowship. BER is
supported by a Hubble Fellowship grant, program number
HST-HF-51262.01-A provided by NASA from the Space Telescope Science
Institute, which is operated by the Association of Universities for
Research in Astronomy, Incorporated, under NASA contract
NAS5-26555. DPS acknowledges support from the STFC through the award
of a Postdoctoral Research Fellowship.  RAAB acknowledges the support
provided by a European Research Council Advanced Grant PhD
studentship. The authors would like to acknowledge St\'{e}phane
Charlot for providing the CB07 stellar population models.

\newpage

\begin{appendix}
\section{IRAC deconfusion algorithm}
As previously discussed in Section 2.3, deep Spitzer IRAC imaging data
is available for all three of the fields analysed in this paper
($3.6\mu$m+$4.5\mu$m for the HUDF+ERS fields,  and $3.6\mu$m for
HUDF09-2) with total integration times ranging from 23 to 46 hours.
Fully exploiting the information provided by the IRAC imaging is vital
for analysing the high-redshift galaxy population for two fundamental
reasons.  Firstly, given that even the $H_{160}$ imaging data is
sampling rest-frame wavelengths of $\lambda \leq 2200$\AA\, at
$z\geq6$, the information long-ward of the 4000\AA\,break provided by
the IRAC imaging is crucial for providing constraints on the stellar
mass and the contribution of any older stellar population (see Section
5). Secondly, inclusion of the IRAC photometry in the SED fitting
process is very effective at breaking photometric redshift
degeneracies between genuine $z\geq 6$ candidates, and interlopers
at both $z\simeq 2$ and $z\simeq 5$ (see SED plots in Appendix B).

Although the availability of ultra-deep IRAC imaging is potentially
hugely beneficial, because the IRAC data is heavily confused,
obtaining accurate flux measurements for faint, high-redshift galaxies
is technically challenging. Several approaches to solving this problem have been presented in the literature, all of which rely on using a 
higher resolution image (ideally as close as possible in wavelength) as prior information to deconfuse the lower-resolution IRAC data. 
One approach, which has been recently applied to the ultra-deep IRAC data in the GOODS-N field, uses
a modified version of the CLEAN algorithm (H\"{o}gbom 1974), traditionally used in radio astronomy, 
together with model templates extracted from high-resolution, ground-based, $K-$band imaging to 
estimate IRAC fluxes via an iterative subtraction scheme (Wang et al. 2010). 
An alternative approach relies on building two-dimensional,
axisymmetric models of each the galaxy based on the available
high-resolution imaging (e.g. Labb\'{e} et al. 2006). The
two-dimensional models are then convolved to the spatial resolution of
the IRAC imaging and a $\chi^{2}$ minimization procedure is employed to
determine the individual galaxy fluxes which best reproduce the
observed IRAC image. A distinct advantage of this approach is that it
allows the precise centroiding of the individual galaxies to be
included as a free parameter in the fitting process, which can
compensate for slight astrometric differences between the
high-resolution and IRAC images. However, this method obviously has the
disadvantage of relying on axisymmetric galaxy models, and is only
really suitable for deconfusing relatively small areas of IRAC imaging
at a time.

The method adopted in this paper relies on a related, but different,
approach, whereby the actual two-dimensional light distributions of individual
galaxies in the HST imaging are used as the model templates. In this
approach, {\sc sextractor} is used to produce a normalized template of each
individual galaxy based on either the $J_{125}$ or $H_{160}$ WFC3/IR imaging, 
and is transformed to IRAC spatial resolution via convolution with a transfer function:
\begin{equation}
W_{PSF}*T=I_{PSF} 
\end{equation}
\noindent
where $T$ is the transfer function and $W_{PSF}$ and $I_{PSF}$ are the
WFC3/IR and IRAC PSFs respectively. The fundamental assumption behind
this technique is that it is possible to reproduce the
observed IRAC image using a linear combination of these galaxy
templates. The key advantage of this approach is that,
provided the astrometry match between the WFC3/IR and IRAC images is sufficiently 
accurate that the centroid of each galaxy template can be held fixed, 
the amplitude of each template can be uniquely (and analytically)
determined by $\chi^{2}$ minimisation, as follows:
\begin{equation}
\chi^{2}=\displaystyle\sum_{ij}\frac{\left[D_{ij}-\sum_{k=1}^{n}a_{k}M_{ij}^{k}\right]^{2}}{\sigma_{ij}^{2}}
\end{equation}
\noindent
where $D_{ij}$ is the $(i,j)$-th pixel of the IRAC image, $\sigma_{ij}$ is the
corresponding uncertainty and $M_{ij}^{k}$ is the $(i,j)$-th pixel of the $k$-th galaxy template. The minimum
$\chi^{2}$ occurs when:
\begin{equation}
\frac{\delta \chi^{2}}{\delta a_k}=2\displaystyle\sum_{ij}
\frac{\left[D_{ij}-\sum_{k=1}^{n}a_{k}M_{ij}^{k}\right]M_{ij}^{m}}{\sigma_{ij}^{2}}=0
\end{equation}
\noindent
which can be rearranged as:
\begin{equation}
\displaystyle\sum_{k=1}^{n}\left[\displaystyle\sum_{ij}\frac{M_{ij}^{k}M_{ij}^{m}}{\sigma_{ij}^{2}}\right]a_{k}
=\displaystyle\sum_{ij}\frac{D_{ij}M_{ij}^{m}}{\sigma_{ij}^{2}}
\end{equation}
\noindent
which describes a set of $n$ linear equations, which can be re-cast as
a matrix equation:

\begin{equation}
\mathbf{A\underbar{a}=\underbar{b}}
\end{equation}
\noindent
where\\
$\mathbf{A_{mk}}=\displaystyle\sum_{ij}\frac{M_{ij}^{m}M_{ij}^{k}}{\sigma_{ij}^{2}}$ \,\,\,\,and\,\,\,\,
$b_{k}=\displaystyle\sum_{ij}\frac{D_{ij}M_{ij}^{k}}{\sigma_{ij}^{2}}$

\medskip

\noindent
From equation A5 it is possible to find $\mathbf{\underbar{a}}$, the $n$-dimensional vector of galaxy template
amplitudes, using standard matrix inversion techniques. Moreover, the
variance of the fitted amplitudes are automatically provided by the
diagonal terms of the inverse matrix:
\begin{equation}
\sigma^{2}(a_{k})=\mathbf{A}^{-1}_{kk}
\end{equation}
\noindent
We should note that the basic algorithm outlined here is very
similar to that employed by the publicly available TFIT (Laidler et al. 2007) and ConvPhot (De Santis et al. 2007) software packages.

\section{Candidate Photometry, SED fits and postage-stamp images}
In Tables B1, B2 \& B3 we provide the photometry for each member of
our final robust sample. In Figures B1, B2 \& B3 we illustrate the
results of our SED-fitting analysis by showing the best-fitting
high-redshift galaxy template, the best-fitting low-redshift
alternative solution and the dependence of $\chi^{2}$ on photometric
redshift. This information is included to allow the reader to judge
for themselves the robustness of each high-redshift
candidate. In Figure B4 we show the results of fitting the twenty-one object sub-sample
detected at 3.6$\mu$m with SED models featuring nebular continuum and line emission.
Finally, in Figures B5, B6 \& B7 we also provide 3\asec
$\times$ 3\asec postage-stamp images of each high-redshift candidate
in the  $z_{850}, Y_{105}/Y_{098}, J_{125} \,\&\, H_{160}$ filters.

\begin{table*}
\caption{Basic observational properties of the final high-redshift
galaxy sample in the HUDF field. Columns one to three list the
candidate IDs and coordinates. The remaining columns list the
photometry of each candidate in the F775W, F850LP, F105W, F160W, IRAC1
($3.6\mu$m) and IRAC2 ($4.5\mu$m) filters, as measured in a 0.6\asec
diameter aperture, along with their corresponding uncertainties. The
magnitudes listed here are derived from the actual fluxes used in the
SED fitting and are not corrected to total, but have
been corrected for galactic extinction and the relative aperture losses between ACS and
WFC3/IR. All candidates are undetected in filters at shorter wavelengths than F775W,
and all detections which are significant at less than the $2\sigma$
level are listed as $2\sigma$ upper limits. It should be noted that because
the IRAC photometry is derived via a deconfusion process the
uncertainties and $2\sigma$ limits are highly position dependent.}
\begin{tabular}{lccccccccc}
\hline
ID & RA(J2000)&DEC(J2000)&F775W & F850LP & F105W & F125W & F160W &IRAC1&IRAC2\\
\hline
HUDF\_1344 &03:32:36.63&$-27$:47:50.1&$>$30.28&27.85$^{+0.16}_{-0.14}$&27.66$^{+0.11}_{-0.10}$&27.70$^{+0.09}_{-0.08}$&27.77$^{+0.09}_{-0.08}$&$>$26.83&$>$26.78\\[2ex]

HUDF\_1016 &03:32:35.06&$-27$:47:40.2&$>$30.21&27.83$^{+0.10}_{-0.09}$&27.49$^{+0.08}_{-0.07}$&27.20$^{+0.08}_{-0.07}$&27.27$^{+0.08}_{-0.07}$&26.94$^{+0.37}_{-0.28}$&$>$27.38\\[2ex]

HUDF\_522  &03:32:36.47&$-27$:46:41.4&28.88$^{+0.12}_{-0.11}$&26.54$^{+0.08}_{-0.07}$&26.18$^{+0.08}_{-0.07}$&26.16$^{+0.08}_{-0.07}$&26.20$^{+0.08}_{-0.07}$&25.62$^{+0.24}_{-0.20}$&26.45$^{+0.45}_{-0.32}$\\[2ex]

HUDF\_2622 &03:32:36.64&$-27$:47:50.2&$>$30.28&28.45$^{+0.30}_{-0.23}$&28.21$^{+0.18}_{-0.15}$&28.24$^{+0.14}_{-0.13}$&28.33$^{+0.14}_{-0.13}$&$>$26.64&$>$26.52\\[2ex]

HUDF\_796  &03:32:37.46&$-27$:46:32.8&$>$29.90&27.72$^{+0.08}_{-0.07}$&27.08$^{+0.08}_{-0.07}$&27.02$^{+0.08}_{-0.07}$&26.88$^{+0.08}_{-0.07}$&$>$25.36&25.00$^{+0.47}_{-0.33}$\\[2ex]

HUDF\_2836 &03:32:35.05&$-27$:47:25.8&$>$30.55&29.03$^{+0.38}_{-0.28}$&28.44$^{+0.25}_{-0.20}$&28.92$^{+0.43}_{-0.31}$&28.99$^{+0.24}_{-0.19}$&$>$26.64&$>$26.52\\[2ex]

HUDF\_1692 &03:32:43.03&$-27$:46:23.6&$>$30.07&28.03$^{+0.17}_{-0.15}$&27.62$^{+0.11}_{-0.10}$&27.88$^{+0.11}_{-0.10}$&28.14$^{+0.16}_{-0.14}$&$>$27.26&$>$27.03\\[2ex]

HUDF\_2743 &03:32:36.52&$-27$:46:42.0&$>$30.21&29.37$^{+0.58}_{-0.38}$&28.90$^{+0.33}_{-0.25}$&28.75$^{+0.28}_{-0.22}$&29.01$^{+0.42}_{-0.30}$&$>$26.64&$>$26.52\\[2ex]

HUDF\_2316 &03:32:44.31&$-27$:46:45.2&$>$30.76&28.89$^{+0.38}_{-0.28}$&28.38$^{+0.28}_{-0.22}$&28.33$^{+0.20}_{-0.17}$&28.74$^{+0.34}_{-0.26}$&$>$27.14&$>$26.97\\[2ex]

HUDF\_2281 &03:32:39.79&$-27$:46:33.7&$>$29.90&29.10$^{+0.37}_{-0.28}$&28.40$^{+0.16}_{-0.14}$&28.62$^{+0.30}_{-0.23}$&28.70$^{+0.25}_{-0.21}$&$>$27.04&$>$26.86\\[2ex]

HUDF\_1442 &03:32:42.19&$-27$:46:27.8&$>$30.01&28.77$^{+0.26}_{-0.21}$&27.93$^{+0.10}_{-0.09}$&27.83$^{+0.09}_{-0.09}$&27.87$^{+0.08}_{-0.07}$&$>$26.83&26.13$^{+0.39}_{-0.29}$\\[2ex]

HUDF\_2324 &03:32:41.60&$-27$:47:04.5&$>$29.95&29.18$^{+0.35}_{-0.26}$&28.39$^{+0.18}_{-0.16}$&28.55$^{+0.20}_{-0.17}$&28.75$^{+0.23}_{-0.19}$&$>$27.07&$>$26.92\\[2ex]

HUDF\_2672 &03:32:37.80&$-27$:47:40.4&$>$30.21&29.18$^{+0.49}_{-0.34}$&28.38$^{+0.20}_{-0.17}$&28.60$^{+0.23}_{-0.19}$&28.59$^{+0.21}_{-0.17}$&$>$26.64&$>$26.52\\[2ex]

HUDF\_1818 &03:32:36.38&$-27$:47:16.3&$>$30.21&29.18$^{+0.47}_{-0.33}$&27.95$^{+0.12}_{-0.11}$&28.22$^{+0.20}_{-0.17}$&28.28$^{+0.14}_{-0.13}$&$>$27.19&$>$27.02\\[2ex]

HUDF\_1473 &03:32:36.77&$-27$:47:53.6&$>$29.90&28.98$^{+0.41}_{-0.29}$&27.78$^{+0.09}_{-0.09}$&27.96$^{+0.10}_{-0.09}$&27.91$^{+0.08}_{-0.07}$&$>$27.34&$>$27.14\\[2ex]

HUDF\_1730 &03:32:43.78&$-27$:46:33.7&$>$30.07&29.16$^{+0.48}_{-0.33}$&28.04$^{+0.12}_{-0.11}$&28.03$^{+0.10}_{-0.09}$&28.18$^{+0.17}_{-0.15}$&$>$26.78&$>$26.65\\[2ex]

HUDF\_1632 &03:32:37.44&$-27$:46:51.2&$>$30.07&29.10$^{+0.32}_{-0.24}$&27.97$^{+0.10}_{-0.09}$&28.06$^{+0.12}_{-0.11}$&28.08$^{+0.14}_{-0.13}$&$>$27.21&$>$27.04\\[2ex]

HUDF\_2084 &03:32:40.57&$-27$:46:43.6&$>$30.45&29.57$^{+0.50}_{-0.34}$&28.19$^{+0.18}_{-0.16}$&28.49$^{+0.16}_{-0.14}$&28.54$^{+0.14}_{-0.13}$&$>$27.42&$>$27.17\\[2ex]

HUDF\_1995 &03:32:39.58&$-27$:46:56.5&$>$30.01&29.29$^{+0.72}_{-0.43}$&28.24$^{+0.20}_{-0.17}$&28.12$^{+0.18}_{-0.15}$&28.45$^{+0.15}_{-0.13}$&$>$27.00&$>$26.90\\[2ex]

HUDF\_658  &03:32:42.56&$-27$:46:56.6&$>$30.37&28.13$^{+0.20}_{-0.17}$&26.73$^{+0.08}_{-0.07}$&26.53$^{+0.08}_{-0.07}$&26.54$^{+0.08}_{-0.07}$&25.65$^{+0.24}_{-0.20}$&26.06$^{+0.29}_{-0.23}$\\[2ex]

HUDF\_2701 &03:32:41.82&$-27$:46:11.3&$>$30.14&29.79$^{+0.70}_{-0.42}$&28.62$^{+0.31}_{-0.24}$&28.66$^{+0.23}_{-0.19}$&28.96$^{+0.30}_{-0.24}$&$>$26.64&$>$26.52\\[2ex]

HUDF\_860  &03:32:38.81&$-27$:47:07.2&$>$30.55&$>$29.75&27.61$^{+0.10}_{-0.09}$&27.14$^{+0.08}_{-0.07}$&27.01$^{+0.08}_{-0.07}$&25.69$^{+0.24}_{-0.20}$&26.12$^{+0.26}_{-0.21}$\\[2ex]

HUDF\_1102 &03:32:39.55&$-27$:47:17.5&$>$29.84&$>$29.49&28.08$^{+0.15}_{-0.13}$&27.53$^{+0.08}_{-0.07}$&27.42$^{+0.09}_{-0.09}$&$>$26.30&$>$26.20\\[2ex]

HUDF\_1419 &03:32:43.13&$-27$:46:28.5&$>$30.28&$>$29.75&28.34$^{+0.16}_{-0.14}$&28.00$^{+0.13}_{-0.12}$&27.84$^{+0.09}_{-0.09}$&$>$26.69&$>$26.62\\[2ex]

HUDF\_2641 &03:32:39.73&$-27$:46:21.3&$>$30.14&$>$29.39&28.96$^{+0.38}_{-0.28}$&28.41$^{+0.19}_{-0.16}$&28.92$^{+0.29}_{-0.23}$&$>$26.64&$>$26.52\\[2ex]

HUDF\_1962 &03:32:38.36&$-27$:46:11.9&$>$30.21&$>$29.53&28.69$^{+0.33}_{-0.25}$&28.14$^{+0.20}_{-0.17}$&28.42$^{+0.17}_{-0.15}$&$>$26.55&$>$26.61\\[2ex]

HUDF\_1173 &03:32:44.70&$-27$:46:44.3&30.45$^{+0.75}_{-0.44}$&$>$29.57&28.07$^{+0.31}_{-0.24}$&27.45$^{+0.10}_{-0.10}$&27.53$^{+0.10}_{-0.09}$&26.92$^{+0.66}_{-0.41}$&$>$26.87\\[2ex]

HUDF\_2664 &03:32:33.13&$-27$:46:54.5&$>$29.95&$>$29.53&29.18$^{+0.43}_{-0.31}$&28.55$^{+0.26}_{-0.21}$&29.18$^{+0.63}_{-0.40}$&$>$26.64&$>$26.52\\[2ex]

HUDF\_1660 &03:32:37.21&$-27$:48:06.2&$>$30.14&$>$29.75&28.59$^{+0.20}_{-0.17}$&27.99$^{+0.09}_{-0.09}$&28.12$^{+0.13}_{-0.12}$&$>$27.24&$>$27.01\\[2ex]

HUDF\_1679 &03:32:42.88&$-27$:46:34.5&$>$30.14&$>$29.49&29.14$^{+0.24}_{-0.20}$&28.14$^{+0.16}_{-0.14}$&28.13$^{+0.10}_{-0.09}$&$>$27.32&$>$27.04\\[2ex]

HUDF\_2003 &03:32:38.13&$-27$:45:54.0&$>$30.28&$>$29.79&$>$29.66&28.40$^{+0.20}_{-0.17}$&28.46$^{+0.19}_{-0.16}$&$>$27.38&$>$27.10\\

\hline\hline
\end{tabular}
\end{table*}

\begin{table*}
\caption{Basic observational properties of the final high-redshift
galaxy sample in the ERS field. Columns one to three list the
candidate IDs and coordinates. The remaining columns list the
photometry of each candidate in the F775W, F850LP, F098M, F160W, IRAC1
($3.6\mu$m) and IRAC2 ($4.5\mu$m) filters, as measured in a 0.6\asec
diameter aperture, along with their corresponding uncertainties. The
magnitudes listed here are derived from the actual fluxes used in the
SED fitting and are not corrected to total, but have
been corrected for galactic extinction and the relative aperture losses between ACS and
WFC3/IR. All candidates are undetected in filters at shorter wavelengths than F775W,
and all detections which are significant at less than the $2\sigma$
level are listed as $2\sigma$ upper limits. It should be noted that because
the IRAC photometry is derived via a deconfusion process the
uncertainties and $2\sigma$ limits are highly position dependent.}
\begin{tabular}{lccccccccc}
\hline
ID & RA(J2000)&DEC(J2000)&F775W & F850LP & F098M & F125W & F160W &IRAC1&IRAC2\\
\hline
ERS\_7086 &03:32:34.75&$-27$:40:35.1  &$>$28.28&27.25$^{+0.21}_{-0.18}$&26.78$^{+0.12}_{-0.11}$&26.89$^{+0.12}_{-0.11}$&26.99$^{+0.17}_{-0.14}$&26.82$^{+0.63}_{-0.40}$&$>$26.79\\[2ex]

ERS\_6066 &03:32:07.86&$-27$:42:17.8&$>$28.19&27.27$^{+0.48}_{-0.33}$&26.55$^{+0.15}_{-0.13}$&26.68$^{+0.08}_{-0.08}$&26.50$^{+0.13}_{-0.11}$&24.86$^{+0.24}_{-0.20}$&24.99$^{+0.24}_{-0.20}$\\[2ex]

ERS\_9869 &03:32:15.40&$-27$:43:28.6  &$>$28.04&27.52$^{+0.32}_{-0.25}$&26.99$^{+0.23}_{-0.19}$&27.35$^{+0.16}_{-0.14}$&27.53$^{+0.34}_{-0.26}$&$>$26.45&$>$26.39\\[2ex]

ERS\_8668 &03:32:27.96&$-27$:41:19.0  &$>$28.41&27.63$^{+0.48}_{-0.33}$&27.11$^{+0.15}_{-0.14}$&27.17$^{+0.16}_{-0.14}$&27.56$^{+0.36}_{-0.27}$&$>$26.65&$>$26.58\\[2ex]

ERS\_9100 &03:32:20.24&$-27$:43:34.3  &$>$28.45&27.87$^{+0.58}_{-0.38}$&27.16$^{+0.10}_{-0.09}$&27.24$^{+0.12}_{-0.11}$&27.20$^{+0.14}_{-0.13}$&26.39$^{+0.31}_{-0.24}$&26.52$^{+0.43}_{-0.31}$\\[2ex]

ERS\_7225 &03:32:36.31&$-27$:40:15.0  &$>$28.12&$>$27.88&27.00$^{+0.18}_{-0.15}$&26.92$^{+0.08}_{-0.08}$&26.82$^{+0.16}_{-0.14}$&24.88$^{+0.24}_{-0.20}$&25.23$^{+0.24}_{-0.20}$\\[2ex]

ERS\_6438 &03:32:25:28&$-27$:43:24.2  &$>$27.99&$>$27.82&26.77$^{+0.15}_{-0.13}$&26.76$^{+0.10}_{-0.09}$&26.57$^{+0.11}_{-0.10}$&24.83$^{+0.30}_{-0.23}$&25.73$^{+0.58}_{-0.38}$\\[2ex]

ERS\_6263 &03:32:06.83&$-27$:44:22.2  &$>$28.17&27.42$^{+0.40}_{-0.29}$&26.71$^{+0.17}_{-0.15}$&26.73$^{+0.14}_{-0.12}$&26.95$^{+0.18}_{-0.15}$&26.25$^{+0.39}_{-0.29}$&$>$26.77\\[2ex]

ERS\_7776 &03:32:03.77&$-27$:44:54.4  &$>$28.29&27.95$^{+0.50}_{-0.34}$&27.12$^{+0.20}_{-0.17}$&27.03$^{+0.12}_{-0.11}$&27.28$^{+0.16}_{-0.14}$&26.36$^{+0.31}_{-0.24}$&$>$26.98\\[2ex]

ERS\_5847 &03:32:16.00&$-27$:43:01.4  &$>$28.52&27.68$^{+0.28}_{-0.22}$&26.60$^{+0.10}_{-0.09}$&26.63$^{+0.08}_{-0.07}$&26.73$^{+0.19}_{-0.16}$&25.61$^{+0.69}_{-0.42}$&$>$25.65\\[2ex]

ERS\_8987 &03:32:16.01&$-27$:41:59.0&$>$28.06&$>$28.16&27.62$^{+0.46}_{-0.32}$&27.22$^{+0.16}_{-0.14}$&27.17$^{+0.19}_{-0.16}$&$>$25.36&24.58$^{+0.35}_{-0.26}$\\[2ex]

ERS\_3679 &03:32:22.66&$-27$:43:00.7  &$>$28.24&27.28$^{+0.22}_{-0.18}$&26.19$^{+0.09}_{-0.08}$&25.94$^{+0.08}_{-0.07}$&25.91$^{+0.08}_{-0.07}$&25.07$^{+0.26}_{-0.21}$&25.67$^{+0.50}_{-0.34}$\\[2ex]

ERS\_7412 &03:32:09.85&$-27$:43:24.0  &$>$28.18&$>$27.87&26.76$^{+0.13}_{-0.11}$&26.96$^{+0.10}_{-0.10}$&26.66$^{+0.12}_{-0.11}$&26.89$^{+0.75}_{-0.44}$&$>$26.61\\[2ex]

ERS\_6427 &03:32:24.09&$-27$:42:13.9  &$>$28.38&$>$27.89&27.14$^{+0.26}_{-0.21}$&26.76$^{+0.10}_{-0.09}$&26.87$^{+0.09}_{-0.08}$&25.67$^{+0.30}_{-0.23}$&25.92$^{+0.26}_{-0.21}$\\[2ex]

ERS\_8858 &03:32:16.19&$-27$:41:49.8&$>$28.12&$>$27.70&27.47$^{+0.36}_{-0.27}$&27.20$^{+0.14}_{-0.13}$&27.40$^{+0.14}_{-0.13}$&$>$27.15&$>$26.94\\[2ex]

ERS\_7376 &03:32:29.54&$-27$:42:04.5  &$>$28.07&$>$28.19&27.17$^{+0.23}_{-0.19}$&26.95$^{+0.10}_{-0.09}$&27.05$^{+0.27}_{-0.22}$&26.44$^{+0.69}_{-0.42}$&$>$26.41\\[2ex]

ERS\_8176 &03:32:23.15&$-27$:42:04.7  &$>$28.01&$>$28.25&27.06$^{+0.22}_{-0.19}$&27.09$^{+0.13}_{-0.12}$&27.50$^{+0.24}_{-0.19}$&$>$27.06&$>$26.83\\[2ex]

ERS\_7672 &03:32:10.03&$-27$:45:24.6  &$>$28.06&$>$28.01&27.07$^{+0.23}_{-0.19}$&27.01$^{+0.21}_{-0.18}$&27.10$^{+0.14}_{-0.12}$&$>$26.94&$>$26.60\\[2ex]

ERS\_7475 &03:32:32.81&$-27$:42:38.5  &$>$28.49&$>$28.28&27.42$^{+0.37}_{-0.28}$&26.97$^{+0.11}_{-0.10}$&26.94$^{+0.11}_{-0.10}$&$>$26.47&$>$26.34\\[2ex]

ERS\_7236 &03:32:11.51&$-27$:45:17.1  &$>$28.07&$>$27.96&27.67$^{+0.34}_{-0.26}$&26.93$^{+0.14}_{-0.12}$&27.25$^{+0.13}_{-0.12}$&$>$26.40&$>$26.19\\[2ex]

ERS\_9041 &03:32:23.37&$-27$:43:26.5  &$>$28.39&$>$27.84&$>$28.48&27.23$^{+0.10}_{-0.09}$&27.91$^{+0.38}_{-0.28}$&$>$27.01&$>$26.82\\[2ex]

ERS\_10288&03:32:35.44&$-27$:41:32.7&$>$28.46&$>$27.97&$>$28.76&27.40$^{+0.12}_{-0.11}$&27.47$^{+0.16}_{-0.14}$&$>$27.19&$>$26.89\\[2ex]

ERS\_8584 &03:32:02.99&$-27$:43:51.9&$>$28.16&$>$28.41&$>$28.52&27.16$^{+0.22}_{-0.18}$&27.08$^{+0.17}_{-0.14}$&$>$26.32&25.79$^{+0.54}_{-0.36}$\\[2ex]

\hline\hline
ERS\_8496 &03:32:29.69&$-27$:40:49.9&$>$28.63&$>$27.81&27.18$^{+0.27}_{-0.21}$&27.14$^{+0.18}_{-0.15}$&27.41$^{+0.33}_{-0.25}$&25.36$^{+0.24}_{-0.20}$&25.91$^{+0.37}_{-0.28}$\\[2ex]

ERS\_9923 &03:32:10.06&$-27$:45:22.6  &$>$28.00&$>$27.92&26.82$^{+0.20}_{-0.17}$&27.36$^{+0.28}_{-0.22}$&27.29$^{+0.16}_{-0.14}$&$>$26.95&$>$26.71\\[2ex]

\hline\hline
\end{tabular}
\end{table*}

\begin{table*}
\caption{Basic observational properties of the final high-redshift
galaxy sample in the HUDF09-2 field. Columns one to three list the
candidate IDs and coordinates. The remaining columns list the
photometry of each candidate in the F775W, F850LP, F105W, F160W and IRAC1
($3.6\mu$m) filters, as measured in a 0.6\asec
diameter aperture, along with their corresponding uncertainties. The
magnitudes listed here are derived from the actual fluxes used in the
SED fitting and are not corrected to total, but have
been corrected for galactic extinction and the relative aperture losses between ACS and
WFC3/IR. All candidates are undetected in filters at shorter
wavelengths than F775W, and all detections which are significant at less than the $2\sigma$
level are listed as $2\sigma$ upper limits. It should be noted that because
the IRAC photometry is derived via a deconfusion process the
uncertainties and $2\sigma$ limits are highly position dependent.}
\begin{tabular}{lcccccccc}
\hline
ID & RA(J2000)&DEC(J2000)&F775W & F850LP & F105W & F125W & F160W &IRAC1\\
\hline
 HUDF09-2\_2459 &03:33:06.30&$-27$:50:20.2&$>$30.07&27.88$^{+0.15}_{-0.13}$&27.83$^{+0.16}_{-0.14}$&27.73$^{+0.10}_{-0.09}$&27.94$^{+0.12}_{-0.11}$&$>$25.08\\[2ex]
 HUDF09-2\_2613 &03:33:06.52&$-27$:50:34.6&$>$29.14&27.92$^{+0.20}_{-0.17}$&27.74$^{+0.13}_{-0.11}$&27.91$^{+0.15}_{-0.13}$&28.20$^{+0.21}_{-0.18}$&$>$25.89\\[2ex]
 HUDF09-2\_2638 &03:33:06.65&$-27$:50:30.2&$>$29.17&28.50$^{+0.33}_{-0.25}$&28.02$^{+0.16}_{-0.14}$&28.03$^{+0.14}_{-0.12}$&28.06$^{+0.16}_{-0.14}$&$>$26.17\\[2ex]
 HUDF09-2\_1543 &03:33:01.18&$-27$:51:22.3&$>$29.02&27.08$^{+0.08}_{-0.07}$&26.62$^{+0.08}_{-0.07}$&26.66$^{+0.08}_{-0.07}$&26.77$^{+0.08}_{-0.07}$&26.31$^{+0.42}_{-0.30}$\\[2ex]
 HUDF09-2\_605  &03:33:01.95&$-27$:52:03.2&$>$29.07&28.20$^{+0.23}_{-0.19}$&27.53$^{+0.10}_{-0.09}$&27.51$^{+0.08}_{-0.08}$&27.56$^{+0.11}_{-0.10}$&$>$24.87\\[2ex]
 HUDF09-2\_2587 &03:33:04.20&$-27$:50:31.3&$>$29.84&27.47$^{+0.09}_{-0.09}$&26.73$^{+0.08}_{-0.07}$&26.60$^{+0.08}_{-0.07}$&26.63$^{+0.08}_{-0.07}$&25.71$^{+0.45}_{-0.32}$\\[2ex]
 HUDF09-2\_1660 &03:33:01.10&$-27$:51:16.0&29.12$^{+0.73}_{-0.43}$&27.65$^{+0.16}_{-0.14}$&26.82$^{+0.08}_{-0.07}$&26.64$^{+0.08}_{-0.07}$&26.71$^{+0.08}_{-0.07}$&25.71$^{+0.41}_{-0.30}$\\[2ex]
 HUDF09-2\_1745 &03:33:01.19&$-27$:51:13.3&$>$29.14&28.90$^{+0.57}_{-0.37}$&27.70$^{+0.13}_{-0.12}$&27.74$^{+0.09}_{-0.09}$&27.72$^{+0.11}_{-0.10}$&$>$25.80\\[2ex]
 HUDF09-2\_1620 &03:33:05.40&$-27$:51:18.8&$>$29.07&$>$28.78&27.88$^{+0.14}_{-0.13}$&28.15$^{+0.22}_{-0.19}$&28.07$^{+0.13}_{-0.12}$&$>$24.37\\[2ex]
 HUDF09-2\_1721 &03:33:01.17&$-27$:51:13.9&$>$29.20&28.75$^{+0.46}_{-0.32}$&27.69$^{+0.15}_{-0.13}$&27.23$^{+0.08}_{-0.07}$&27.24$^{+0.08}_{-0.07}$&$>$25.26\\[2ex]
 HUDF09-2\_2455 &03:33:09.65&$-27$:50:50.8&$>$29.20&28.17$^{+0.16}_{-0.14}$&26.58$^{+0.08}_{-0.07}$&26.64$^{+0.08}_{-0.07}$&26.62$^{+0.08}_{-0.07}$&$>$22.60\\[2ex]
 HUDF09-2\_1584 &03:33:03.79&$-27$:51:20.4&$>$29.29&$>$28.80&27.20$^{+0.10}_{-0.09}$&26.67$^{+0.08}_{-0.07}$&26.58$^{+0.08}_{-0.07}$&24.96$^{+0.44}_{-0.31}$\\[2ex]
 HUDF09-2\_2814 &03:33:07.05&$-27$:50:55.5&$>$28.88&$>$28.86&28.09$^{+0.24}_{-0.20}$&27.56$^{+0.09}_{-0.08}$&27.67$^{+0.09}_{-0.08}$&$>$26.21\\[2ex]
 HUDF09-2\_1596 &03:33:03.76&$-27$:51:19.7&$>$29.39&$>$28.95&27.35$^{+0.09}_{-0.08}$&26.83$^{+0.08}_{-0.07}$&26.82$^{+0.08}_{-0.07}$&$>$24.52\\[2ex]
 HUDF09-2\_2000 &03:33:04.64&$-27$:50:53.0&$>$29.70&$>$29.79&28.24$^{+0.26}_{-0.21}$&27.53$^{+0.08}_{-0.07}$&27.58$^{+0.10}_{-0.09}$&$>$26.75\\[2ex]
 HUDF09-2\_2765 &03:33:07.58&$-27$:50:55.0&$>$29.29&$>$28.97&$>$29.26&27.85$^{+0.12}_{-0.11}$&27.52$^{+0.08}_{-0.07}$&$>$27.18\\[2ex]
\hline\hline
HUDF09-2\_799 &03:33:09.15&$-27$:51:55.4&$>$29.23&$>$29.06&27.45$^{+0.09}_{-0.08}$&27.72$^{+0.14}_{-0.13}$&27.63$^{+0.16}_{-0.13}$&$>$25.69\\
\hline\hline
\end{tabular}
\end{table*}

\clearpage

\begin{figure*}
\centerline{\psfig{file=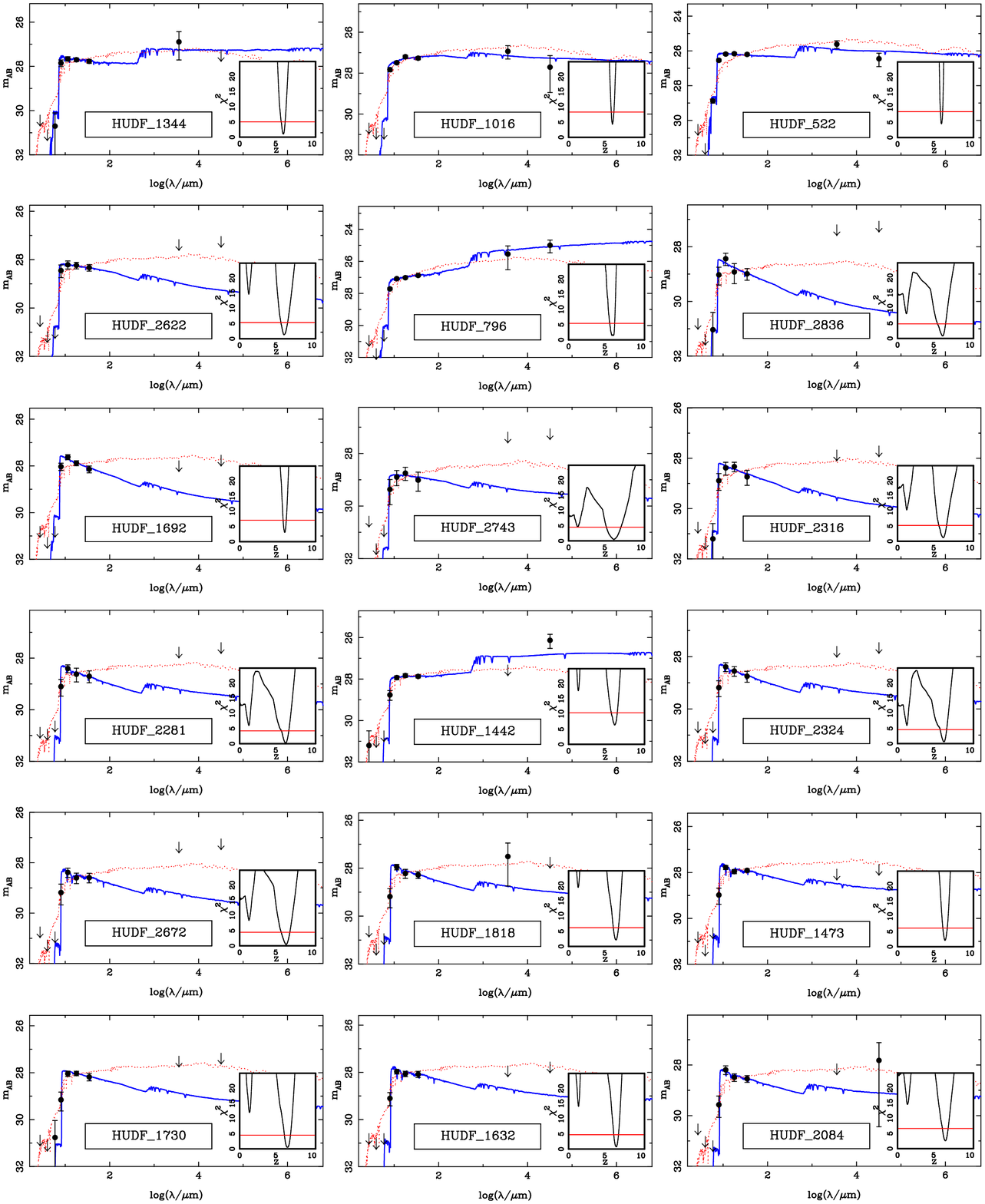,width=18.0cm,angle=0}}
\caption{SED fits for each member of the final HUDF sub-sample. In
each plot the solid (blue) line is the best-fitting $z\geq 6$ galaxy
SED template, and the dotted (red) line is the best-fitting alternative low-redshift solution ($z\leq 2.5$). 
All data points which are detected at less
than $1\sigma$ significance are shown as $1\sigma$ upper limits. In
each case the inset panel shows $\chi^{2}$ versus redshift, produced
after marginalizing over all other free parameters. The 
horizontal red line in the inset is set a ($\chi^{2}_{min}+4$), and
indicates the threshold for determining whether or not the competing
low-redshift solution can be excluded at the $95\%$ confidence level.}
\end{figure*}

\setcounter{figure}{0}
\begin{figure*}
\centerline{\psfig{file=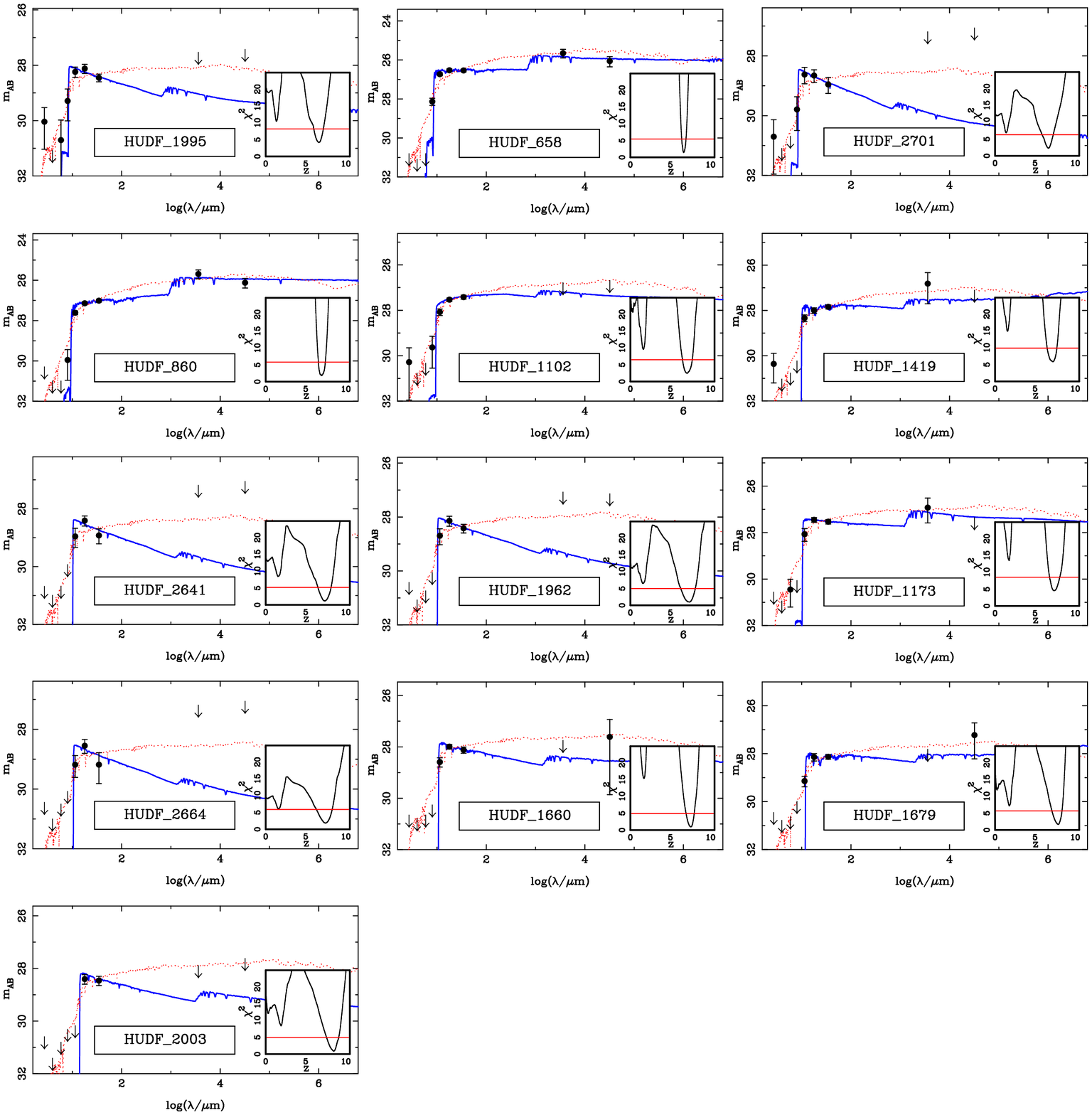,width=18.0cm,angle=0}}
\caption{Continued.}
\end{figure*}

\setcounter{figure}{1}
\begin{figure*}
\centerline{\psfig{file=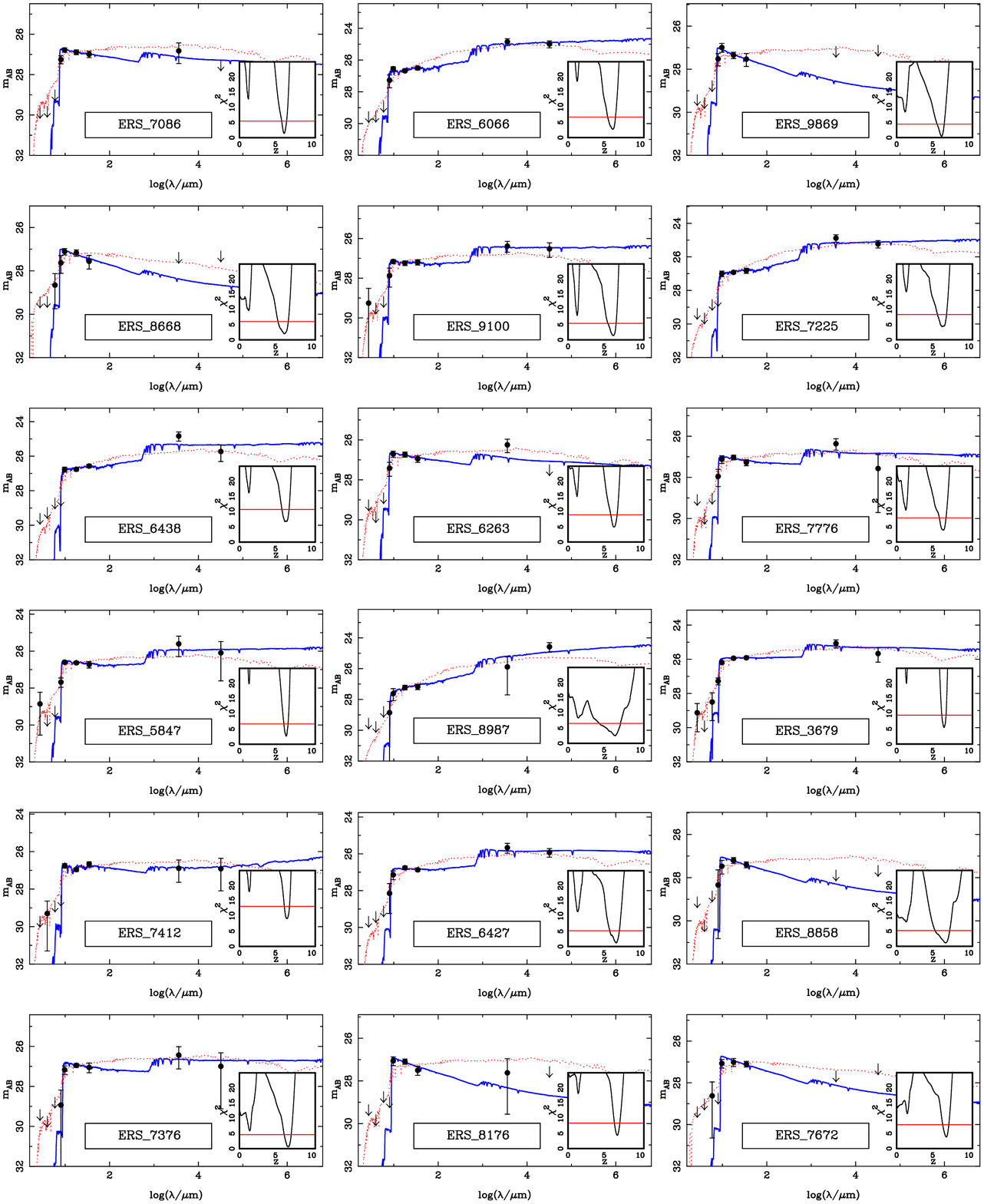,width=18.0cm,angle=0}}
\caption{SED fits for each member of the final ERS sub-sample. In
each plot the solid (blue) line is the best-fitting $z\geq 6$ galaxy
SED template, and the dotted (red) line is the best-fitting alternative low-redshift solution ($z\leq 2.5$). 
All data points which are detected at less
than $1\sigma$ significance are shown as $1\sigma$ upper limits. In
each case the inset panel shows $\chi^{2}$ versus redshift, produced
after marginalizing over all other free parameters. The 
horizontal red line in the inset is set a ($\chi^{2}_{min}+4$), and
indicates the threshold for determining whether or not the competing
low-redshift solution can be excluded at the $95\%$ confidence level.}
\end{figure*}

\setcounter{figure}{1}
\begin{figure*}
\centerline{\psfig{file=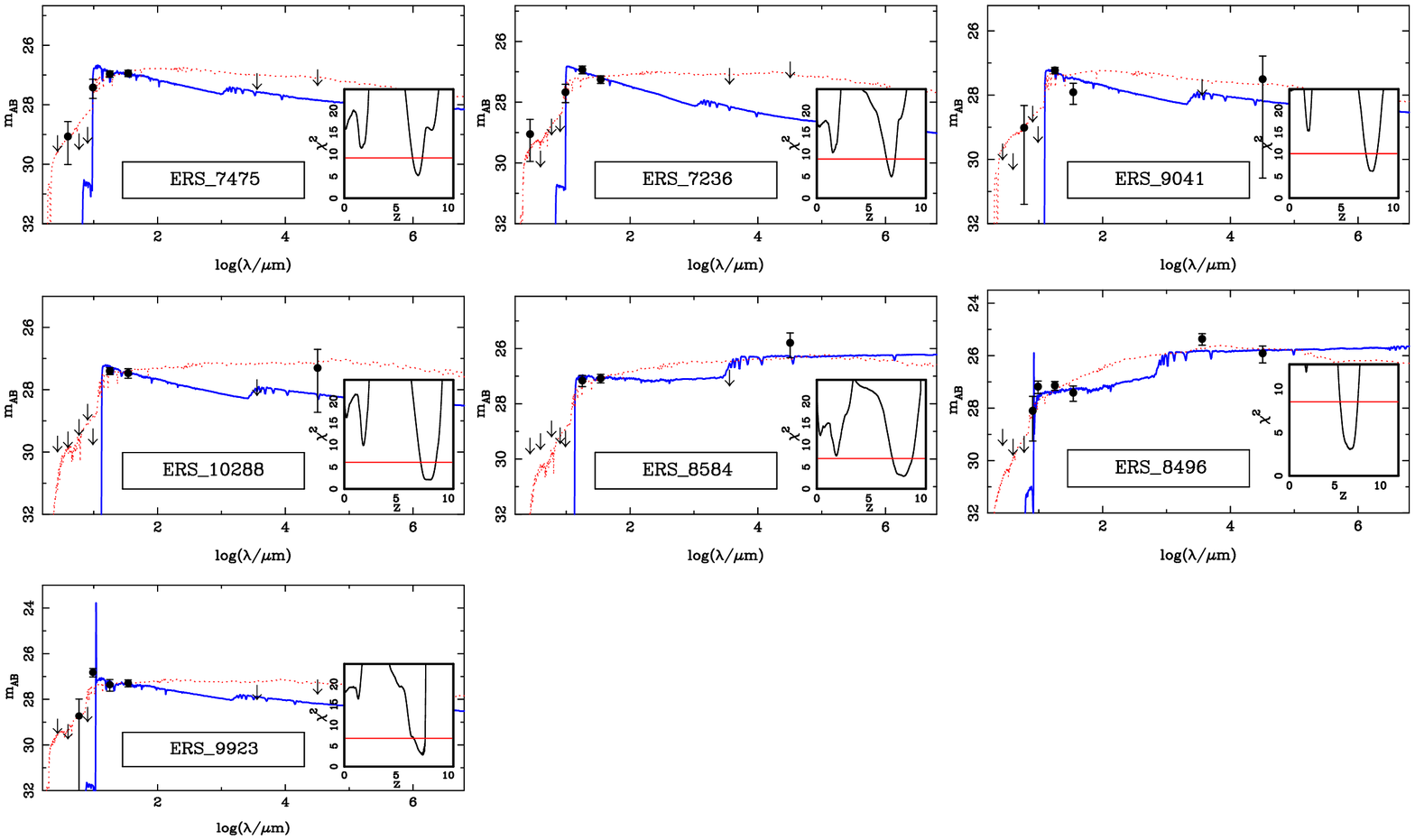,width=18.0cm,angle=0}}
\caption{Continued.}
\end{figure*}

\setcounter{figure}{2}
\begin{figure*}
\centerline{\psfig{file=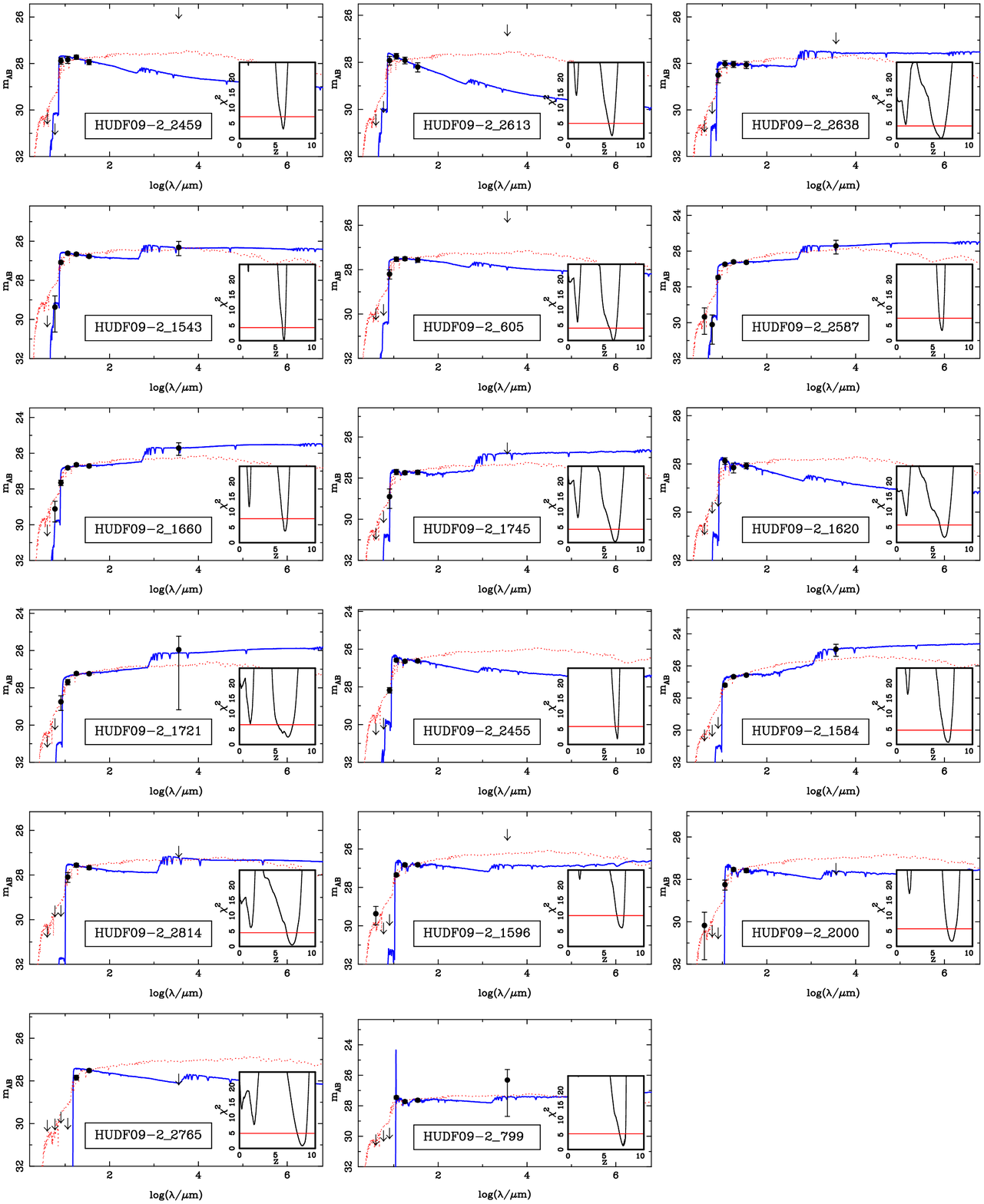,width=18.0cm,angle=0}}
\caption{SED fits for each member of the final HUDF09-2 sub-sample. In
each plot the solid (blue) line is the best-fitting $z\geq 6$ galaxy
SED template, and the dotted (red) line is the best-fitting alternative low-redshift solution ($z\leq 2.5$). 
All data points which are detected at less
than $1\sigma$ significance are shown as $1\sigma$ upper limits. In
each case the inset panel shows $\chi^{2}$ versus redshift, produced
after marginalizing over all other free parameters. The 
horizontal red line in the inset is set a ($\chi^{2}_{min}+4$), and
indicates the threshold for determining whether or not the competing
low-redshift solution can be excluded at the $95\%$ confidence level.}
\end{figure*}

\begin{figure*}
\centerline{\psfig{file=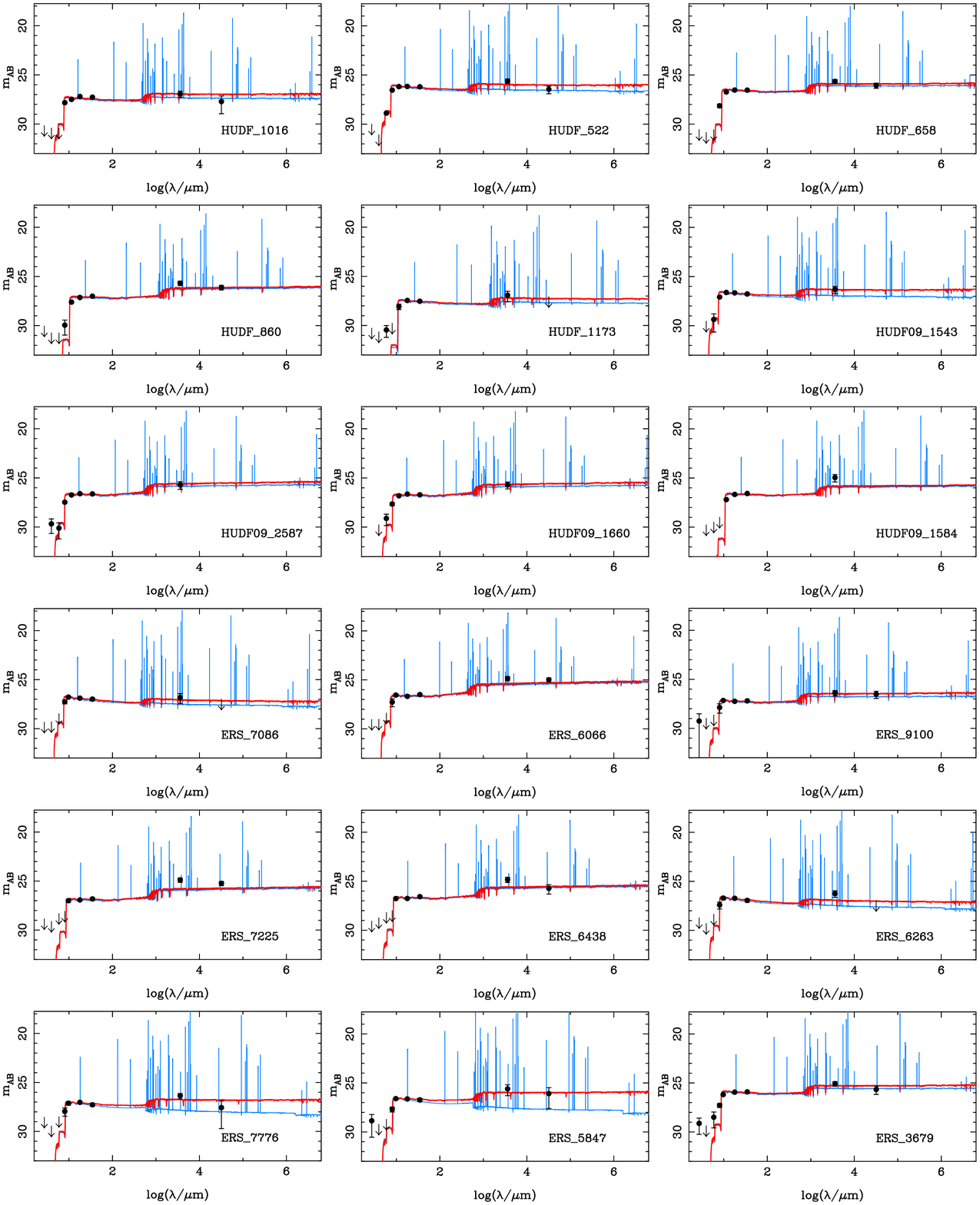,width=18.0cm,angle=0}}
\caption{SED fits featuring nebular continuum and line emission for each member of 
the twenty-one object sub-sample detected at 3.6$\mu$m. In each plot the thick red line is the 
best-fitting $\tau=0.4$ Gyr stellar population model (with no nebular emission) and the
thin blue line is the best-fitting nebular model (see Section 5.3 for a full discussion).
All data points which are detected at less than $1\sigma$ significance are shown as $1\sigma$ upper limits.}
\end{figure*}

\setcounter{figure}{3}
\begin{figure*}
\centerline{\psfig{file=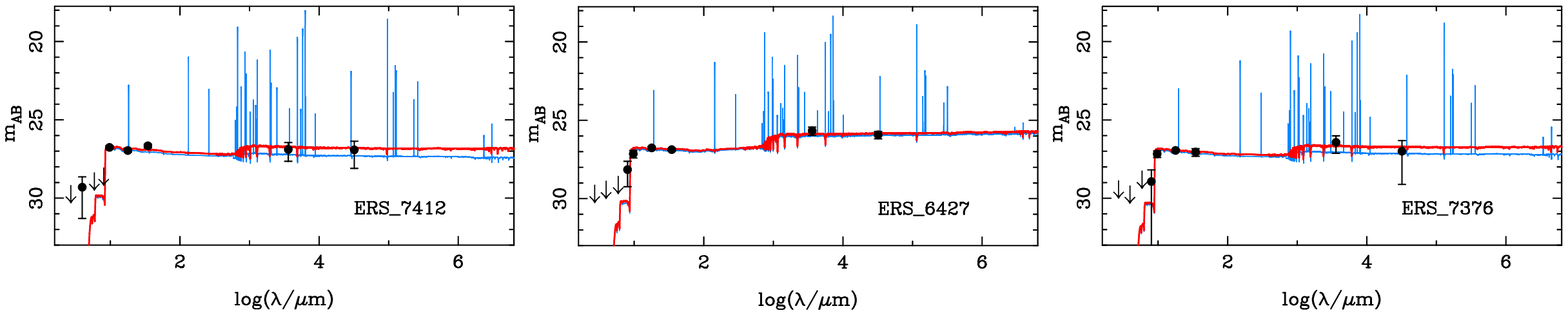,width=18.0cm,angle=0}}
\caption{Continued.}
\end{figure*}

\begin{figure*}
\centerline{\psfig{file=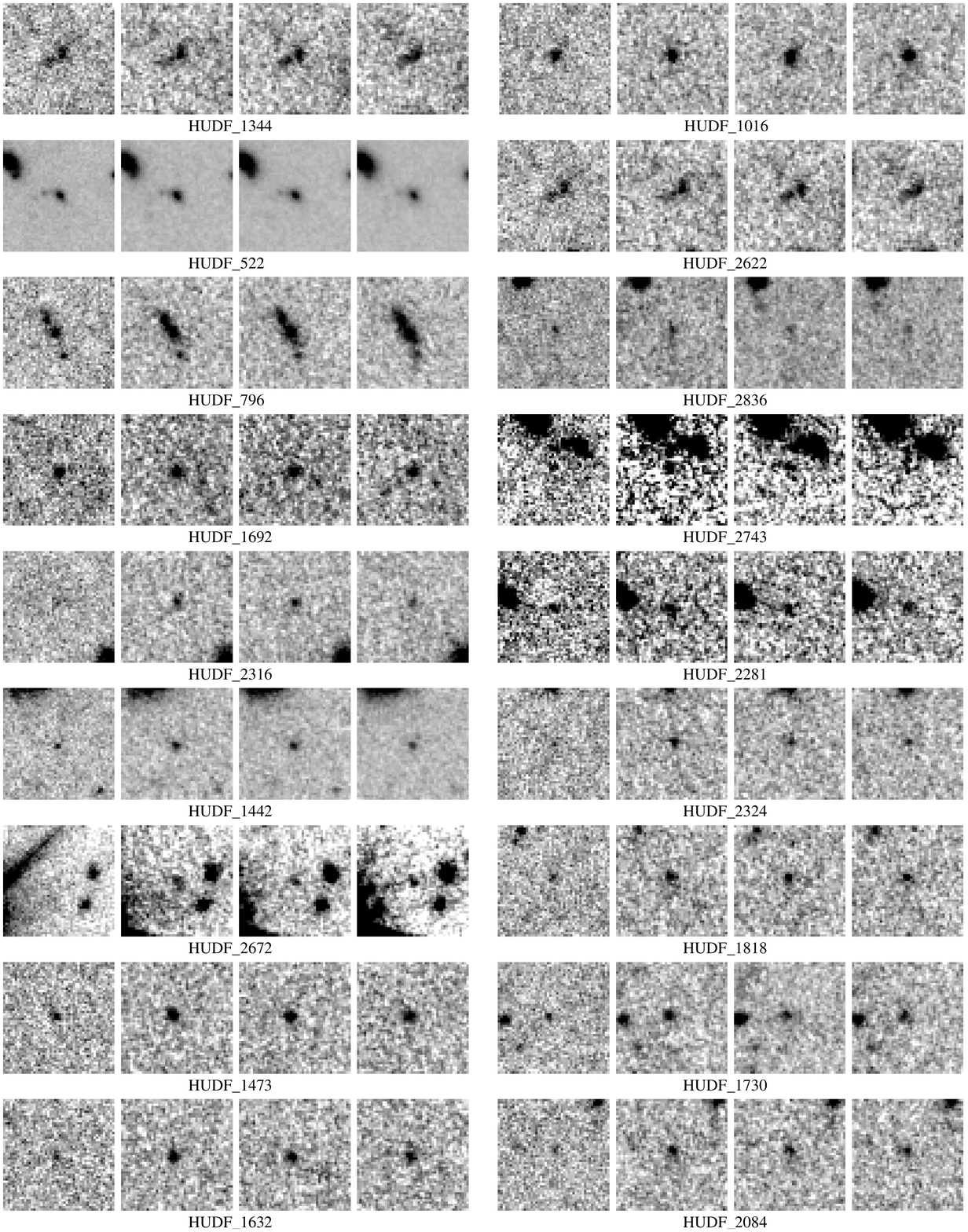,width=18.0cm,angle=0}}
\caption{3\asec$\times$ 3\asec postage-stamp images of the members of
the final HUDF sub-sample in the $z_{850}, Y_{105}, J_{125} \,\&\,
H_{160}$ filters (left to right). The postage-stamps are orientated such that North is top and East is left.}
\end{figure*}

\setcounter{figure}{4}
\begin{figure*}
\centerline{\psfig{file=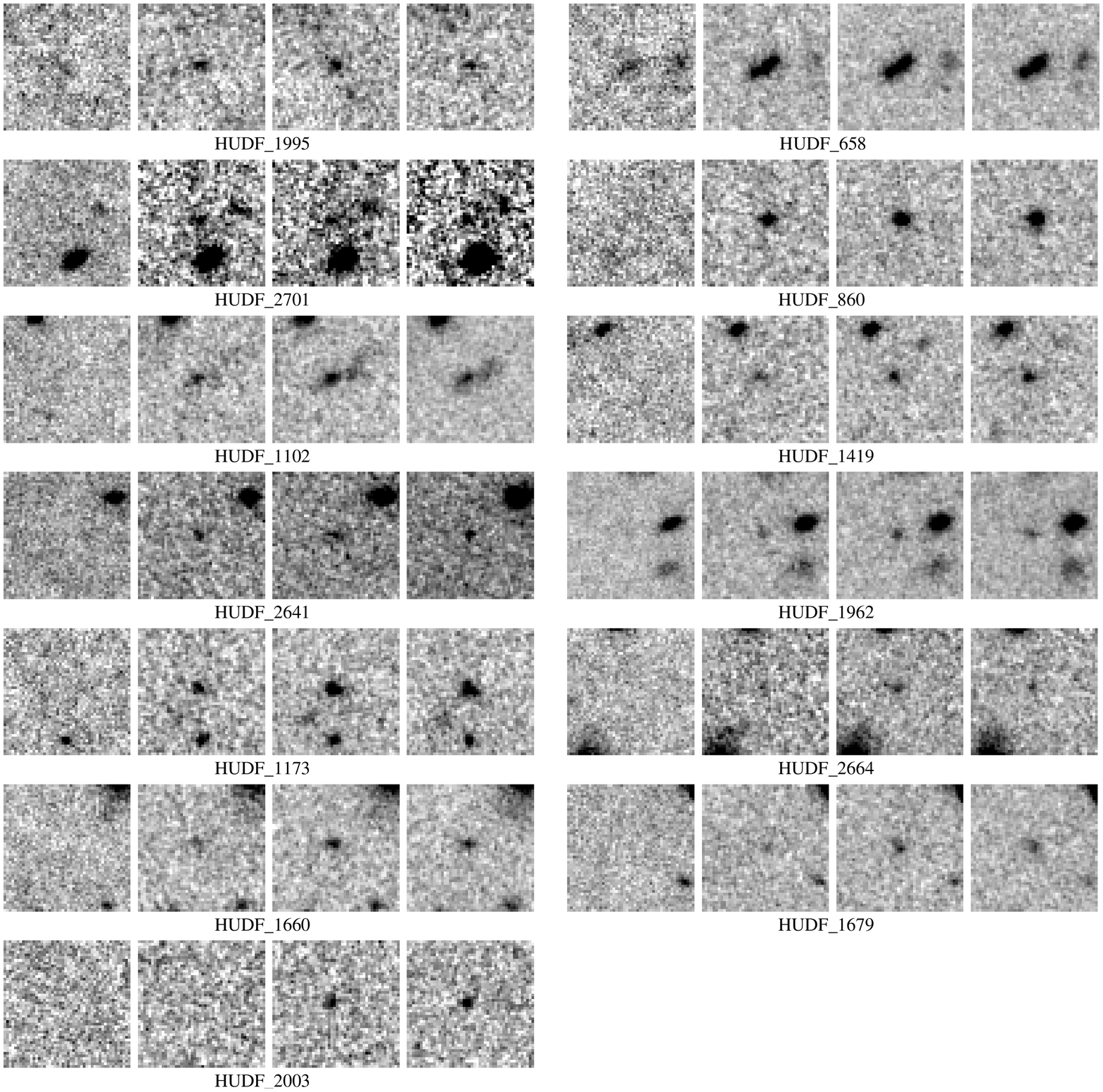,width=18.0cm,angle=0}}
\caption{Continued.}
\end{figure*}

\begin{figure*}
\centerline{\psfig{file=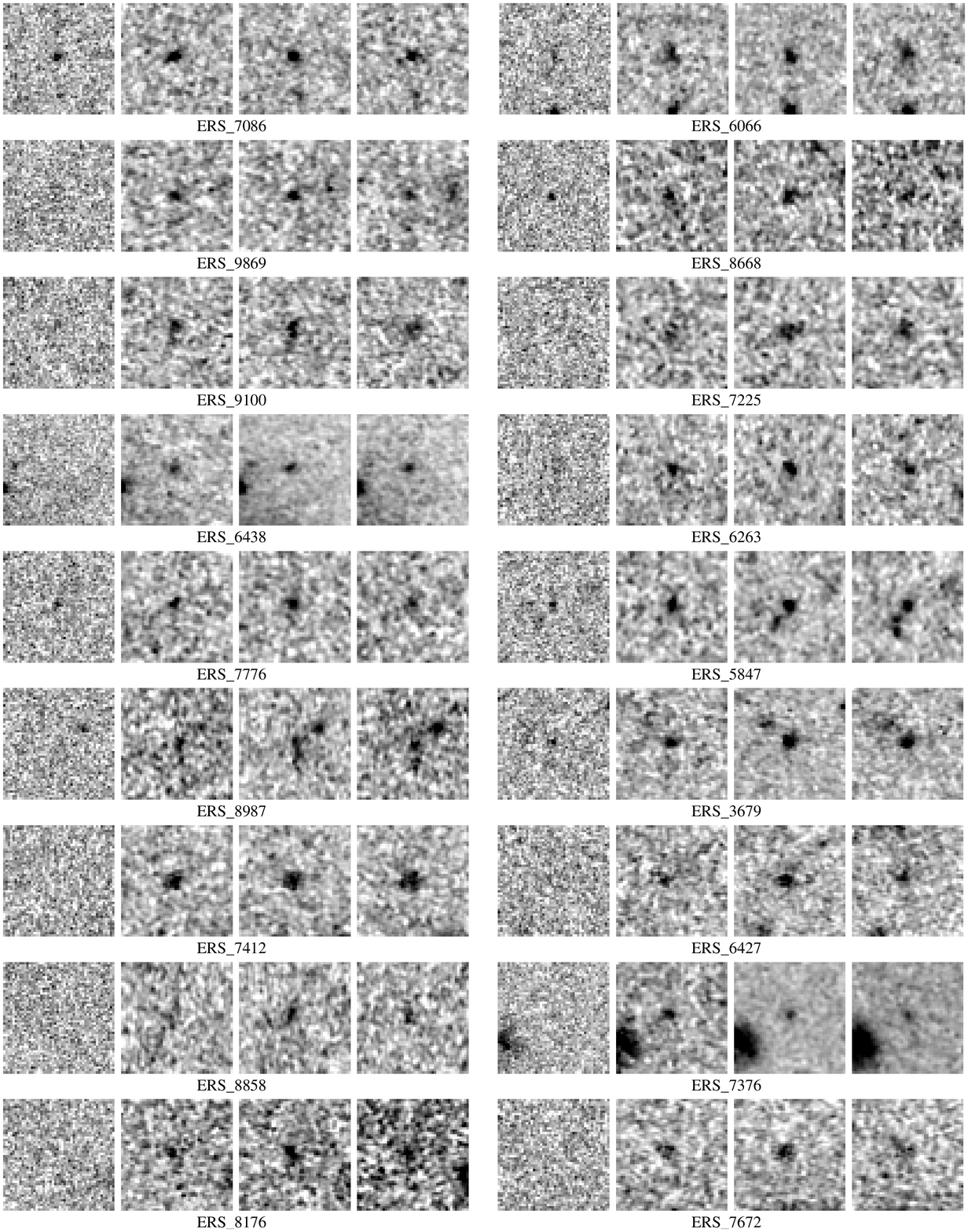,width=18.0cm,angle=0}}
\caption{3\asec$\times$ 3\asec postage-stamp images of the members of
the final ERS sub-sample in the $z_{850},
Y_{098}, J_{125} \,\&\, H_{160}$ filters (left to right). The postage-stamps are orientated such that North is top and East is left}
\end{figure*}

\setcounter{figure}{5}
\begin{figure*}
\centerline{\psfig{file=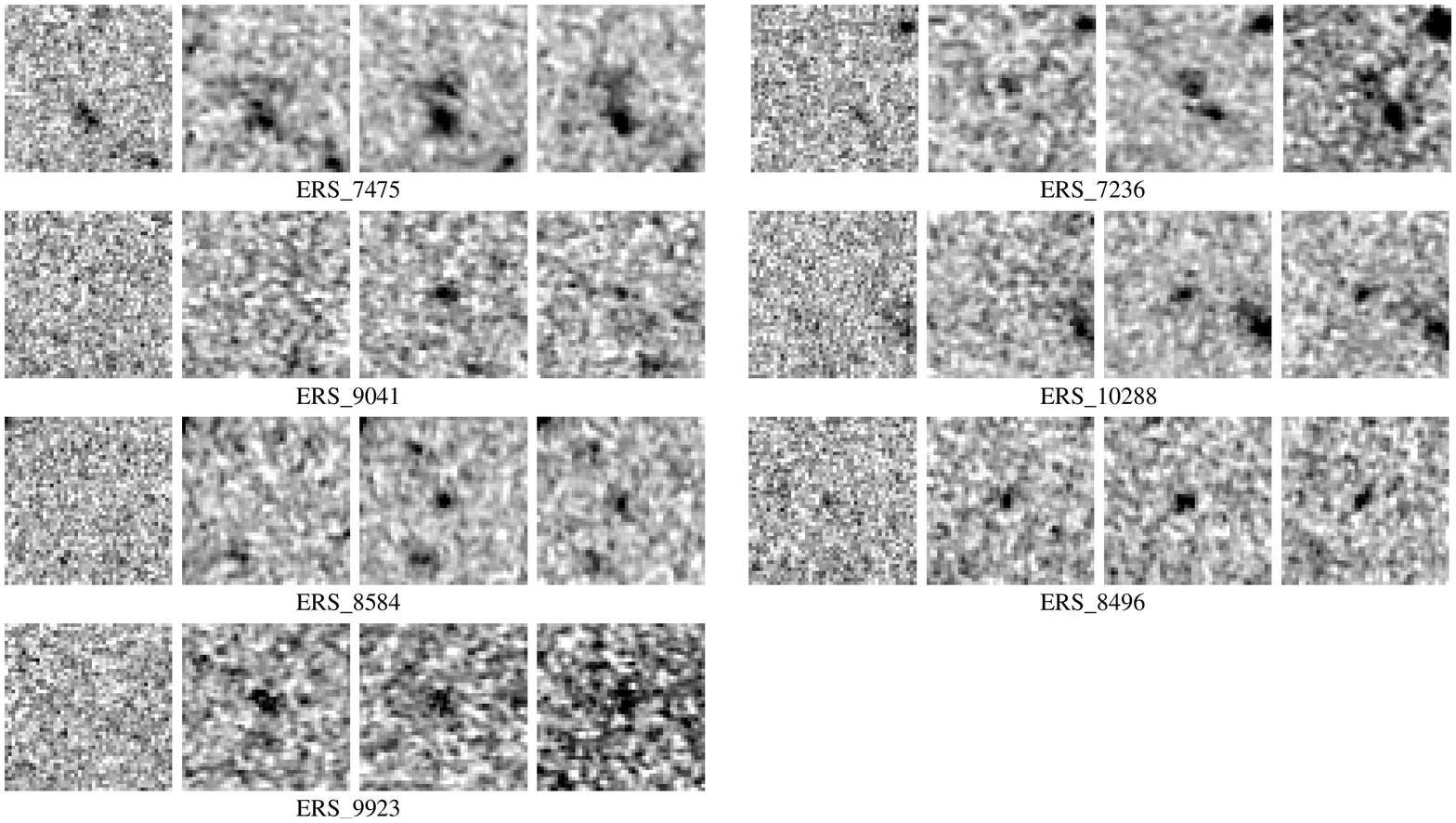,width=18.0cm,angle=0}}
\caption{Continued.}
\end{figure*}

\begin{figure*}
\centerline{\psfig{file=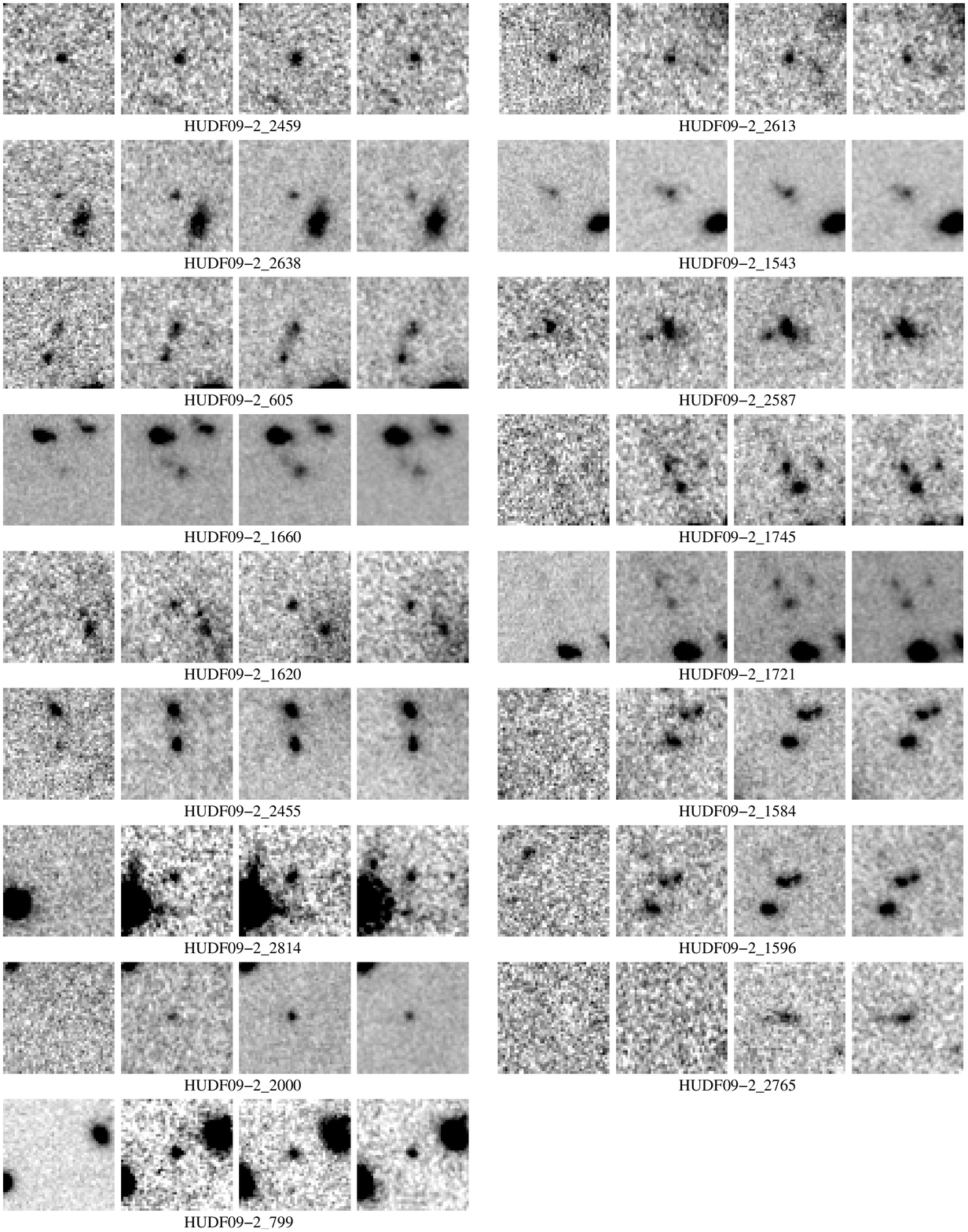,width=18.0cm,angle=0}}
\caption{3\asec$\times$ 3\asec postage-stamp images of the members of
the final HUDF09-2 sub-sample in the $z_{850},
Y_{105}, J_{125} \,\&\, H_{160}$ filters (left to right).The postage-stamps are orientated such that North is top and East is left }
\end{figure*}

\end{appendix}
\end{document}